\documentclass[a4paper,11pt]{article}

\usepackage{acronym}
\usepackage{amsfonts}
\usepackage{amsmath}
\usepackage{amsthm} 
\usepackage{amssymb}
\usepackage{bbm} 
\usepackage{bm}
\usepackage{booktabs}
\usepackage{cases}
\usepackage{color}
\usepackage{comment}
\usepackage{dcolumn} 
\usepackage{enumerate}
\usepackage{enumitem}
\usepackage{epsfig}
\usepackage{etoolbox}
\usepackage{fancyref}
\usepackage{fancyhdr}
\usepackage[draft,author=,nomargin,inline]{fixme}
\usepackage[T1]{fontenc} 
\usepackage{footnote}
\usepackage{graphicx}
\usepackage{indentfirst}
\usepackage{jcappub} 
\usepackage{latexsym}
\usepackage{lineno}
\usepackage{mathrsfs}
\usepackage{mciteplus}
\usepackage{multirow}
\usepackage{ragged2e}
\usepackage{rotating}
\usepackage{siunitx} 
\usepackage{scrextend}
\usepackage{soul}
\usepackage{subcaption} 
\usepackage{verbatim}

\DeclareMathOperator{\sech}{sech}

\makesavenoteenv{table} 
\fxusetheme{color}
\FXRegisterAuthor{ml}{aml}{Michele}
\FXRegisterAuthor{ama}{aama}{Arnau}
\FXRegisterAuthor{cfs}{acfs}{Carlos}

\DeclareSIUnit{\solarmass}{\ensuremath{\mathit{M_{\odot}}}}
\DeclareSIUnit{\parsec}{pc}

\title{Korteweg-de Vries Integrals for Modified Black Hole Potentials: Instabilities and other Questions}

\author[a,b]{Michele Lenzi}
\author[a,c]{Arnau Montava Agudo}
\author[a,b]{Carlos F. Sopuerta}

\affiliation[a]{Institut de Ci\`encies de l'Espai (ICE,~CSIC), Campus UAB, Carrer de Can Magrans~s/n, Cerdanyola del Vall\`es~08193, Spain}
\affiliation[b]{Institut d'Estudis Espacials de Catalunya (IEEC), Edifici RDIT, C/ Esteve~Terradas, 1, desp.~212, Castelldefels~08860, Spain}
\affiliation[c]{Departament de F\'{\i}sica, Universitat de les Illes Balears, IAC3–IEEC, Crta. Valldemossa km 7.5, E-07122 Palma, Spain}

\emailAdd{lenzi@ice.csic.es}
\emailAdd{montava@ice.csic.es}
\emailAdd{carlos.f.sopuerta@csic.es}

\date{\today}

\abstract{
Quasi-normal modes (QNMs) and greybody factors are some of the most characteristic features of the dynamics of black holes (BHs) and represent the basis for a number of fundamental physics tests with gravitational wave observations. It is therefore important to understand the properties of these quantities, naturally introduced within BH perturbation theory, in particular the stability properties under modifications of the BH potential. Instabilities in the QNMs have been recently shown to appear in the BH pseudospectrum under certain circumstances. In this work, we give a novel point of view based on the existence of some recently discovered hidden symmetries in BH dynamics and the associated infinite series of conserved quantities, the Korteweg-de Vries (KdV) integrals. We provide different motivations to use the KdV integrals as indicators of some crucial BH spectral properties. In particular, by studying them in different scenarios described by modified BH barriers, we find strong evidence that the KdV conserved quantities represent a useful tool to look for instabilities in the BH spectrum of QNMs and in their greybody factors.
}


\arxivnumber{2503.09918} 

\begin{document}

\maketitle

\flushbottom

\section{Introduction}
\label{S:introduction}

The Black Holes (BHs) of General Relativity (GR)~\cite{Einstein:1915by,Einstein:1915ca,Einstein:1916vd} have received a lot of attention (see, 
e.g.~\cite{Hawking:1973uf,Wald:1984cw,Novikov:1989sz,Chandrasekhar:1992bo} for general accounts) since the times when their structure and properties were clarified, essentially when the notion of BH horizon was first put on a firm ground~\cite{Finkelstein:1958zz}. BHs are very interesting and remarkable objects. First of all, they are simple as in GR they are fully characterized in terms of three quantities: mass, spin and electric charge. Another radical prediction of GR is that BHs can have any mass, or size, from the microscopic to the macroscopic world, being the only requirement for their existence to have a viable physical formation mechanism. As a consequence, they have become the target of many investigations in astrophysics, cosmology and even in fundamental physics, where they have become excellent testbeds for understanding gravity both from the theoretical and experimental points of view, and are one of the main avenues to make incursions into the land of quantum gravitational phenomena. 

Regarding observations, the advent of gravitational wave astronomy~\cite{LIGOScientific:2018mvr,LIGOScientific:2021usb,KAGRA:2021vkt} together with cutting-edge astronomical observatories covering a significant fraction of the electromagnetic spectrum, are providing us with better and better opportunities to test whether the observed compact astrophysical objects comply with the general relativistic model of a BH. In this respect, it is interesting to mention that we have already reached the point in which the most conservative description for many of these systems, as for instance Sgr* at the center of our galaxy~\cite{Schodel:2002py,Schodel:2003gy}, is that they are general relativistic BHs. Indeed, alternative descriptions often require invoking exotic matter models (see, for a review, e.g.~\cite{Cardoso:2019rvt}) for which there is currently no observational evidence. In this sense, there are great expectations for tests of the nature of BHs, both in GR and alternative theories, with the next decade gravitational-wave observatories, both third-generation ground-based detectors~\cite{Sathyaprakash:2012jk,Evans:2021gyd} and space-based detectors~\cite{LISA:2017pwj,Amaro-Seoane:2022rxf,LISACosmologyWorkingGroup:2022jok,LISA:2022kgy}. 

One of the main tools we have to study the physics of BHs is relativistic perturbation theory. Indeed, BH perturbation theory (BHPT) is the best tool we have to describe scattering processes around BHs as well as their quasinormal mode (QNM) oscillations. This is mainly due to the fact that the background solution in BHPT is an exact solution of the full non-linear Einstein's equations. 
Regarding the scattering of particles and/or fields off BHs, it is a process of great relevance for both astrophysics and fundamental physics, from where we can obtain a lot of information about the dynamics around BHs (see, e.g.~\cite{Vishveshwara:1970zz,Starobinskil:1974nkd,Sanchez:1976fcl,Sanchez:1976xm,Sanchez:1977vz,Futterman:1988ni,Andersson:2000tf,Castro:2013lba,Folacci:2019vtt}). Moreover, using BHPT in GR we can also get important insights into what is known as semiclassical gravity, where quantum matter fields are considered over curved spacetimes (see~\cite{Birrell:1982ix,Wald:1995yp,Hollands:2014eia} for general accounts). Of particular relevance are the Unruh effect~\cite{Unruh:1976db} and Hawking radiation~\cite{Hawking:1975vcx}, which are commonly used as bridges towards the study of gravitational quantum phenomena.

In this paper, we restrict ourselves to the particular case of Schwarzschild (non-rotating) BHs. The spherical symmetry of the exterior of a Schwarzschild BH allows us to expand the metric perturbation in scalar, vector, and tensor spherical harmonics. The crucial point is that the perturbative Einstein equations decouple for each harmonic mode and for each of the two possible parities. Nevertheless, the equations for each harmonic constitute a set of coupled partial differential equations (PDEs) on a two-dimensional Lorentzian manifold.  These equations can be decoupled in terms of \emph{master functions} (linear combinations of the metric perturbations and their derivatives) that satisfy 1+1 wave-like equations, the \emph{master equations}, with a potential term that describes the response of the BH to excitations~\cite{Regge:1957td,Cunningham:1978cp,Cunningham:1979px,Cunningham:1980cp,Zerilli:1970la,Zerilli:1970se,Moncrief:1974vm} (see also Ref.~\cite{Chandrasekhar:1992bo}). 

Once the master equations are solved, given that the master functions are gauge-invariant functions that describe the degrees of freedom of the perturbations, any observable can be computed in terms of them. Of special interest for gravitational wave astronomy is the computation of fluxes of energy, and linear and angular momentum (see, e.g.~\cite{Martel:2005ir}).  In the particular case of scattering processes, we can obtain the transmission and reflection probabilities from the analysis of the master equations. Some low- and high-frequency limits for the greybody factors have been obtained in the literature in~\cite{Page:1976df,Unruh:1976fm,Starobinsky:1973aij,Starobinskil:1974nkd} and in~\cite{Sanchez:1976fcl,Sanchez:1976xm,Sanchez:1977vz}, respectively. Other methods to obtain greybody factors are: WKB approximations~\cite{Schutz:1985km,Iyer:1986np,Iyer:1986nq} (see also~\cite{Konoplya:2003ii,Matyjasek:2017psv,Konoplya:2019hlu}); bounding of Bogoliubov coefficients~\cite{Visser:1998ke,Boonserm:2008zg}; monodromy techniques~\cite{Neitzke:2003mz,Castro:2013lba,Harmark:2007jy}, etc.

In this paper, we study modifications of the BH potential from a different point of view, the one we adopted recently~\cite{Lenzi:2022wjv,Lenzi:2023inn} to compute BH greybody factors in a different way.  This new approach comes from the study of the structure of the space of master functions and master equations~\cite{Lenzi:2021wpc}, and on the symmetries that one can identify in that space~\cite{Lenzi:2021njy}. It turns out that the space of master functions and equations is infinite, beyond the well-known master functions of Regge-Wheeler~\cite{Regge:1957td} (odd-parity) and Zerilli-Moncrief~\cite{Zerilli:1970la,Moncrief:1974vm} (even-parity). However, in vacuum, all the different possible master equations describe the same physics. Behind this fact there are \emph{hidden} symmetries~\cite{Lenzi:2021njy}. First, it was shown that all master equations can be connected via Darboux transformations, which means that all of them have the same spectral properties, in particular the same set of quasinormal modes and the same reflection and transmission coefficients. Second, the master equations in the frequency domain admit an infinite number of symmetries generated by the flow of the Korteweg-de Vries (KdV) equation~\cite{doi:10.1137/1018076} and the associated infinite hierarchy of non-linear PDEs. This allow us to introduce techniques from inverse scattering and integrable systems~\cite{Miura:1968JMP.....9.1202M, Gardner:1967wc, Faddeev:1976xar, Deift:1979dt, Novikov:1984id, Ablowitz:1981jq} for the study of the dynamics of perturbed BHs. The KdV symmetries lead to conservation laws, which in turn lead to an infinite set of conserved quantities~\cite{Miura:1968JMP.....9.1204M, Zakharov:1971faa, Lax:1968fm}, known as the \emph{KdV integrals}, which are integrals of differential polynomials in the BH potential. Moreover, these symmetries are isospectral deformations of the master equations that preserve both the scattering transmission coefficient and the quasi-normal mode frequencies. Finally, the two sets of symmetries, Darboux and KdV symmetries, are related by the fact that the infinite set of KdV conserved quantities are the same for all the possible BH potentials (for all master equations), and hence they are invariant under Darboux transformations.

Recent studies of the spectrum of BH QNMs that use the concept of pseudospectrum indicate that this spectrum could be unstable under small changes in the BH potential~\cite{Jaramillo:2020tuu,Jaramillo:2021tmt}, which has generated an active area of research (see, e.g.~\cite{Cheung:2021bol,Destounis:2021lum,Destounis:2023nmb,Boyanov:2023qqf,Courty:2023rxk}). These studies have carried beyond the QNM spectrum into the greybody factors and, in particular, their stability under modifications of the BH potential has been considered recently~\cite{Oshita:2022pkc,Oshita:2023cjz,Okabayashi:2024qbz,Oshita:2024fzf,Rosato:2024arw}.
Due to the relevance of these results for the understanding of BH dynamics and the implications for GW physics, in this work we have looked at these questions from a different point of view, using the machinery of hidden symmetries of BH dynamics described above. In particular, we analyze in a number of situations whether the KdV conserved quantities can be used as an indicator of the appearance of instabilities when the BH potential is modified due to different reasons. The different scenarios we have considered are: (i) The P\"oschl-Teller potential, for which the Schr\"odinger equation can be integrated in closed form, fitted to the standard BH potentials. This fit is well-known to provide a good approximation for the BH QNM modes in the eikonal limit. (ii) Astrophysically-motivated environmental corrections to the BH potential. (iii) Potentials with small-scale (oscillatory) modifications. (iv) Potentials coming from Effective Field Theory (EFT) modifications of GR. For all these cases, we analyze in details the properties of the KdV integrals as we modify the different physical parameters that characterize these modified potentials. In particular, the size of the modifications, their main location, and their width.  The outcome of this study is that KdV integrals are a very interesting tool to characterize and predict the possibility of introducing instabilities with the modification of the BH potentials. Furthermore, the importance of KdV integrals in BH spectral studies is even clearer once one realizes that the BH greybody factors are completely and uniquely determined by the KdV integrals, as shown in Refs.~\cite{Lenzi:2022wjv, Lenzi:2023inn}. This gives a solid mathematical handle to investigate the stability of greybody factors. While our study is of a phenomenological character, it motivates future more in-depth studies.

~

\noindent{\em Structure of the Paper}. 
In Section~\ref{S:master-landscape} we introduce the basic elements of BHPT that we need in this work. In particular, the structure of the space of master functions and equations (Sec.~\ref{Ss:master-equations}), 
how isospectrality arises from Darboux covariances (Sec.~\ref{Ss:Isospectrality-Darboux}), 
and the emergence of integrable structures that lead to the KdV symmetries and conserved quantities (Sec.~\ref{Ss:KdV-isospectrality}). 
In Sec.~\ref{S:GF-from-KdV} we explain how the KdV integrals completely determine the BH greybody factors via a moment problem. 
In Sec.~\ref{S:PT-fit}, we fit the BH potentials (here we consider both the Regge-Wheeler and Zerilli potentials for odd- and even-parity perturbations, respectively) to the P\"oschl-Teller potential, as an approximation to the \emph{exact} potentials (Sec.~\ref{Ss:Fit-to-PT}). 
Then, we study some criteria, based on the KdV integrals, to characterize how good as an approximation are these fits (Sec.~\ref{Ss:KdV-fit}). 
Finally, we compare the previous criteria with the computation of QNMs both for the fits and comparing with the actual values (Sec.~\ref{Ss:PT-fit-QNM}).  
In Sec.~\ref{S:KdV-integrals-modified-potentials} we compute the KdV integrals for modified BH potentials with the aim of studying the connection between stability under these modifications and the behaviour of the KdV integrals. We consider different classes of BH potentials: 
Potentials that represent (astrophysical) environmental modifications (Sec.~\ref{Ss:PT-bump});
phenomenological oscillatory potentials (Sec.~\ref{Ss:oscillatory-potentials}); 
and potentials coming from EFT modifications of GR (Sec.~\ref{Ss:EFT}). 
In Sec.~\ref{S:GF-stability} we discuss the question of stability in terms of the greybody factors. 
Sec.~\ref{S:Conclusions-and-Discussion} is devoted to conclusions and discussions of the future prospects of this work.  
We have some appendices where we include some details of the scattering problem for the P\"oschl-Teller potential, for which the master equation in frequency domain is exactly solvable (Appendix~\ref{App:PT}). 
In Appendix~\ref{App:KdV-PT} we provide details for the derivation of the general formula for KdV integrals for the P\"oschl-Teller potential shown in Sec.~\ref{Ss:KdV-fit}. 
In Appendix~\ref{App:KdV-densities} we present general and simplied expressions for the KdV densities by using essentially the integration by parts method. 
Finally, in Appendix~\ref{App:EFT-coefficients}, we include the values of the coefficients that determine the BH potentials from the EFTs of Sec.~\ref{Ss:EFT}.
Throughout this paper, otherwise stated, we use geometric units in which $G = c = 1\,$.

\section{Perturbed BHs: Master Equations and Hidden Symmetries}
\label{S:master-landscape}

The theory of BH perturbations offers a simplified framework tailored to describe various physical processes around BHs, ranging from the study of quasi-normal oscillations of BHs to the scattering of test fields and particles, of any spin ($s=0,1/2,1,2$) by a BH. Despite the interactions are linearized over a fixed background, the fact that the background itself is a solution of the full nonlinear Einstein equations leads to nontrivial results, sometimes going beyond its regime of application. Some examples are: 1) the existence of new hidden symmetries associated to integrable systems~\cite{Lenzi:2021njy,Lenzi:2022wjv, Jaramillo:2024nvr}; 2) the fact that some of the results provided by perturbation theory are, in certain situations, valid beyond the regime of validity of the approximation, as for instance, in applications of the close limit approximation~\cite{Price:1994pm,Pullin:1999rg,Sopuerta:2006wj,Sopuerta:2006et}.

In the case of a spherically symmetric BH, the (static) background is the Schwarzschild metric
\begin{equation}
d{s}{}^2= {g}^{}_{\mu\nu}dx^{\mu}dx^{\nu}=-f(r)\,dt^2+\frac{dr^2}{f(r)}+r^2d\Omega^2\,,
\quad
f(r) = 1 - \frac{r^{}_{s}}{r}\,, 
\label{schwarzschild-metric}
\end{equation}
where $r_s$ is the Schwarzschild radius, $r_{s} = 2GM/c^{2}=2M\,$, providing the location of the event horizon. One of the main advantages of BH perturbation theory resides in the possibility of decoupling the linearized Einstein equations for the perturbations into an infinite set of master equations. Exploiting the spherical symmetry of the background, one can decompose every perturbative quantity in scalar, vector and tensor harmonics\footnote{These terms may have different meaning depending on the context, e.g. when studying perturbations in higher dimensions~\cite{Kodama:2003jz}.} so that the Einstein equations decouple for each harmonic $(\ell,m)$ and parity\footnote{Under parity transformation $(\theta,\phi)$ $\rightarrow$ $(\pi-\theta, \phi+\pi)$ (i.e. spatial inversion), an even-parity harmonic ${\cal O}^{\ell m}$ transforms as ${\cal O}^{\ell m} \rightarrow (-1)^{\ell}{\cal O}^{\ell m}$, while an odd-parity one transforms as ${\cal O}^{\ell m} \rightarrow (-1)^{\ell+1}{\cal O}^{\ell m}$.}. Manipulations of such equations can lead to further simplification (thanks also to variable separation), so that the Einstein equations for the perturbations, for each harmonic and parity, reduce to a simple {\em master equations} (see, e.g.~\cite{Martel:2005ir,Nagar:2005ea,Lenzi:2021wpc}):
\begin{equation}
\left(-\frac{\partial^2}{\partial t^2} + \frac{\partial^2}{\partial x^2} - V^{\rm even/odd}_\ell  \right)\Psi^{\rm even/odd}_{\ell m} = 0\,.
\label{master-wave-equation}
\end{equation}
Here, $x$ is the {\em tortoise} coordinate, defined by $dx/dr = 1/f(r)$; $V^{\rm even/odd}_\ell(r)$ is a potential (barrier) which only depends on the Schwarzschild background curvature, the harmonic number $\ell$ and the mode parity; the functions $\Psi^{\rm even/odd}_{\ell m}$ are the so called \emph{master functions}, i.e. linear combinations of the metric perturbations and their first-order derivatives, that decouple the system of linearized Einstein PDEs for the perturbations\footnote{In what follows, for the sake of simplicity, we drop the harmonic numbers $(\ell,m)$ from the master functions.}.

\subsection{Infinite Hierarchy of Master Equations}
\label{Ss:master-equations}

With the purpose of performing the decoupling into master equations in a general and systematic way, in Ref.~\cite{Lenzi:2021wpc} all the possible master equations with master functions linear in the metric perturbations and their first-order derivatives were determined (an approach to this problem from curvature wave equations has recently appeared~\cite{Mukkamala:2024dxf}).
The outcome is an infinite hierarchy of master equations which were organized into two branches: the {\it standard branch} and the {\it Darboux branch}. Here, we will only give explicit expressions for the potentials, while we reserve other details to Refs.~\cite{Lenzi:2021wpc, Lenzi:2024tgk}. The standard branch is the sector where we find the known potentials of the Schwarzschild BH, namely the Regge-Wheeler potential~\cite{Regge:1957td} for odd-parity perturbations
\begin{equation}
V_{\ell}^{\mathrm{RW}}(r) = \left(1-\frac{r^{}_s}{r}\right)  \left(\frac{\ell(\ell+1)}{r^{2}} - \frac{3r^{}_{s}}{r^{3}} \right)  \,,
\label{RW-potential}
\end{equation}
and the Zerilli potential~\cite{Zerilli:1970la} for even-parity perturbations 
\begin{eqnarray}
V^{\rm Z}_\ell(r) & = & \frac{f}{\lambda^{2}}\left[ \frac{(\ell-1)^{2}(\ell+2)^{2}}{r^{2}}\left( \ell(\ell+1) + \frac{3r^{}_{s}}{r} \right)  +  \frac{9r^{2}_{s}}{r^{4}}\left( (\ell-1)(\ell+2) + \frac{r^{}_{s}}{r} \right) \right] \,,
\label{Z-potential}
\end{eqnarray}
where $\lambda$ is a function of $r$ given by
\begin{eqnarray}
\lambda(r) =  (\ell-1)(\ell+2) + \frac{3r^{}_{s}}{r}\,.
\label{lambda-def-sch}
\end{eqnarray}

It turns out that these master equations can be extended for the case of test fields of diverse spin that propagate on the exterior of the Schwarzschild BH background. This is frequently used to study small perturbations generated by these fields whose own gravitational field is neglected to some level of approximation. 
In this way, one can study, in particular,  the dynamics of scalar (spin zero) and electromagnetic fields (spin 1)  on the geometry of a Schwarzschild BH (see, e.g.~\cite{Chandrasekhar:1992bo}). This results in master equations where the potential is a slight modification of the Regge-Wheeler potential for gravitational odd-parity perturbations, which depend on the spin of the field. The generalized expression for the potential in the master equations for spin values $s =0,1,2$ is:
\begin{equation}
V^{}_{\mathrm{RW}}(r) = \left( 1- \frac{r^{}_s}{r} \right) \left(\frac{\ell (\ell +1)}{r^2} + (1-s^2)\frac{r^{}_s}{r^3} \right) \,.
\label{potential-scalar-vector-grav}
\end{equation}
In this sense, it is worth pointing out that it is not always possible to find a consistent description of the dynamics for a field of a given spin on the Schwarzschild BH geometry, as pointed out by Price in~\cite{Price:1972pw}. Indeed, while in flat spacetime there is a systematic procedure to build field equations for fields of an arbitrary spin, there is no a unique way of doing the same in curved spacetime. The most common approach is to assume the flat spacetime equations are valid locally and then, exchange partial derivatives with covariant derivatives, the so-called \emph{minimal coupling} prescription. However, as Price shows in~\cite{Price:1972pw}, only the resulting equations for spins $s=0,1$ have a consistent dynamics. The case $s=2$ comes from the perturbative Einstein equations in the odd-parity case. 
The case of spin $1/2$ fields, following the Dirac equation in curved spacetime, was first studied in~\cite{Brill:1957fx} and later in~\cite{Page:1976df,Unruh:1976fm} (see also~\cite{Chandrasekhar:1992bo,Berti:2009kk}). It turns out that the master equations that appear in this case have potentials with a form different from the Regge-Wheeler potential in Eq.~\eqref{potential-scalar-vector-grav}. There are also studies of the Rarita-Schwinger equation ($s=3/2$) on algebraically-special spacetimes, including the Schwarzschild spacetime~\cite{Kamran:1985tp,TorresDelCastillo:1989gf,TorresDelCastillo:1989hk}. 

Going back to the question of the master functions, in Ref.~\cite{Lenzi:2021wpc} it was shown that the most general odd-parity master function is a linear combination of the Regge-Wheeler~\cite{Regge:1957td} and the Cunningham-Price-Moncrief~\cite{Cunningham:1978cp,Cunningham:1979px,Cunningham:1980cp} master functions, $\Psi_{\mathrm{RW}}$ and $\Psi_{\mathrm{CPM}}$ respectively. For the even-parity sector, it was found that the most general master function is a linear combination of the Zerilli-Moncrief master function~\cite{Zerilli:1970la,Moncrief:1974vm}, $\Psi_{\mathrm{ZM}}$, and another master function that appears to be new~\cite{Lenzi:2021wpc}.

On the other hand, the Darboux branch described in~\cite{Lenzi:2021wpc} contains an infinite class of master equations of the form of Eq.~\eqref{master-wave-equation}. This is a consequence of the fact that the analysis does not fix one single potential for each parity but it just selects an infinite family of possible potentials, determined by a non-linear ordinary differential equation. As a consequence, the master functions in this branch depend on the potential. Since we only focus on the underlying hidden symmetries associated to this hierarchy, and we do not need the explicit expressions for these master functions, we refer to~\cite{Lenzi:2021wpc} for details (see also~\cite{Lenzi:2024tgk}).

\subsection{Isospectrality of the Master Equations: Darboux covariance}
\label{Ss:Isospectrality-Darboux}

The structure of this infinite landscape of master equations and functions was investigated in~\cite{Lenzi:2021njy}. It was shown that all the pairs $(V,\Psi)$ are connected by Darboux transformations (DTs). A DT connecting two pairs, $(\Phi,v)$ and $(\Psi,V)$, is defined in a general sense as follows:
\begin{eqnarray}
\left( -\partial^2_t + \partial^2_x   - v \right) \Phi = 0
~\longrightarrow~
\left\{
\begin{array}{l}
\Psi =  \Phi^{}_{,x}  + W\,\Phi 
\\[2mm]
V = v + 2\,W^{}_{,x}
\\[2mm]
W^{}_{,x}- W^2 + v = {\mathcal C} 
\end{array}
\right.
\!\!
\longrightarrow
~
\left( -\partial^2_t + 
\partial^2_x   - V \right) \Psi = 0\,,
\label{DT}
\end{eqnarray}
where  $W$ is the DT generating function, which satifies a Ricatti equation, and ${\mathcal C} $ is an arbitrary constant. Therefore, we have: i) a standard branch odd sector; ii) a standard branch even sector; iii) a Darboux branch odd sector (containing infinite master equations) and iv) a Darboux branch even sector (again containing infinite master equations). All these sectors are connected between them by DTs of the form given in Eq.~\eqref{DT}. 
From a physical perspective, the consequence is that each master equation in the infinite hierarchy found in Ref.~\cite{Lenzi:2021wpc} describes the same physics. Indeed, DTs are isospectral, which means that they preserve the spectrum of the frequency domain operator associated with the wave-like master equation, as well as the reflection and transmission coefficients~\cite{Glampedakis:2017rar,Lenzi:2021njy}. This becomes much clearer by considering single frequency solutions, i.e. 
\begin{equation}
\Psi(t,r) = e^{ik t}\,\psi(x;k)\,,    
\end{equation}
so that the master equation~\eqref{master-wave-equation} becomes a time-independent Schr\"odinger equation of the form
\begin{equation}
\psi^{}_{,xx} - V\psi = -k^2 \psi \,.
\label{schrodinger}
\end{equation}
Then, as shown in~\cite{Lenzi:2021njy}, DTs map a time-independent Schr\"odinger equation to a physically equivalent one, even if potential barrier and master function may appear very different. Isospectrality of the Regge-Wheeler and Zerilli potentials, already known to Chandrasekhar~\cite{1980RSPSA.369..425C}, is part of this bigger structure and it is explained in terms of the DT between the standard branch odd and even sectors (see also~\cite{Glampedakis:2017rar}). Indeed, starting from the so called \textit{algebraically special} solution~\cite{Couch:1973zc,Chandrasekhar:1984:10.2307/2397739}, which we denote by $\psi_0$, one can define the DT generating function
\begin{eqnarray}
W^{}_{0}(r) &=& -(\ln{\psi^{}_0})^{}_{,x}
=
i k^{}_0 + \frac{3\,r^{}_s\, f(r)}{\lambda(r)r^2} \,,
\label{generating-function-RWZ}
\end{eqnarray}
which automatically satisfies the Riccati equation in Eq.~\eqref{DT} with $\mathcal{C} = k^{2}_{0}$, where the specific value of the frequency is
\begin{equation}
k^{}_0 = -i\frac{(\ell+2)(\ell+1)\ell(\ell-1)}{6\,r^{}_s}\,.
\label{algebraically-special-frequency}
\end{equation}
This frequency is the eigenvalue associated to $\psi_0\,$.
Then, it is quite easy to check that the Regge-Wheeler and Zerilli potentials, despite being apparently very different, they just  differ by a total derivative term, i.e.
\begin{equation}
V^{\mathrm{RW}}_{\mathrm{Z}}
= V^{\mathrm{Z}}_{\mathrm{RW}} \mp 2 W^{}_{0,x} =  \mp W^{}_{0,x}  + W_{0}^2 + k_0^2\,.
\label{RWZ-superpotential}
\end{equation}
%

\subsection{Integrable structures: KdV isospectral symmetry}
\label{Ss:KdV-isospectrality}

So far, we have briefly described how to simplify and separate, through symmetry reduction, the perturbative Einstein equations into infinite physically equivalent master equations (wave-like equations with a potential term) for a master function. The potential is determined by the BH geometry and describes its response to external perturbations, while the master functions describes the (linearized) gravitational radiative degrees of freedom. We have reviewed the connection between the infinite hierarchy of master equations and the presence of a hidden symmetry in the BH perturbative dynamics, Darboux covariance~\cite{Lenzi:2021njy}. However, one can pull the strings even further by noticing that the DT is strictly related to the construction of soliton solutions of linear and nonlinear partial differential equations~\cite{Matveev:1991ms}. In particular, it provides a surprising connection between our spectral problem in Eq.~\eqref{schrodinger} and the Korteweg-de Vries (KdV) equation and the associated infinite hierarchy of KdV equations. This connection introduces another type of isospectral symmetry for the perturbed BH dynamics. Indeed, if we deform the frequency-domain master equations~\eqref{schrodinger} by introducing a parameter $\tau$, i.e.
\begin{equation}
\psi(x)\rightarrow\psi(\tau,x)\,,\;\;
V(x)\rightarrow V(\tau,x)\,,\;\;
k\rightarrow k(\tau)\,,
\end{equation}
in such a way that the potential $V(\tau,x)$ follows the KdV equation with respect to $\tau$
\begin{equation}
V^{}_{,\tau} - 6 VV^{}_{,x} + V^{}_{,xxx}=0 \,,
\label{kdv-equation}
\end{equation}
we find that the spectrum is conserved~\cite{Lenzi:2021njy} (see also~\cite{Lenzi:2022wjv}) in the sense that 
\begin{equation}
(k^2)_{,\tau}=0\,.    
\end{equation}
This result holds for the continuous spectrum, for bound-states (discrete spectrum) and resonances (or QNMs). Although in the case of Schwarzschild BHs, where the potential is positive everywhere in the BH exterior, there is no discrete spectrum.

Isospectrality of the master equations points once more to a hidden symmetry in our system. Such symmetry is deeply related to the integrable structure of the KdV equation and its associated conserved quantities. The study of the complete integrability of the KdV equation, through the inverse scattering transform~\cite{Gardner:1967wc}, paved the way to a whole new field of research, the one of integrable systems and solitons. As we already mentioned, integrability usually comes associated with an infinite number of first integrals for the dynamics of the system. Among the numerous ways to find such conserved quantities for the KdV equation, the one which is probably more intuitive in physics is the one that shows that one can find action-angle variables for the KdV equation and cast it in Hamiltonian form~\cite{Zakharov:1971faa} (see also Refs.~\cite{gardner1971korteweg,FadeevTakhtajan:1987lda} or for a more recent account Refs.~\cite{Lenzi:2022wjv, Jaramillo:2024nvr}). The underlying idea is that one can generate an infinite hierarchy of (higher-order) KdV equations such that each of them can be cast in Hamiltonian form.  This shows that each KdV equation of the KdV hierarchy constitutes a symmetry for the other ones. Therefore, this introduces infinite conservation laws whose associated densities can be obtained from the following recurrence relation
\begin{eqnarray}
\kappa^{}_1(x) & = & V(x)\,, \\
\kappa^{}_n(x) & = & - \frac{d}{dx} \kappa^{}_{n-1}(x) - \sum_{k=1}^{n-1} \kappa^{}_{n-k-1}(x) \kappa^{}_k(x)\,, 
\quad (n = 2, \ldots ) \,,
\label{kdv-densities-recurrence}
\end{eqnarray}
where clearly the densities $\kappa_n(x)$ are nonlinear differential polynomials in the potential $V$ (that is, nonlinear polynomials in $V$ and its derivatives).
Then, the conserved quantities, the KdV integrals, read
\begin{equation}
\mathcal{K}^{}_n  = \int^{\infty}_{-\infty} dx\, \kappa^{}_n(x) \,. 
\label{kdv-integrals}
\end{equation}

In summary, the KdV deformations constitute an infinite set symmetries of the time-independent Schr\"odinger equation~\eqref{schrodinger}, the time-independent master equation.  The associated conserved quantities, the KdV integrals, are functionals of the BH potential that appears in the master equation.  This structure is much richer in the GR description of BHs and their perturbations. In fact, as we have seen in the previous sections, the perturbative Einstein equations can be reduced to infinite different master equations, that is, to master equations with an infinite number of different potentials, apart from the standard Regge-Wheeler~\eqref{RW-potential} and Zerilli~\eqref{Z-potential} ones, and all these potentials can be connected by DTs. Therefore, the picture up to now contains, on the one hand, a set of infinite potentials related by DTs and, on the other hand, a continuum of infinite potentials obtained by parametric deformations following the flow of the KdV equation. While it is obvious that the KdV deformed potentials have the same integrals as the initial ones (by construction), it is less obvious that the value of the integrals for DT related potentials is equal to the values for the starting potential, say for instance the Regge-Wheeler potential. However, in Ref.~\cite{Lenzi:2021njy} (see also~\cite{Lenzi:2022wjv} for more details) it was found that the KdV integrals for any allowed physically equivalent BH potential (starting from the Regge-Wheeler one, either a Darboux related or KdV deformed potential), are the same\footnote{Indeed, that the first few KdV integrals where equal, just for the Regge-Wheeler and Zerilli potentials, has been first noticed by Chandrasekhar in Refs.~\cite{1980RSPSA.369..425C,Chandrasekhar:1992bo}}. 

At this point, let us clarify some subtle points regarding the domain of applicability of these hidden symmetries. While the Darboux covariance between odd and even parity modes, tightly related to the algebraically special mode, is a special feature of GR in spacetime dimension\footnote{Actually, this isospectrality is generally broken in most modifications of the gravitational interactions or in $d>4$ (see, e.g.~\cite{Chen:2021pxd,del-Corral:2022kbk,Moura:2022gqm,Zhao:2023jiz,Li:2023ulk,Cano:2024wzo}).}  $d=4$, the Darboux covariance within each parity sector is expected to be more general. The same holds in general for the interplay with the KdV integrable structure, highlighting that GR in $d=4$ has a rich and special structure whose breaking in alternative descriptions can still provide useful analytical handles on possible modifications of the theory. The main ingredient required for what we are going to show in this paper is that the chosen theory admits a perturbative scheme that allows for a reduction of the perturbative equations to master wave-like equations [or to Schr\"odinger equations in the frequency domain~\eqref{schrodinger}]. 

As a final comment, let us mention that the KdV equation has a much richer structure when one realizes that it actually possesses a bi-Hamiltonian description~\cite{Magri:1977gn}, as most integrable systems. It was shown that such second Hamiltonian structure is associated to the semiclassical realization of the Virasoro algebra~\cite{Gervais:1985fc, GERVAIS1985279}. Consequences of this algebraic structure have been analyzed in the context of BH perturbation theory in Refs.~\cite{Lenzi:2022wjv, Jaramillo:2024nvr, Jaramillo:2024qjz}.

\section{BH Greybody Factors and Korteweg-de Vries Integrals}
\label{S:GF-from-KdV}

Integrable systems have many facets as it is a subject that has been developed through different perspectives in mathematics and mathematical physics. The KdV system is no exception and, as we briefly mentioned before, we can investigate this integrable system in an analytic setting or in a group theoretic/algebraic one, with the first associated to KdV deformations while the latter providing the connection with the Virasoro algebra. In Ref.~\cite{Lenzi:2021njy}, we first explored the analytic part of these integrable structures in BHPT. Within the context of perturbed BHs, one has access to several physical systems whose dynamical behavior is correctly described by perturbations over an idealized BH background. Some notable examples are the scattering of matter fields off BHs, the ringdown signal and QNMs oscillations, as well as extreme-mass-ratio inspirals (EMRIs) (see, e.g.~\cite{Berry:2019wgg,Cardenas-Avendano:2024mqp}) and the close-limit approximation~\cite{Price:1994pm,Pullin:1999rg,Sopuerta:2006wj,Sopuerta:2006et}. A fascinating question to explore is whether these integrable structures are connected to any of these physical scenarios, and if so, in what way. This is what motivated the work in Ref.~\cite{Lenzi:2022wjv} where, in the context of BH scattering, it was found that the KdV integrals determine completely the BH greybody factors. This has been referred to as the {\it BH moment problem}: given a BH potential barrier, to find the BH greybody factors from the KdV integrals.

In this context, let us consider the typical scattering through a (BH) potential barrier, i.e. a wave coming from $x\rightarrow-\infty$ (the BH horizon) which scatters through the BH barrier. In Ref.~\cite{Lenzi:2022wjv}, the BH barrier was considered to be any of the KdV deformed or Darboux transformed barriers and it was shown that the greybody factors are equal for each of them, as it happens for the QNM spectra (see Sec.~\ref{Ss:KdV-isospectrality} and~\cite{Lenzi:2021njy}). These spectral results, together with the fact that the infinite hierarchy of KdV integrals is equal for every physically equivalent BH barrier, strongly suggests a deep connection between KdV integrals and spectral quantities. Since we know that all the possible master equations are physically equivalent, we can just consider here the RW potential~\eqref{RW-potential}. The scattering set up, in mathematical terms, can be expressed in the following way:
\begin{eqnarray}
\label{plane-wave-ab}
\psi(x,k) = \left\{ \begin{array}{lc}
a(k) e^{i k x} + b(k)e^{-i k x}  &  \mbox{for~}x\to -\infty\,,\\[5mm]
e^{i k x}  &  \mbox{for~}x \to \infty\,,
\end{array} \right.
\end{eqnarray}
where, after interacting with the potential barrier, part of the wave is transmitted, goes to $x\rightarrow\infty$, and part is reflected, going back to $x\rightarrow -\infty$. The coefficients $a(k)$ and $b(k)$, usually known as the Bogoliubov coefficients, completely determine the BH scattering matrix as well as the reflection and transmission coefficients as follows
\begin{eqnarray}
t(k) = \frac{1}{a(k)} \,, \quad r(k) = \frac{b(k)}{a(k)} \,.
\label{reflection-transmission-coefficient}
\end{eqnarray}
The transmission and reflection probabilities, or greybody factors, are given by the modulus square of the corresponding coefficients, i.e.
\begin{equation}
T(k)=|t(k)|^2\,, \quad R(k)=|r(k)|^2 \,.
\label{rt-probabilities}
\end{equation}
For real $k$ they satisfy the unitarity condition
\begin{eqnarray}
T(k) + R(k) = 1 \,.
\end{eqnarray}
In Ref.~\cite{Lenzi:2022wjv}, it was shown that the KdV integrals and the BH greybody factors are related by an infinite set of integral relations, the so-called trace identities~\cite{Zakharov:1971faa}.
The starting point for finding these identities is the following expansion for the logarithm of the Bogoliubov coefficient, $\ln a(k)$, for $|k| \rightarrow \infty$, with $k \in \mathbb{C}$:
\begin{align}
\label{ln-a}
\ln a(k) = \sum_{n=1}^{\infty} \frac{c_n}{k^n}\,.
\end{align}
Then, evaluating the coefficients $c_n$ in two different ways and comparing them~\cite{Zakharov:1971faa, Lenzi:2022wjv}, we can obtain the expressions for these identities
\begin{eqnarray}
\mu^{}_{2n} = \int_{-\infty}^{\infty} dk\, k^{2 n}\, p(k) \,.
\label{moments-hamburger}
\end{eqnarray}
Here, the $\mu_{n}$'s are positive constants proportional to the KdV integrals, which can be interpreted as the moments of the (distribution) function $p(k)$, which only depend on the greybody factors, i.e.
\begin{eqnarray}
\mu^{}_{2n} = (-1)^{n}\,\frac{\mathcal{K}^{}_{2n+1}}{2^{2n+1}} \,,\quad p(k) = - \frac{\ln T(k)}{2\pi} \,.
\label{moments-hamburger-kdv}
\end{eqnarray}
where the minus sign guarantees that we are expressing everything in terms of positive quantities. The coefficients $c_n$ are then given by
\begin{align}
c_{2n} = 0\,, \quad
c_{2n+1} = i(-1)^{n} \frac{\mathcal{K}_{2n+1}}{2^{2n+1}} \,.
\end{align}
The connection between the KdV integrable structure and BH scattering is therefore represented by Eq.~\eqref{moments-hamburger}, which is equivalent to the so called \textit{moment problem}~\cite{Shohat1943ThePO,akhiezer1965classical,schmudgen2017moment}, a longstanding mathematical problem that appears in a wide range of physical contexts\footnote{Notice that this is in principle true for any potential barrier without bound states.}. The moment problem consists in three steps: i) existence of the solution; ii) uniqueness of the solution and iii) obtaining the solution. All three steps only depend on the moments/KdV integrals associated to the Regge-Wheeler potential. In Ref.~\cite{Lenzi:2022wjv}, we checked the first two conditions hold and in Ref.~\cite{Lenzi:2023inn} the moment problem was actually solved and the greybody factors for any potential barrier were found only in terms of the associated KdV integrals. 
We can summarize the method shown there for constructing approximations for the BH greybody factors in a schematic way as an algorithmic list of steps:
\begin{enumerate}
\item Evaluate the first $n$ KdV integrals for the chosen potential.

\item Obtain the moments from the KdV integrals and construct the moment generating function, $M(t)$, by using the following expansion at order $n$:
\begin{equation}
 M(t) =
\sum_{n=0}^{\infty} \frac{m_n}{n!}(-t)^n
\,,
\label{MGF-asymptotic}
\end{equation}
where $m_n = \sigma^{2n+1}\mu^{}_{2n}$, with $\sigma = 1/\alpha$ for P\"oschl-Teller and $\sigma = r^{}_s$ for Regge-Wheeler.

\item Construct Pad\'e approximants of order $[K/L]$, with $K+L \leq n$.

\item Evaluate the poles $t_i$ and residues $\lambda_i$ of the Pad\'e approximants.

\item Apply the Laplace inversion formula to finally obtain the following approximation for the greybody factors.
\end{enumerate} 
\begin{eqnarray}
T(k) & \simeq & 
\exp{\left(-2\,\pi\, \sigma\,k\; 
\sum_{i = 1}^{L} \lambda^{}_i e^{-t^{}_i\,\sigma^2 k^2}\right)}  \equiv  T^{}_{[K/L]}(k) \,.
\label{greybody-pade}
\end{eqnarray}
The residues and poles only depend on the KdV integrals, i.e.
\begin{equation}
t_i = t_i(\left\{\mathcal{K}_n \right\})\,,\quad \lambda = \lambda_i(\left\{\mathcal{K}_n \right\})
\end{equation}
For details on the construction and the precision of the approximation see Ref.~\cite{Lenzi:2023inn}.

\section{Fitting the BH Potential Barrier to a P\"oschl-Teller Potential: KdV Integrals and Quasinormal Modes}
\label{S:PT-fit}

The Pöschl-Teller potential~\cite{Poschl:1933zz} has been widely used in the literature as a mimicker for the true BH potentials (see e.g.~\cite{Ferrari:1984ozr, Ferrari:1984zz}) for two main reasons. On the one side, it is qualitatively similar to the Regge-Wheeler (or Zerilli) potential. On the other side, it is extremely useful as an analytical testing ground for many properties of the real BH potential, due to the analytical solvability of the associated scattering problem, while giving away some piece of information (e.g. power law tails and algebraically special solution) because of the different decay rate towards spatial infinity. The Pöschl-Teller potential can be expressed as follows:

\begin{equation}
V_{\rm PT}(x) = \frac{V_0}{\cosh^2{\left[\alpha\left(x-x_0\right)\right]}} = \frac{\alpha^2 \left(\beta^2+\frac{1}{4}\right)}{\cosh^2{\left[\alpha\left(x-x_0\right)\right]}},
\label{Vpt}
\end{equation}
where here, $x$ is going to be identified with the BH tortoise coordinate, $x_0$ is the location of the maximum value of the potential, $V_0$ ($V'_{\rm PT}(x_0)=0$, $V_{\rm PT}(x_0)=V_0$), and $\alpha$ is related to the potential curvature via the following expression:
\begin{equation}
\alpha^2 = -\frac{V''(x_0)}{2\,V_0}.
\label{alpha}
\end{equation}
The parameter $\beta$ is linked to $V_0$ and $\alpha$ by
\begin{equation}
\beta = \pm \frac{\sqrt{4\,V_0 - \alpha^2}}{2\,\alpha}\,.
\label{l0}
\end{equation}
We choose the positive sign as $\beta$ is positive definite.
From the analytical solution of the scattering problem with P\"oschl-Teller potential (see Appendix~\ref{App:PT}), we can easily obtain the corresponding greybody factors. In particular, the transmission coefficient is: 
\begin{eqnarray}
T^{}_{\rm PT} = \frac{\sinh^2\left(\frac{\pi k}{\alpha}\right)}{\cosh^2\left(\frac{\pi k}{\alpha}\right) + \sinh^2\left(\pi \beta\right)}
\,,
\label{pt-transmission}
\end{eqnarray}
and the two branches of QNMs are obtained as poles of the transmission coefficient
\begin{eqnarray}
k^{\rm PT}_n = \alpha \left[ \pm\beta - i \left( n + \frac{1}{2} \right) \right] \,.
\label{QNM-PT}
\end{eqnarray}
In this section, we first review the possibility of fitting the Regge-Wheeler potential to a Pöschl-Teller one and expand these results showing the alternative fit of the Zerilli even parity potential to a Pöschl-Teller one. We then use this set up to study the QNMs and the KdV integrals associated to these approximations. Indeed, among other analytical advantages of the Pöschl-Teller case, we will show that the generic $n$-th KdV integral for a Pöschl-Teller potential can be found in closed form. Furthermore, we consider the Regge-Wheeler potential for generic spin as in Eq.~\eqref{potential-scalar-vector-grav}.

\subsection{Fits to the P\"oschl-Teller Barrier}
\label{Ss:Fit-to-PT}

In Refs.~\cite{Ferrari:1984ozr, Ferrari:1984zz}, the P\"oschl-Teller approximation to the Regge-Wheeler potential has been used to evaluate the QNMs. Here, we repeat the steps and extend them also to the case of the Zerilli potential to assess possible differences coming from the fit to the even parity potential, instead of the odd one, and explore the implications in the study of spectral properties. The method consists in replacing the potential of the system under study with a Pöschl-Teller potential, fitting the maximum point and the curvature there.

The starting point is to set $x_0$ equal to the maximum of the Regge-Wheeler potential in Eq.~\eqref{potential-scalar-vector-grav}, i.e. $V_{\mathrm{RW}}'(r_0)=0$. The maximum of the odd parity BH potential is found in the following form:
\begin{eqnarray}
r^{}_0 = \frac{r^{}_s}{4}\left[3(1-\zeta) + \left(9 + 14 \zeta + 9 \zeta^2 \right)^{1/2} \right]\,, 
\quad \mbox{with} \quad \zeta = 
\frac{1-s^2}{\ell(\ell +1)}\,.
\label{r0}
\end{eqnarray}
The maximum $r^{}_0$ is related to $x^{}_0$ by the tortoise coordinate definition in our spacetime ($dx/dr = f(r)^{-1}$).
Introducing, for convenience, the dimensionless variable $y^{}_0 = 2 r^{}_0/r^{}_s$, the maximum $V^{}_0$ and the curvature $\alpha$ [given by Eq.~\eqref{alpha}] tuned to the Regge-Wheeler values read:
\begin{eqnarray}
V^{}_0 = V^{}_{\mathrm{RW}}(y^{}_0) = \frac{4\ell (\ell +1)}{r^2_s\, y_0^4} \left[(1-\zeta)\,y^{}_0 +4\,\zeta \right]
\,,
\label{V0}
\end{eqnarray}
and 
\begin{eqnarray}
\alpha = -\frac{f^2(y^{}_0)\,V_{\mathrm{RW}}''(y^{}_0)}{r^{}_s\,V^{}_0} =
\frac{2(y^{}_0 -2)}{r^{}_s\,{y_0}^2}
\left[\frac{3\,(1-\zeta)\,y^{}_0 +16\,\zeta}{(1-\zeta)\,y^{}_0 +4\,\zeta} \right]^{1/2} \,.
\label{alpha0}
\end{eqnarray}
In order to obtain these results, we eliminated the appearance of $y_0^2$ by using the following expression:
\begin{equation}
y_0^2=3\,(1-\zeta)\,y_0 +8\,\zeta \,,
\end{equation}
which can be easily obtained from Eq.~\eqref{r0}.

The same procedure can be followed for the case of the Zerilli potential, with the caveat that its more complicate expression makes it harder to write down meaningful analytical expressions.
For instance, the equation that determines the maximum location of the Zerilli potential is a fifth-degree polynomial equation for r, which expressed in terms of the dimensionless variable $\hat{r}\equiv r/r_s$ reads:
\begin{eqnarray}
&& -108 -27 (-13 + 5 \ell + 5 \ell^2) \hat{r} 
-9 (-1 + \ell) (2 + \ell) (-25 + 7 \ell + 7 \ell^2) \hat{r}^2 \nonumber\\[2mm]
&&-3 (-1 + \ell)^2 (2 + \ell)^2 (-23 + 5 \ell + 5 \ell^2)\hat{r}^3 
-3 (-1 + \ell)^3 (2 + \ell)^3 (-3 + \ell + \ell^2) \hat{r}^4 \nonumber \\[2mm]
&& + 2 (-1 + \ell) (8 + 8 \ell + 4 \ell^2 - 14 \ell^3 
- 9 \ell^4 + 5 \ell^5 + 5 \ell^6 + \ell^7) \hat{r}^5 = 0\,. 
\end{eqnarray}
It can be easily solved with the algebra system \textit{Mathematica}~\cite{Mathematica}, but we omit here the explicit expressions  because they are long and they would not contribute significantly to the present discussion.

\begin{figure}[h!]
\centering
\includegraphics[width=0.8\textwidth]{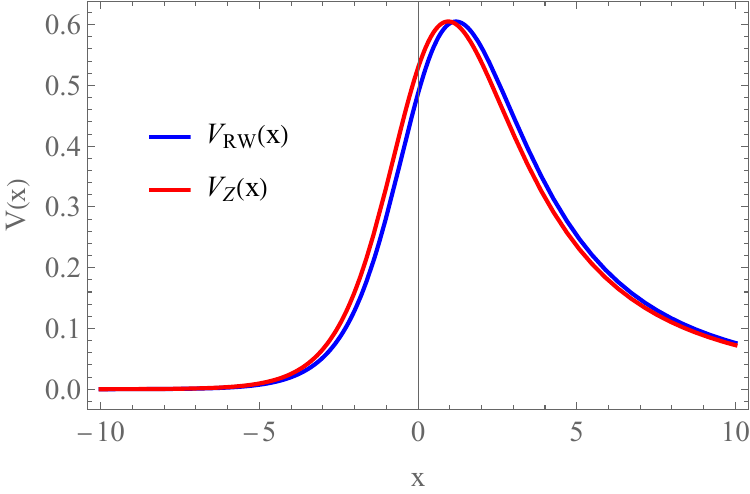}
\caption{Regge-Wheeler (blue) and Zerilli (red) potentials as functions of the tortoise coordinate $x$ ($-\infty< x < \infty$) in units of the Schwarzschild radius $r^{}_s$, for angular momentum harmonic number $\ell=2$.}
\label{Fig:VRW_VZ}
\end{figure}

\begin{figure}[h!]
\centering
\includegraphics[width=0.8\textwidth]{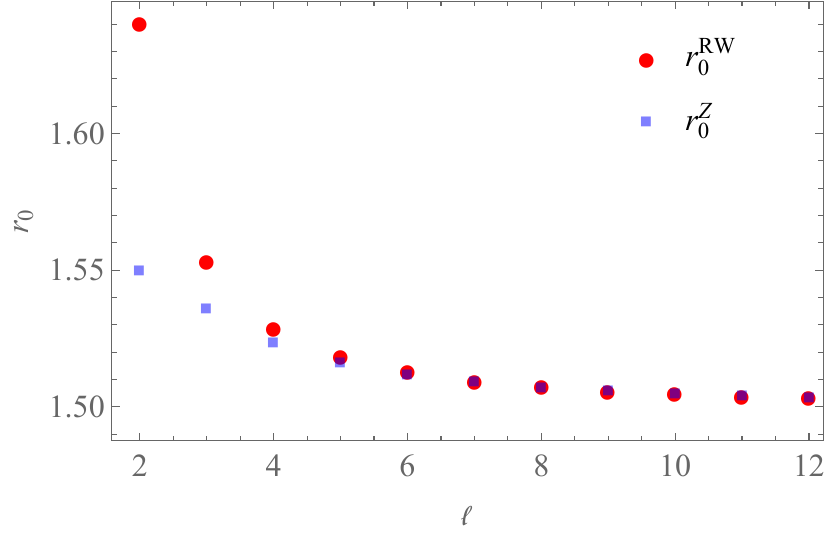}
\caption{Position of the maximum of the Regge-Wheeler potential (red) and the Zerilli potential (blue), in units of the Schwarzschild radius $r^{}_s$, as a function of the angular momentum harmonic number ($\ell$).}
\label{1 rmax vs l}
\end{figure}

Let us first review some general known features of the Regge-Wheeler and Zerilli potential, which are plotted in Fig.~\ref{Fig:VRW_VZ} in terms of the tortoise coordinate and for $\ell=2$. Notice that, the bigger the $\ell$ the more they overlap with each other. In Fig.~\ref{1 rmax vs l}, we plot the location of the maxima of the two potentials for different values of $\ell$. 

\begin{figure}[h!]
\centering
\includegraphics[width=0.49\textwidth]{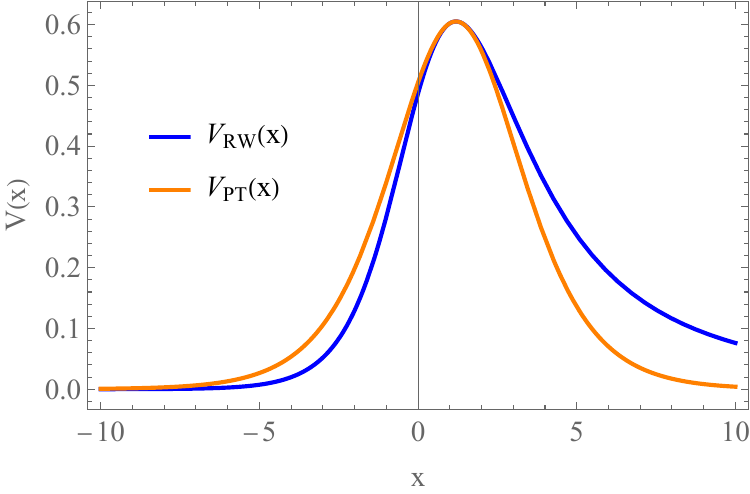}
\includegraphics[width=0.49\textwidth]{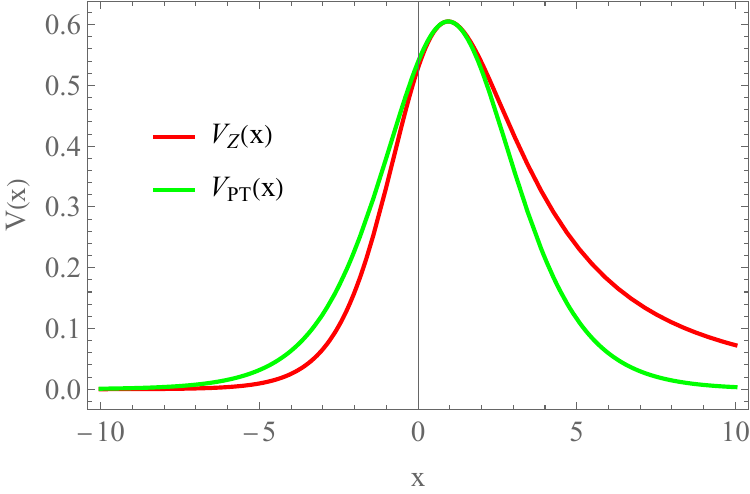}
\caption{Comparison of the Pöschl-Teller potential fits to the Regge-Wheeler (left) and Zerilli (right) potentials in units of the Schwarzschild radius $r^{}_s$.}
\label{Fig:VRW_VZ_VPT}
\end{figure}

\begin{figure}[h!]
\centering
\includegraphics[width=0.49\textwidth]{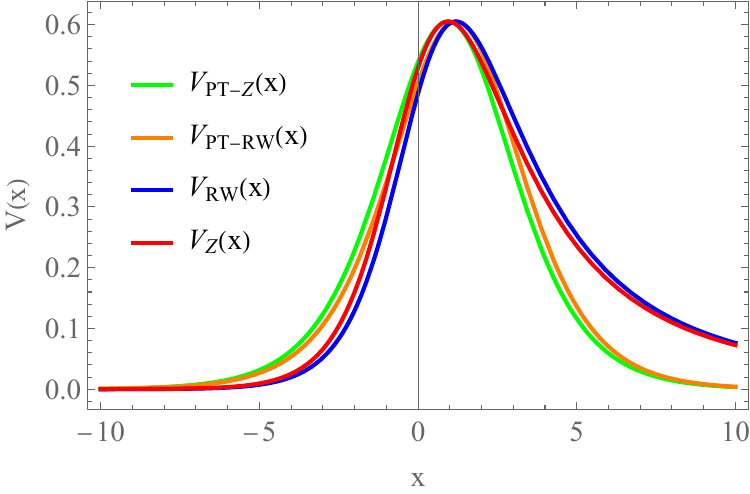}
\includegraphics[width=0.49\textwidth]{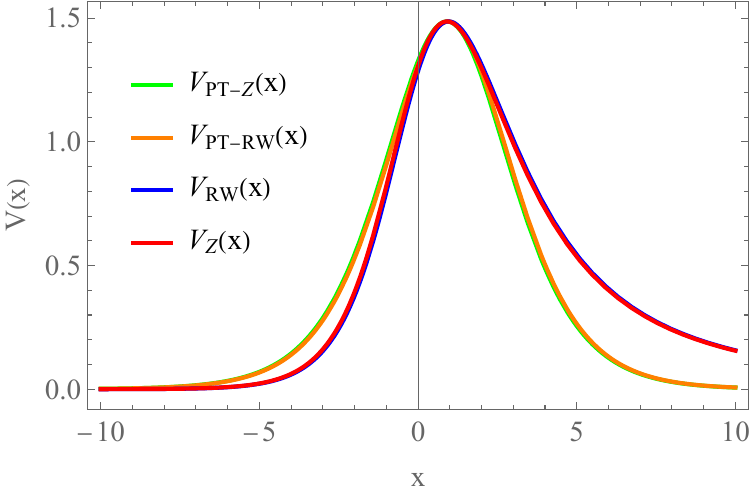}
\caption{Comparison of the four potentials considered in this study (Regge-Wheeler, Zerilli and the two Pöschl-Teller fits) for the cases $\ell=2$ (left) and $\ell=3$ (right) in units of the Schwarzschild radius $r^{}_s$.}
\label{Fig:VRW_VZ_VPT_L2_L3}
\end{figure}

In this picture, we observe that the location of the maximum, while different for low $\ell$, asymptotically approaches the same value (the light-ring radius value) as $\ell$ increases. We expect that the two different P\"oschl-Teller fits should converge as $\ell$ increases, yielding similar results for both the KdV integrals and the QNM frequencies. This expectation is going to be reinforced by the results we present in this work, where it is also shown that even for $\ell = 2$, the results of the two fits are nearly identical. In Fig.~\ref{Fig:VRW_VZ_VPT} we plot the fits of the odd- and even-parity BH potentials to a P\"oschl-Teller potential barrier for $\ell=2$, showing that the P\"oschl-Teller approximation is particularly good near the maximum and close to the horizon while it fails to reproduce the asymptotic behaviour at spatial infinity (exponential decay versus a  power-law decay). Indeed, the P\"oschl-Teller potential decays exponentially for both $x=\pm\infty$, while the BH potentials do so only close to the horizon, while they decay as $r^{-2}$ at spatial infinity (the $\ell$-dependent centrifugal barrier term). In Fig.~\ref{Fig:VRW_VZ_VPT_L2_L3}, we plot the four potentials (Regge-Wheeler, Zerilli and their fits to P\"oschl-Teller) all together for $\ell=2$ (top panel) and for $\ell=3$ (bottom panel), to highlight the anticipated behavior for growing $\ell$.

\subsection{KdV Integrals for the P\"oschl-Teller Potential: Application to the BH Potential Fits}
\label{Ss:KdV-fit}

While we have a clear recurrence relation to obtain all the KdV densities [see Eq.~\eqref{kdv-densities-recurrence}], no relation of this kind can be found in general for the KdV integrals, each of which then has to be evaluated separately. However, in the case of the Pöschl-Teller potential, the existence of an analytical expression for the Bogoliubov coefficient $a(k)$ [see Eq.~\eqref{ak-pt} in Appendix~\ref{App:PT}], enables us to obtain a closed form for the $n$-th term in the series expansion of $\ln a(k)$ and, consequently, an analytical expression for the $n$-th KdV integral of a Pöschl-Teller potential for the first time (see Appendix \ref{App:KdV-PT} for some more details on the derivation). We find that the generic $n$-th KdV integrals for the Pöschl-Teller potential can be cast in the following form:
\begin{eqnarray}
\frac{\mathcal{K}_{2n-1}}{\alpha^{2n-1}} & = & i(-1)^{n}2^{2n-1}\mu_{2n-1} = \frac{1}{2n}+\frac{4^{n-1}}{n(2n-1)}\left(B_{2n}-n+1\right) 
\nonumber \\[2mm]
& + & \sum_{m=1}^{n}\frac{B_{2m}(2n-2)!2^{2n}}{(2m)!}\left[\frac{1}{2(2n-2m)!}-\sum_{j=0}^{n-m}\frac{(-1)^{j+m+n}\beta^{2n-2m-2j}}{2^{2j}(2j)!(2n-2m-2j)!}\right] \nonumber \\[2mm]
& + & \sum_{m=0}^{n-1}\frac{(2\beta)^{2n-2m}(-1)^{m+n}(2n-2)!(2m-1)}{(2m)!(2n-2m)!} \,, \quad n=1,2,3,\ldots
\label{pt-kdv-exact}
\end{eqnarray}
where $B_{n}$ are the Bernoulli numbers.
Analytical solutions to the moment problem are rare and, in general, this depends on the possibility of finding a closed-form expression for the sequence of moments. One of the few analytically-solvable known cases that one can find in the literature happens when the sequence of moments corresponds to the series of Catalan numbers~\cite{bostan:hal-02425917}.  This sequence of moments appear in a very different setup, the \emph{Wigner semicircle distribution}, where the (probability) distribution is defined in a finite domain (Hausdorff moment problem) and the density (probability) distribution is a scaled semicircle. This distribution is relevant in the context of the study of random matrices~\cite{Wigner:1958:rm,anderson2010introduction}.

In Ref.~\cite{Lenzi:2022wjv}, we discussed how the moment problem for the P\"oschl-Teller potential can be solved using the Stieltjes-Perron inversion formula~\cite{AFST_1894_1_8_4_J1_0}. The fact that we are able to find here the sequence of moments in Eq.~\eqref{pt-kdv-exact} can be seen as another way of stating the exact solvability of the moment problem with P\"oschl-Teller potential. In this sense, the result in Eq.~\eqref{pt-kdv-exact} extremely simplifies the evaluation of the KdV integrals and shows once more the usefulness of approximating the BH potentials with a P\"oschl-Teller barrier. Indeed, it reduces the calculation of the KdV integrals for the fits to  P\"oschl-Teller to a simple algebraic evaluation. We profit this result to efficiently study the behavior of the KdV integrals of the P\"oschl-Teller fits to Regge-Wheeler and Zerilli potentials. 
In doing so, we train our analysis and understanding of the behavior of KdV integrals under modifications of the potential, with a particular focus on the stability/instability properties of the KdV integrals and their ability to measure the breaking of isospectrality between the two parities. 

\begin{figure}[h!]
\centering
\includegraphics[width=0.49\textwidth]{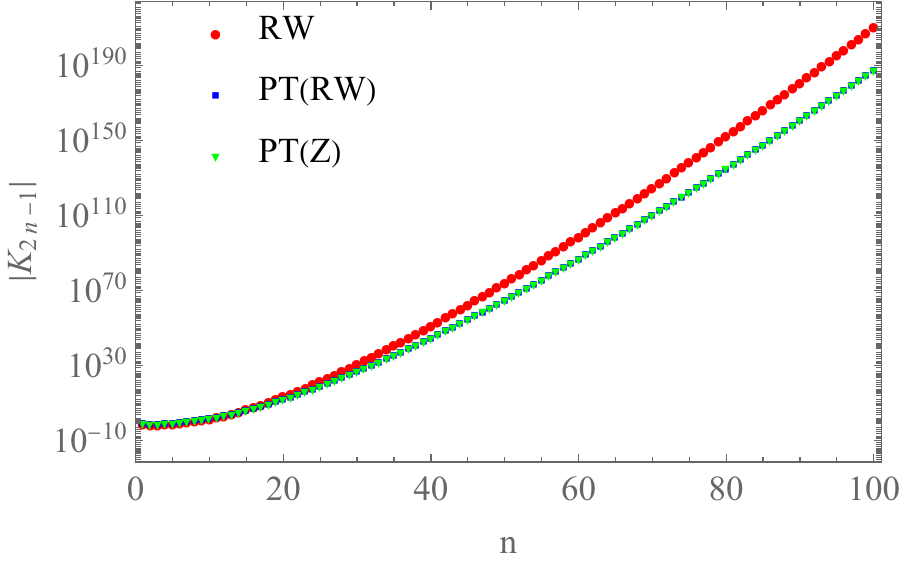}
\includegraphics[width=0.49\textwidth]{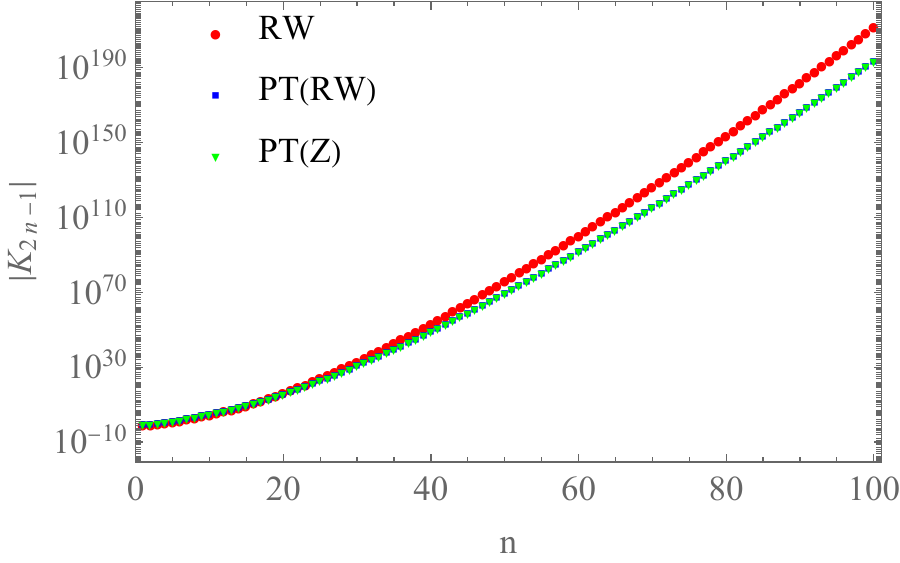}
\caption{Absolute values for the KdV integrals for the Regge-Wheeler potential (RW) and the two Pöschl-Teller fits [PT(RW) and PT(Z)] in logarithmic scale of base $10$. The left figure shows the results for the $\ell=2$ case, whereas the figure on the right shows the results for $\ell=3$. In these plots, each KdV integral $\mathcal{K}_n$ has been multiplied by $r_s^n$ to make them dimensionless.}
\label{Absolute values for the KdV integrals}
\end{figure}

Starting from the analytic expression in Eq.~\eqref{pt-kdv-exact}, we can evaluate an arbitrary number of KdV integrals for the P\"oschl-Teller fits by just tuning the parameters as described in the previous section [Eqs.~\eqref{alpha} and~\eqref{l0}]. In Fig.~\ref{Absolute values for the KdV integrals} we plot the absolute value (since they have alternate signs) of the KdV integrals of the Regge-Wheeler potential and of the odd- and even-parity P\"oschl-Teller fits.

Clearly, the Regge-Wheeler KdV integrals grow much faster than the others independently of $\ell$. Moreover, the discrepancy between the KdV integrals obtained with the fits is almost negligible, as it evident by looking in both plots at the superposition of the green and blue points, representing the fits to the Regge-Wheeler and Zerilli potentials respectively. These two features hold true for any $\ell$, and the two fits become increasingly similar as $\ell$ increases, as expected.

\begin{figure}[h!]
\centering
\includegraphics[width=0.8\textwidth]{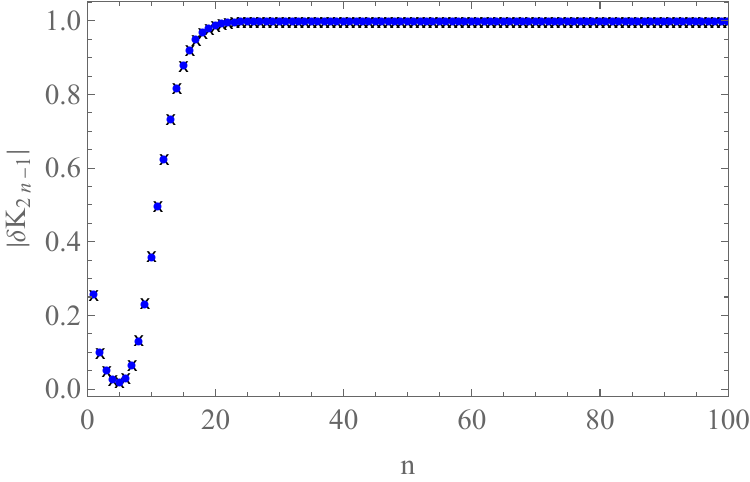}
\caption{Comparison of the dimensionless relative differences [Eq.~\eqref{rel-err-fit}] between the true KdV integrals (the Regge-Wheeler ones) and those obtained with the Pöschl-Teller fit to the Regge-Wheeler potential (black crosses) and to the Zerilli potential (blue dots). These values are computed for the case $\ell=2$.}
\label{PTZ and PTRW KdV relative differences}
\end{figure}

We can get a more detailed understanding of the behavior of the KdV integrals by studying the relative variation with respect to the Regge-Wheeler ones, i.e.
\begin{eqnarray}
 \delta \mathcal{K}_n = \left|\frac{\mathcal{K}_n^{\mathrm{RW}}-\mathcal{K}_n^{\mathrm{fit}}}{\mathcal{K}_n^{\mathrm{RW} }}\right| \,.
 \label{rel-err-fit}
\end{eqnarray}
In Fig.~\ref{PTZ and PTRW KdV relative differences}, we plot the relative error just defined for $\ell=2$. As we can see, the two fits are practically equivalent. Only a close numerical check shows that the Zerilli fit is slightly better even if it does not seem to bring relevant consequences. Moreover, as already mentioned, these differences become even smaller as $\ell$ increases. The behavior of the relative errors shows an initial decreasing trend, followed by an increasing trend until it saturates because of the leading behavior of the Regge-Wheeler KdV integrals. 

We can associate the behavior of the KdV integrals for the fits to P\"oschl-Teller to the moment problem described in Sec.~\ref{S:GF-from-KdV}. Let us recall that the KdV integrals of a potential barrier correspond, up to proportionality constants and sign, to the moments of a logarithmic distribution of the greybody factors of the barrier. As such, they reflect the properties, in the frequency domain, of the distribution itself. Therefore, with the moment problem perspective in mind, we can roughly associate the behavior seen in Fig.~\ref{PTZ and PTRW KdV relative differences} to the differences between the Regge-Wheeler potential and the P\"oschl-Teller fits, exploiting the bridge between the definition of KdV integrals as space integrals~\eqref{kdv-integrals} and as frequency domain integrals~\eqref{moments-hamburger} (for more details see Sec.~\ref{S:GF-stability}). First, we already mentioned that the P\"oschl-Teller potential decays exponentially at spatial infinity while the Regge-Wheeler one only as $1/r^2$. This actually reflects the fact that the first few KdV integrals deviate significantly from the original values. The frequency domain perspective would be that the modification on the asymptotic behavior of the potential affects the low frequency region of the distribution in Eq.~\eqref{moments-hamburger-kdv}. On the other hand, higher KdV integrals are more sensitive to local changes of the potential. Therefore, after a region of “stability” of the “intermediate” integrals, the relative difference with the Regge-Wheeler ones increases. This can be attributed to the fact that the derivative order in the KdV densities~\eqref{kdv-densities-recurrence} grows with $n$, so that the KdV integrals reflect the fact that the P\"oschl-Teller fits are obtained by fixing only some local information, i.e. maximum value and second order derivatives. From the moment problem perspective instead, this seems to signal that local changes in the potential, near the maximum, induce high-frequency instabilities.

\begin{figure}[h!]
\centering
\includegraphics[width=0.49\textwidth]{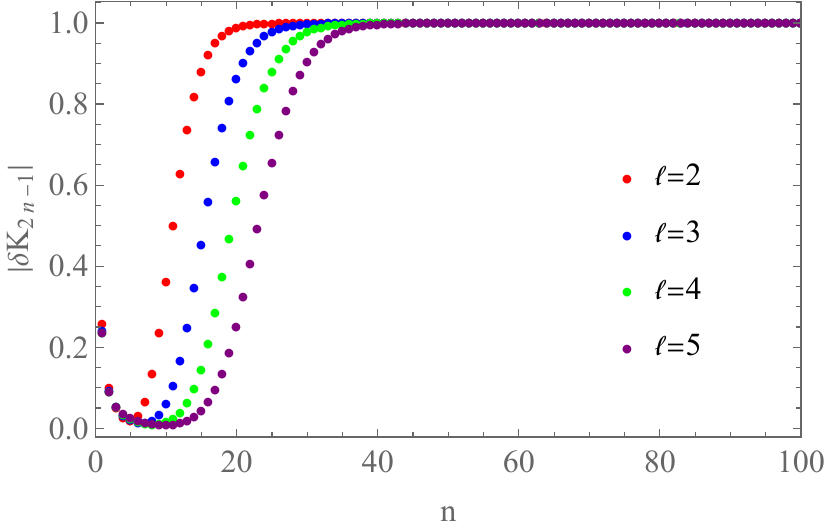}
\includegraphics[width=0.49\textwidth]{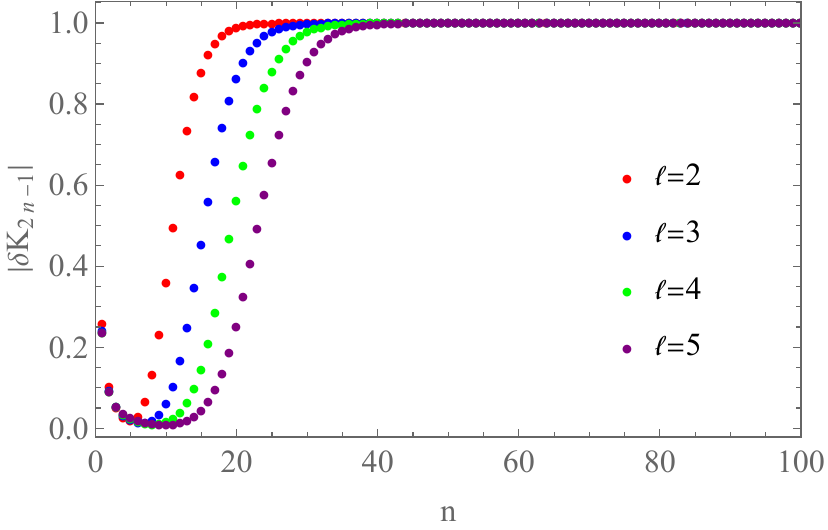}
\caption{Comparison of the dimensionless relative differences [Eq.~\eqref{rel-err-fit}] between the true KdV integrals (the Regge-Wheeler ones) and those obtained with the Pöschl-Teller fits for different values of the angular momentum $\ell$. In the left panel, the fit is to the Regge-Wheeler potential, whereas in the right panel, the fit is to the Zerilli potential.}
\label{KdV different l}
\end{figure}

Next, we study how these relative differences change with $\ell$, and these results are presented in Fig.~\ref{KdV different l}.

In both plots, one for the fit to the Regge-Wheeler potential and the other one for the fit to the Zerilli potential, we clearly see that both fits provide better approximations as $\ell$ increases. Indeed, the discrepancy in the first KdV integrals decrease with increasing $\ell$ and the number of KdV integrals whose error lies in the minimum grows, before the steep rise of the relative error. This is actually consistent with previous results~\cite{Ferrari:1984ozr, Ferrari:1984zz} where the P\"oschl-Teller approximation was used to evaluate the BH QNMs. Indeed, it was found there that this approximation becomes correct in the eikonal limit $\ell \to \infty$, also in accordance with the results obtained with the WKB approximation~\cite{Schutz:1985km,Iyer:1986np} (see also Sec.~\ref{Ss:PT-fit-QNM}).

\begin{figure}[h!]
\centering
\includegraphics[width=0.49\textwidth]{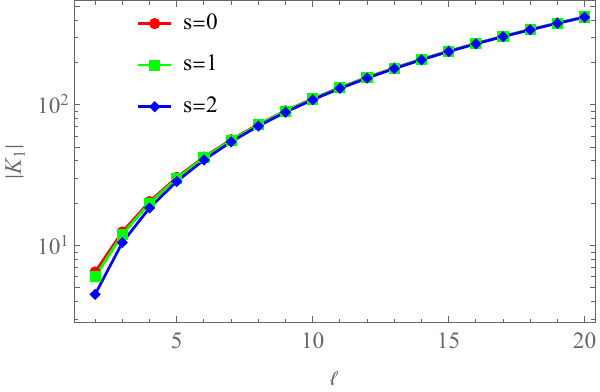}
\includegraphics[width=0.49\textwidth]{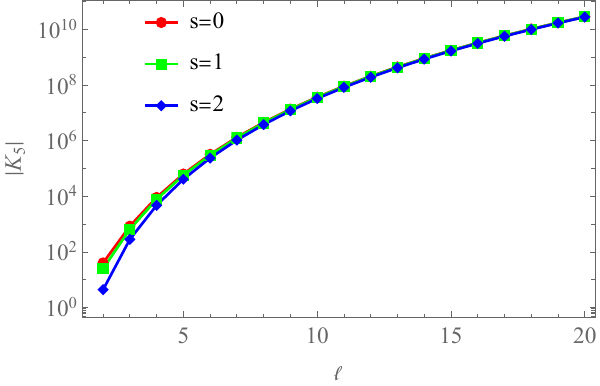}
\caption{Plot of the absolute value of the KdV integral for indices n=1 (left panel) and n=5 (right panel), as a function of angular momentum $\ell$. Each plot displays the results for three spin values: s=0 (red), s=1 (green), and s=2 (blue). The vertical axis is in logarithmic scale of base $10$ and each KdV integral $\mathcal{K}_n$ has been multiplied by $r_s^n$ to make them dimensionless.}
\label{KdV vs l spins 1}
\end{figure}

\begin{figure}[h!]
\centering
\includegraphics[width=0.49\textwidth]{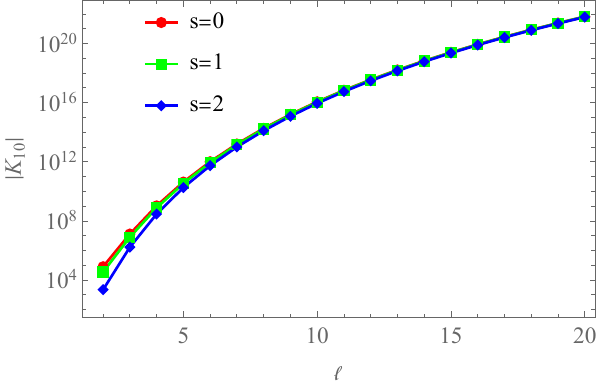}
\includegraphics[width=0.49\textwidth]{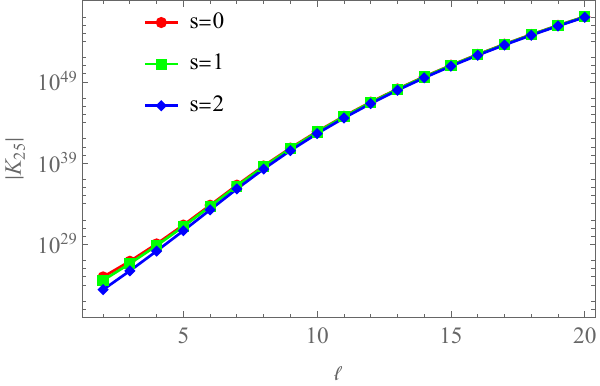}
\caption{Plot of the absolute value of the KdV integral for indices n=10 (left panel) and $n=25$ (right panel), as a function of angular momentum $\ell$. Each plot displays the results for three spin values: $s=0$ (red), $s=1$ (green), and $s=2$ (blue).  The vertical axis is in logarithmic scale of base $10$ and each KdV integral $\mathcal{K}_n$ has been multiplied by $r_s^n$ to make them dimensionless.}
\label{KdV vs l spins 2}
\end{figure}

Finally, we can exploit the fact that perturbations of the Schwarzschild BH with different spin are all described by a master equation with a potential of the form in Eq.~\eqref{potential-scalar-vector-grav}, to study the KdV integrals dependence on the spin.
Figs.~\ref{KdV vs l spins 1} and~\ref{KdV vs l spins 2} illustrate the convergence of different KdV integrals for various spins ($s=0,1 ,2$) as the angular momentum harmonic number $\ell$ increases. Specifically, the integrals corresponding to the indices $n=1, 5, 10, 25$ tend to a common value irrespective of the spin $s$. This reflects the fact that for high $\ell$, the angular barrier in Eq.~\eqref{potential-scalar-vector-grav} dominates the spin term.

\subsection{Quasinormal Modes associated with the Pöschl-Teller Fits}
\label{Ss:PT-fit-QNM}

The QNM frequencies of the Pöschl-Teller potential have been found in analytical form and are given in Eq.~\eqref{QNM-PT}. As it was first shown in Refs.~\cite{Ferrari:1984ozr, Ferrari:1984zz}, one can exploit the qualitative similarity between the Pöschl-Teller and the BH potentials to approximate the QNMs spectrum of the Regge-Wheeler barrier. This is easily done by inserting the fitted values, given by Eqs.~\eqref{V0} and~\eqref{alpha0}, into the definitions in Eqs.~\eqref{QNM-PT} and~\eqref{l0}.

\begin{figure}[h!]
\centering
\includegraphics[width=0.49\textwidth]{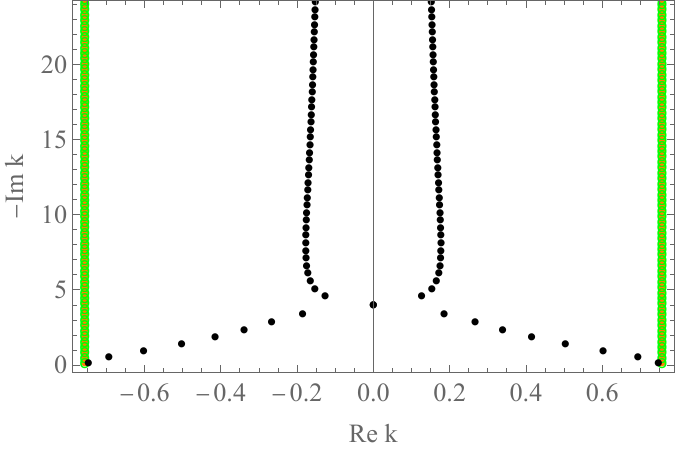}
\includegraphics[width=0.49\textwidth]{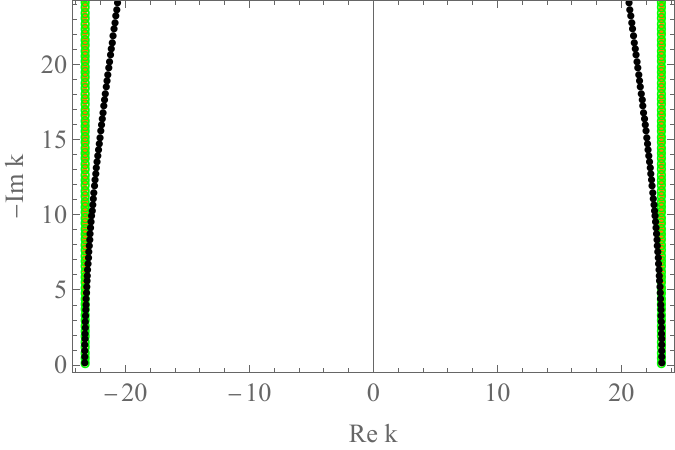}
\caption{Quasinormal mode frequencies obtained for the Regge-Wheeler potential (black dots) and the two Pöschl-Teller fits (orange dots for the fit to Regge-Wheeler and green circles for the fit to Zerilli) for the cases $\ell=2$ (left panel) and $\ell=60$ (right panel). The $\ell=2$ values are taken from Ref.~\cite{BertiWeb} while the $\ell=60$ frequencies are calculated with Mathematica~\cite{Mathematica} using the BH Perturbation Toolkit~\cite{BHPToolkit}. }
\label{Fig:QNM-L2L60}
\end{figure}

The QNM values obtained in this way, that is, by fitting the Regge-Wheeler and Zerilli potentials to a Pöschl-Teller barrier, together with the \emph{true} values of the QNMs associated with the Regge-Wheeler potential, are shown in Fig.~\ref{Fig:QNM-L2L60} for $\ell = 2$ (left panel) and for $\ell = 60$ (right panel). As is well known, the different decay properties of the potential at $r=\infty$ produce different behaviors of the associated QNMs. The QNMs obtained from a Pöschl-Teller potential have a constant real part, fixed by the maximum of the potential $V(x_0)$ (and its curvature near the maximum). On the other hand, the two branches of Regge-Wheeler QNMs cross the imaginary axis and their real part asymptotically approaches $\log(3)/4\pi$. Indeed, the asymptotic BH quasinormal frequencies, $k_{n}^{\mathrm{RW}}$, have the following analytical expression~\cite{Nollert:1993zz,Hod:1998vk,Motl:2003cd}:
\begin{equation}
r^{}_s\, k_{n}^{\mathrm{RW}} \simeq \pm \frac{\log(3)}{4\pi} -\frac{i}{2}\left(n+\frac{1}{2}\right)\,.
    \label{RW qnm asymptotically}
\end{equation}
This expression has been particularly relevant for its possible connection with quantum gravitational results~\cite{Motl:2003cd,Andersson:2003fh}, especially thanks to the coincidence of the term $\log{3}$ with the so-called Barbero-Immirzi parameter in Loop Quantum Gravity (see also~\cite{Maggiore:2007nq}). 

The Pöschl-Teller approximation to the Regge-Wheeler QNMs is however known to be valid only for the least damped QNMs. However, the accuracy improves for growing $\ell$ (see Fig.~\ref{Fig:QNM-L2L60}) and the approximation becomes exact in the eikonal limit $\ell \rightarrow \infty$. This becomes clear by observing the plots for $\ell = 2$ and for $\ell=60$ in Fig.~\ref{Fig:QNM-L2L60}. In the $\ell=60$ plot,  more QNMs in the Schwarzschild spectra line up with the P\"oschl-Teller ones as compared with the $\ell=2$ plot. Indeed, by taking the limit $\ell \rightarrow \infty$ of the fit obtained from Eqs.~\eqref{l0},~\eqref{QNM-PT},~\eqref{V0} and~\eqref{alpha0}, one obtains, at first order, the following expression:
\begin{eqnarray}
 \frac{3\sqrt{3}}{2} r^{}_s\, k_{n}^{\mathrm{RW}} \simeq \pm \left(\ell +\frac{1}{2} \right) - i\left(n +\frac{1}{2} \right)  \,,
\end{eqnarray}
in agreement with the result from the WKB approximation~\cite{Schutz:1985km,Iyer:1986np,Iyer:1986nq, Barreto97distributionof}. 

\begin{figure}[h!]
\centering
\includegraphics[width=0.49\textwidth]{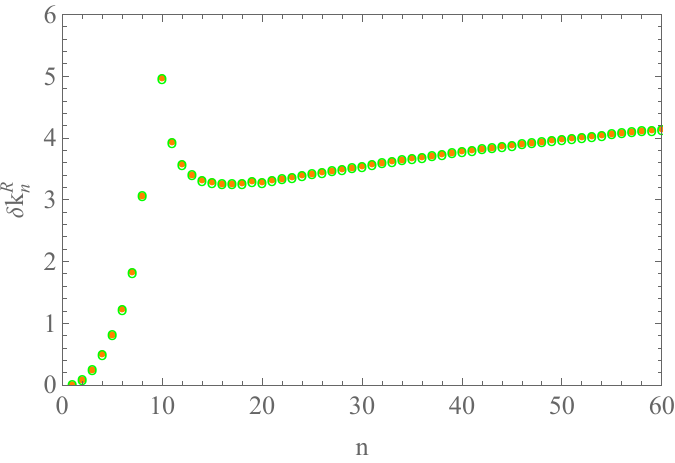}
\includegraphics[width=0.49\textwidth]{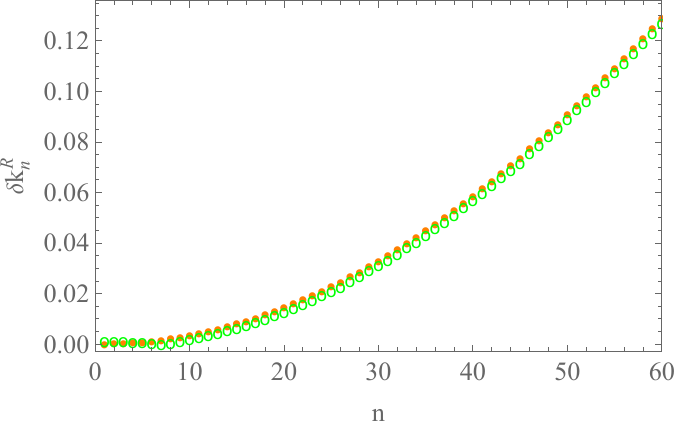}
\caption{Plot of the real parts error in Eq.~\eqref{error-QNM-fit} for $\ell=2$ (left panel) and $\ell=60$ (right panel) for the P\"oschl-Teller fit to Regge-Wheeler (black crosses/orange dots) and Zerilli (blue dots/green circles) potentials.}
\label{Fig:QNM-real-L2L60}
\end{figure}

\begin{figure}[h!]
\centering
\includegraphics[width=0.49\textwidth]{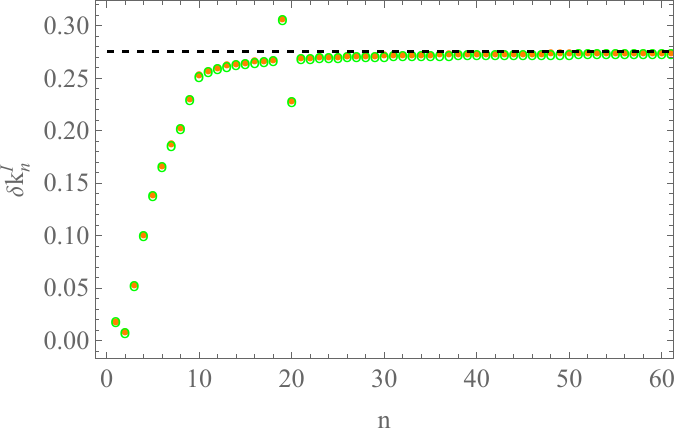}
\includegraphics[width=0.49\textwidth]{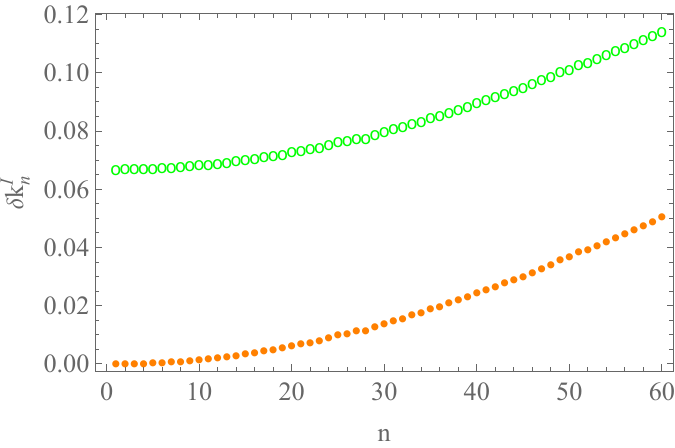}
\caption{Plot of the imaginary parts error in Eq.~\eqref{error-QNM-fit} for $\ell=2$ (left panel) and $\ell=60$ (right panel) for the P\"oschl-Teller fit to Regge-Wheeler (orange dots) and Zerilli (green circles) potentials.}
\label{Fig:QNM-imag-L2L60}
\end{figure}

\begin{figure}[h!]
\centering
\includegraphics[width=0.80\textwidth]{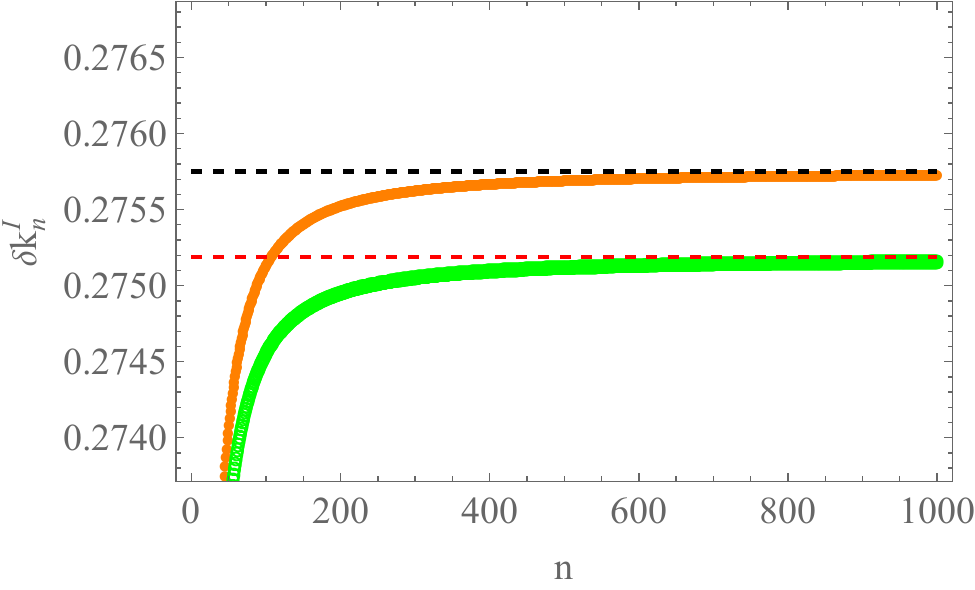}
\caption{Plot of the imaginary parts error in Eq.~\eqref{error-QNM-fit} for $\ell=2$ for the P\"oschl-Teller fit to Regge-Wheeler (orange dots) and Zerilli (green circles) potentials, with a restriction on the range of the plot to show the limiting values given by Eq.~\eqref{error-limit} for the fit to Regge-Wheeler (black dashed line) and Zerilli (red dashed line). }
\label{Fig:QNM_imag_zoom}
\end{figure}

In Figs.~\ref{Fig:QNM-real-L2L60} and~\ref{Fig:QNM-imag-L2L60}, we plot the relative errors
\begin{eqnarray}
\delta k^{\mathrm{R}}_n = \left|\frac{{\mathrm{Re}}(k_{n}^{\mathrm{PT}} - k_{n}^{\mathrm{RW}})}{{\mathrm{Re}}\,k_{n}^{\mathrm{RW}}}\right|
\,,\qquad 
\delta k^{\mathrm{I}}_n = \left|\frac{{\mathrm{Im}}(k_{n}^{\mathrm{PT}} - k_{n}^{\mathrm{RW}})}{{\mathrm{Im}}\,k_{n}^{\mathrm{RW}}}\right| \,,
\label{error-QNM-fit}
\end{eqnarray}
showing the results for the P\"oschl-Teller approximation to both the Regge-Wheeler and Zerilli potentials.
First, let us note that the errors in the real part, shown in Fig.~\ref{Fig:QNM-real-L2L60}, diverge at the algebraically special mode, i.e. at the crossing of the imaginary axis in Fig.~\ref{Fig:QNM-L2L60}. For this reason, it is eliminated from the plot. This point is located at the QNM with $n = 10$ for $\ell = 2$, at $n = 16$ for $\ell = 3$, and so on. 
Moreover, by looking at both Figs.~\ref{Fig:QNM-real-L2L60} and~\ref{Fig:QNM-imag-L2L60} we can confirm the general statement that the approximation works better for lower modes. The two points which lies clearly outside the general trend in Fig.~\ref{Fig:QNM-imag-L2L60} are probably due to the numerical evaluation of the corresponding QNMs and indeed correspond to the only two QNMs which do not follow the increasing trend of the imaginary part when increasing $n$. Apart from this, the error on the imaginary parts tends to saturate approaching a constant value (see the left panel of Fig.~\ref{Fig:QNM-imag-L2L60}). This behavior is related to the choice of normalization of the error in Eq.~\eqref{error-QNM-fit}, which is chosen to be the Regge-Wheeler QNM frequencies. 
This limiting value can be computed from Eqs.~\eqref{QNM-PT} and~\eqref{RW qnm asymptotically} for $n \rightarrow \infty$ to give:
\begin{eqnarray}
\delta k^{\mathrm{I}}_n \simeq \left|2r^{}_s\, \alpha -1 \right| \quad\mathrm{for}\quad n \to\infty  
\label{error-limit}
\end{eqnarray}
This behavior can be observed in Fig.~\ref{Fig:QNM-imag-L2L60} (see also Fig.~\ref{Fig:QNM_imag_zoom} for a more detailed view).

Comparing the left and right panels of Figs.~\ref{Fig:QNM-real-L2L60} and~\ref{Fig:QNM-imag-L2L60}, we can notice the improvement of the P\"oschl-Teller approximation when $\ell$ grows, as dictated by the eikonal limit. In addition, considering high values of $\ell$ discloses an asymmetry between the two fits which could not have been noticed at lower angular momentum values. In fact, while the two approximations are almost equivalent for the real part of the QNMs (see right panel of Fig.~\ref{Fig:QNM-real-L2L60}), the fit to the Regge-Wheeler potential provides significantly better results for the imaginary parts. For example, the error at $n=60$ is $5$\% for the Regge-Wheeler fit while it is around $11$\% for the Zerilli fit (see the right panel of Fig.~\ref{Fig:QNM-imag-L2L60}). The final qualitative statement that can be done is that the two fits to a Pöschl-Teller potential break the isospectrality of the QNMs.

To conclude this section, let us note that the KdV integrals studied in Sec.~\ref{Ss:KdV-fit} share a qualitatively similar behavior with respect to the QNMs. Actually, the same behavior in the eikonal limit is found, i.e. the KdV integrals seem to converge to the correct values for growing $\ell$. This is something which is not obvious due do the high degree of nonlinearity in the expressions for these integrals. Furthermore, the almost indistinguishability of the two approximations is also reflected. Finally, it is interesting to realize that the plot in Fig.~\ref{Fig:QNM-imag-L2L60}, for the relative error in the imaginary parts~\eqref{error-QNM-fit}, shows a similar behavior to the relative variation of the KdV integrals shown in Fig.~\ref{PTZ and PTRW KdV relative differences}, thus increasing the expectations for a clear connection between the KdV integrals and the QNMs, or at least their imaginary part.

\section{KdV Integrals for Modified BH Potentials: (in)Stability Studies}
\label{S:KdV-integrals-modified-potentials}

Until now, we have considered the (non-rotating) BH potentials predicted by GR, represented by the Regge-Wheeler and Zerilli potentials for odd- and even-parity harmonic modes respectively. We have also considered P\"oschl-Teller approximations to them. However, the increasing interest in the search for possible deviations from GR in gravitational-wave (GW) signals, has led to a number of proposals for corrections to the BH potential barrier and/or the corresponding BH background metric descriptions. In most cases, these BH potentials are motivated by phenomenological reasons, frequently motivated for the need to regularize the compact object description~\cite{Rezzolla:2014mua, Cardoso:2019rvt} or to mimic some special astrophysical situation, such as environmental effects~\cite{Barausse:2014tra,Cheung:2021bol,Cardoso:2021wlq}. In a much smaller number of cases, the modified BH potentials can be obtained from “first-principle” derivations, such as in the case of some Effective Field Theory (EFT) corrections, see e.g.~\cite{Silva:2024ffz}. Moreover, most of these potentials naturally raise the relevant issue of spectral instabilities in the QNMs spectra~\cite{Nollert:1996rf, Jaramillo:2020tuu, Jaramillo:2021tmt}.

In what follows, we show some features of the KdV integral that suggest that they can be recognized as useful and simply computable indicators of BH spectral properties, with a particular focus on isospectrality between even- and odd-parity perturbations and QNM instabilities. Let us start from BH potential barriers that can be seen as a small perturbation of the standard BH potentials, coming from any of the above mentioned physical motivations, in the following form
\begin{eqnarray}
V^{\rm odd}_{\rm even} = V^{\mathrm{RW}}_{\mathrm{Z}} + \epsilon\,\delta V^{\mathrm{odd}}_{\mathrm{even}} \,.
\label{modified-V}
\end{eqnarray}
We first study how the KdV integrals of different potentials get modified and address their stability properties. Later, we perform qualitative comparisons between the stability behavior of the KdV integrals and that of the QNMs, which has been found recently in the literature for certain families of potentials. This reveals an interesting qualitative connection between KdV integrals and spectral properties which deserves further analytical investigations.

Since the goal is to assess the stability properties of the KdV integrals under phenomenological or “first-principles” corrections, it is convenient to define the relative error between the Regge-Wheeler values of the KdV integrals and those of the modified potential as
\begin{eqnarray}
\delta \mathcal{K}^{}_n = \left|\frac{\mathcal{K}^{\epsilon}_n -\mathcal{K}^{\mathrm{RW}}_n}{\mathcal{K}^{\mathrm{RW}}_n}\right| \,,
\label{KdV-rel-error}
\end{eqnarray}
where $\mathcal{K}^{\epsilon}_n$ denote the KdV integrals for the potential~\eqref{modified-V}. The stability criterion considered here is 
\begin{eqnarray}
\delta \mathcal{K}^{}_n \lesssim \epsilon\,.
\label{stability-requirement}
\end{eqnarray}
That is, broadly speaking, it requires that the effects of the perturbations on the KdV integrals associated with the modified potentials do not grow bigger than the intensity of the perturbation itself, $\epsilon$. 
The definition~\eqref{KdV-rel-error} holds for both odd- and even-parity modes since the normalization is given by the GR values, which are equal for the Regge-Wheeler and Zerilli potentials~\cite{1980RSPSA.369..425C,Chandrasekhar:1992bo, Lenzi:2021njy}. It was first noted by Chandrasekhar that the first few KdV integrals where equal for the isospectral Regge-Wheeler and Zerilli potentials~\cite{1980RSPSA.369..425C,Chandrasekhar:1992bo}.Recently, this equality has been demonstrated in Refs.~\cite{Lenzi:2021njy,Lenzi:2024tgk} for all the infinite hierarchy of KdV integrals for the Regge-Wheeler and Zerilli potentials, as well as for the DT-related ones. In particular, in~\cite{Lenzi:2021njy,Lenzi:2024tgk} it has been shown that this fact is due to the presence of a new hidden symmetry in the dynamics of the perturbations of vacuum spherically symmetric spacetimes, connected to DTs and the completely integrable flow of the KdV equation (see also Sec.~\ref{S:GF-from-KdV}). 

This motivates even further the connection of the KdV integrals to spectral properties of the system, making them a simple tool to address, at least, isospectrality breaking in modified theories. Actually, the KdV integrals  immediately provide a very simple way to determine the isospectrality (breaking) between two different potentials since they are easy to evaluate and because two non-isospectral potentials give different integrals.
Indeed, this argument has also been used in Ref.~\cite{Silva:2024ffz} for the study of the QNM spectra in an EFT extension of GR, although restricted to the first KdV integral. In this section we also consider this EFT-motivated potential and extend the argument beyond the first KdV integral, which brings a more detailed view of the breaking of isospectrality in terms of KdV integrals. This kind of analysis already puts some constraints on the type of potentials that we are allowed to consider. Clearly, the modified potential~\eqref{modified-V} needs to be integrable and infinitely differentiable. Indeed, if we consider, for instance, piece-wise defined potentials, something that is very common in simple quantum mechanical exercises (like the step or square barrier potentials), they introduce in general jumps in the derivatives of the potentials (described by distributions like the Heaviside theta function or the Dirac delta distribution and its derivatives).  Then, the non-linear character of the KdV densities makes it very difficult to define, even in the sense of distributions, the associated KdV integrals. Therefore, we can think, as a criterion of admissibility of a potential or of a modification of a standard potential,  the condition that it has a well-defined set of KdV integrals.

The KdV integrals themselves are evaluated with the definition in Eq.~\eqref{kdv-integrals}.  Nevertheless,
as outlined in Appendix~\ref{App:KdV-densities}, we can simplify the computation by realizing that the algorithm of Eq.~\eqref{kdv-integrals} produces a number of total derivative terms that do not contribute to the integral since they vanish at the boundaries ($x\rightarrow\pm\infty$ in our problem). It is possible to also create an algorithm to eliminate this terms in a systematic way before we evaluate the integrals, saving a significant amount of computer time. We have implemented such a reduction algorithm using Mathematica~\cite{Mathematica}. This is particularly useful when one wants to evaluate a large number of KdV integrals.  This is illustrated with the expressions shown in Appendix~\ref{App:KdV-densities}. This procedure, apart from  suppressing systematically all the total derivative terms in the KdV densities, it also reduces significantly the maximum derivative order.

\subsection{Environmental Corrections to the BH Potentials}
\label{Ss:PT-bump}

In this section, we consider a P\"oschl-Teller correction to the Regge-Wheeler potential, i.e. a potential of the form prescribed in Eq.~\eqref{modified-V} where the deviation is given by
\begin{eqnarray}
    \delta V = \frac{\sech^2\left[\alpha(x-x^{}_0)\right]}{r^{2}_s} = \frac{1}{r^2_s\,\cosh^2\left[\alpha(x-x^{}_0)\right]}  \,,
    \label{bump-potential-expression}
\end{eqnarray}
where $x_0$ is the location of the maximum of the P\"oschl-Teller correction and $\alpha$ is the width, as in Sec.~\ref{S:PT-fit} (the smaller the $\alpha$ the wider the P\"oschl-Teller potential barrier). Depending on the value of $x_0$ we have a perturbation close to infinity ($x^{}_0/r^{}_s\gg 1$), close to the maximum of the potential ($x^{}_0/r^{}_s \sim 1$), or close to the BH horizon ($x^{}_0/r^{}_s\ll -1$). Modifications of this type have been considered recently in the literature as a toy model for studying the QNM instability in the presence of environmental effects~\cite{Barausse:2014tra,Cheung:2021bol,Cardoso:2021wlq}. Nevertheless, the effects of varying the width of the P\"oschl-Teller potential do not seem to have been investigated in this context. Here, for the sake of simplicity, we only consider the  potential for odd-parity perturbations (as usually done in the references just mentioned). In GR, the isospectrality between the odd- and even-parity sectors is intimately related to the fact that the two potentials in GR are connected by a DT~\cite{1980RSPSA.369..425C, Glampedakis:2017rar, Lenzi:2021njy} (see Sec.~\ref{Ss:Isospectrality-Darboux}). It would be therefore tempting to just translate the same result to the presence of environmental effects, but we believe that this deserves a deeper study, beyond the scope of the present work. Indeed, the isospectrality between odd- and even-parity modes in GR is deeply related to the existence of the so-called algebraically special mode (see Sec.~\ref{Ss:Isospectrality-Darboux}), which is a particular single-frequency solution of the vacuum GR perturbative equations whose frequency is purely imaginary.

\begin{figure}[h!]
\centering
\includegraphics[width=0.8\textwidth]{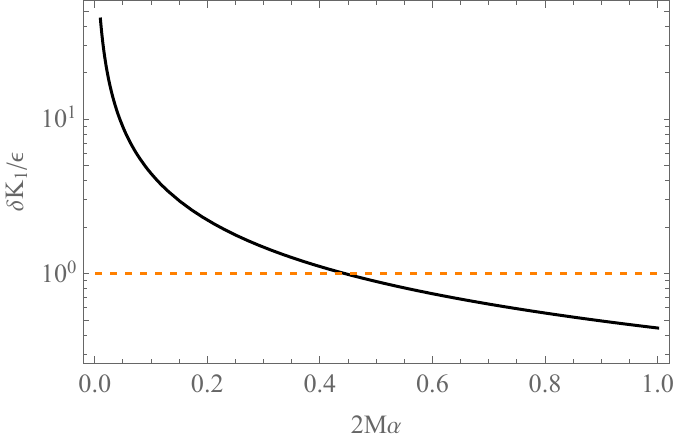}
\caption{Relative error $\delta \mathcal{K}_{1}$~\eqref{KdV-rel-error} divided by $\epsilon$ (black line) and the line corresponding to the threshold value of $\delta \mathcal{K}_{1}=\epsilon$ (orange dashed line), above which we have the instability region for the first KdV integral.}
\label{Fig:KdVbump_1}
\end{figure}

In the framework we have just established, we have introduced two more parameters, in addition to the intensity of the potential modification, namely the position of the maximum $x_0$ and the width of the potential $\alpha$. The variation of these parameters can affect the spectral stability. Therefore, in order to properly address the stability properties of the KdV integrals, one should consider their dependence on the parameters: $(\epsilon, x_0, \alpha)$. Let us start from the first KdV integral for the corrected potential with a P\"oschl-Teller “bump”. Given that the first density is just the potential and hence it is linear, the expression can be easily obtained analytically to give
\begin{eqnarray}
    \mathcal{K}^{\epsilon}_{1} = \mathcal{K}^{\mathrm{RW}}_{1} + \frac{2\epsilon}{r^{2}_s\,\alpha} \,.
    \label{K1-bump-exact}
\end{eqnarray}
It is therefore clear from Eq.~\eqref{stability-requirement}, that the stability of the first KdV integral does not depend neither on the intensity of the perturbation $\epsilon$ nor on the position of the maximum $x_0$, but rather only on the width of the perturbation $\alpha$. Indeed, the stability requirement~\eqref{stability-requirement} simply reads 
\begin{eqnarray}
   r^{2}_s\, \alpha\, \mathcal{K}^{\mathrm{RW}}_{1} \gtrsim 2\,.
\end{eqnarray}
Then, we can find the turning point beyond which the P\"oschl-Teller “bump” destabilizes the first KdV integral at
\begin{equation}
\alpha = \frac{2}{r^{2}_s\,\mathcal{K}^{\mathrm{RW}}_{1}} \,.    
\end{equation}
This is shown in Fig.~\ref{Fig:KdVbump_1}, where we can see that in  order to destabilize the first KdV integral, one has to produce a sufficiently wide perturbation, irrespectively of the location of the peak of the bump.

\begin{figure}[h!]
\centering
\includegraphics[width=0.49\textwidth]{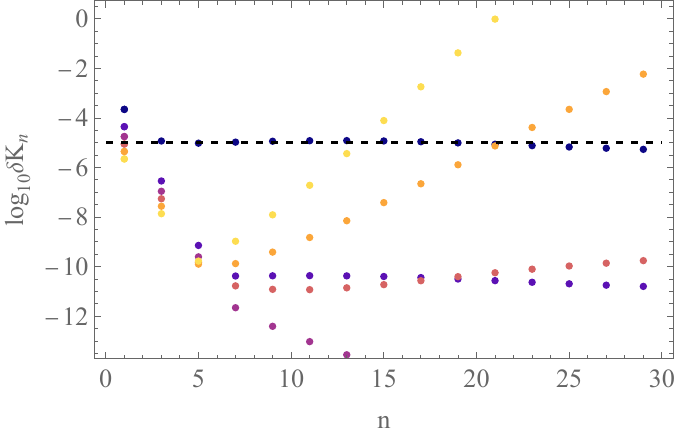}
\includegraphics[width=0.49\textwidth]{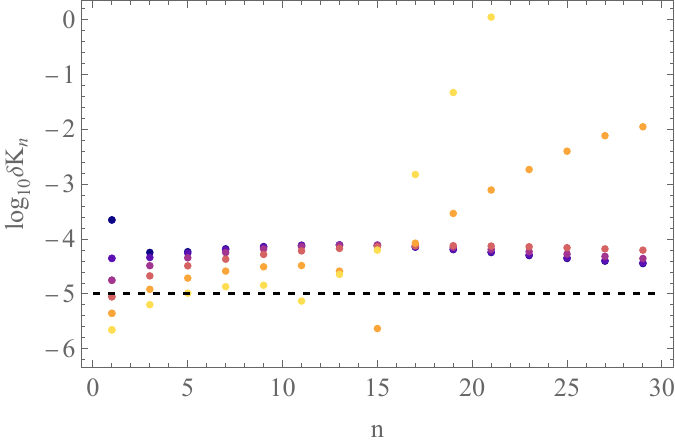}
\caption{Relative error $\delta \mathcal{K}_{n}$~\eqref{KdV-rel-error} for $x_0= 80 r^{}_s$ (left panel) and $x_0=0$ (right panel) for different values of the width of the potential bump, i.e. $r_s\,\alpha = 1/50, 1/10, 1/4, 1/2, 1, 2$, where the progression is represented starting from darker colors to brighter ones. These plots are done by fixing $\epsilon = 10^{-5}$ (black dashed line in both panels).}
\label{Fig:KdVbump_cp_d}
\end{figure}

Going beyond the first KdV integral is a computational matter involving the construction of the KdV densities, which are differential polynomials in the potential. 
In Fig.~\ref{Fig:KdVbump_cp_d}, we show the plot of the first thirty KdV integrals for fixed values of the position of the maximum of the bump while varying its width (again, only integrals with odd $n$ are represented). Already at this stage one notices a general trend upon varying the bump width $\alpha$: the wider the bump the more unstable are the KdV integrals. This is somewhat expected but not trivial because of the high degree of nonlinearity in the definition of the KdV integrals. Moreover, we find that corrections close to the maximum of the Regge-Wheeler potential induce stronger instabilities in the KdV integrals, especially in the lower order ones.

\begin{figure}[h!]
\centering
\includegraphics[width=0.4\textwidth]{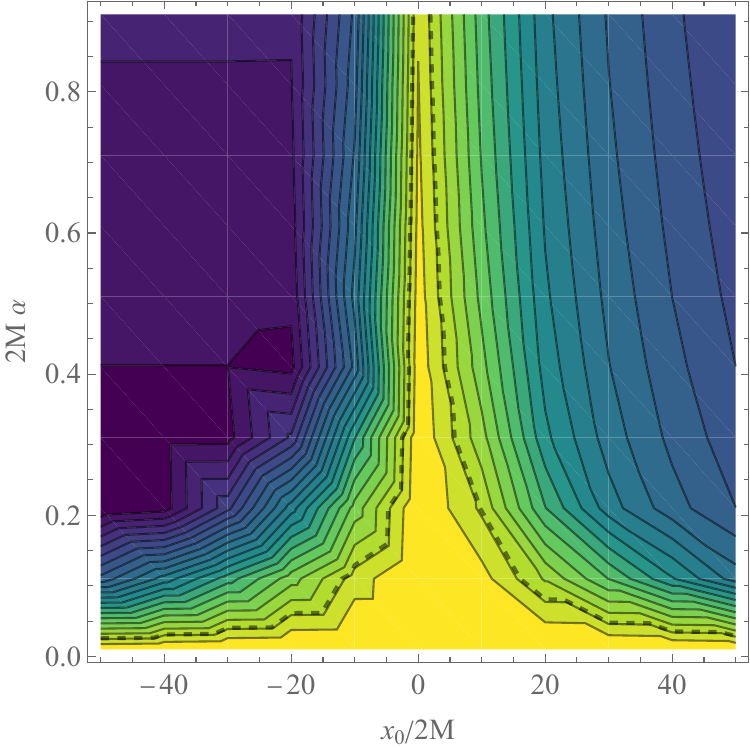}
\includegraphics[width=0.08\textwidth]{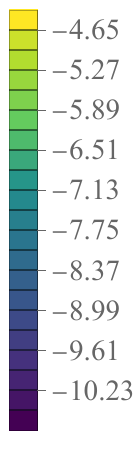}
\includegraphics[width=0.4\textwidth]{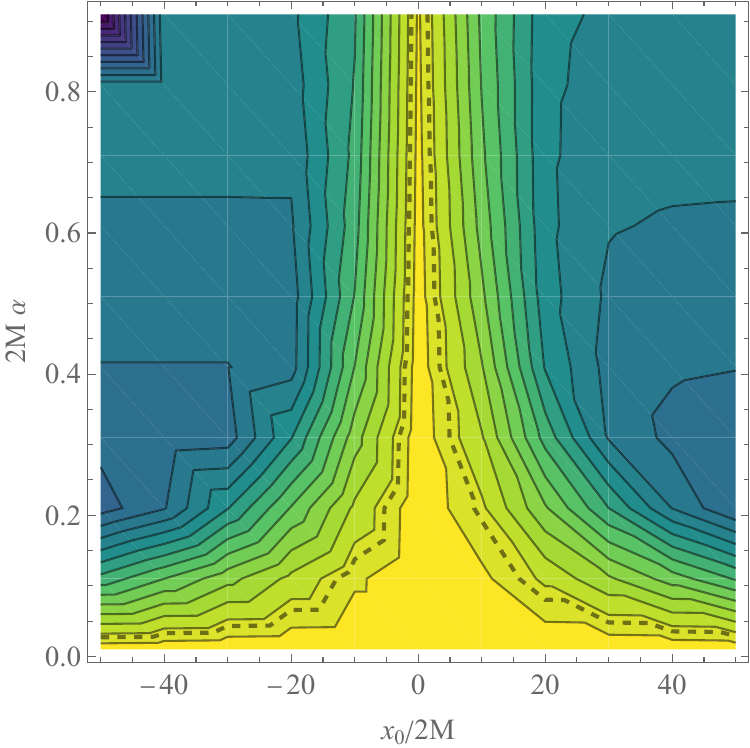}
\includegraphics[width=0.08\textwidth]{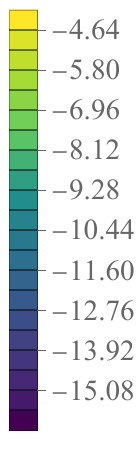}
\caption{Relative error $\log_{10}\delta \mathcal{K}_{5}$~\eqref{KdV-rel-error} (left panel) and $\log_{10}\delta \mathcal{K}_{7}$ (right panel) for $\epsilon = 10^{-5}$ and for varying $x_0$ and $\alpha$. The dashed line correspond to the threshold value $\delta \mathcal{K}_{n}=\epsilon$.}
\label{Fig:KdVbump_x0-alpha}
\end{figure}

In order to understand better the interplay among the free constants that have a relevant role in this study, it is useful to plot the KdV integrals against two parameters at the time. In Fig.~\ref{Fig:KdVbump_x0-alpha} we show two-dimensional color-map plots of the values of two selected KdV integrals with respect to the bump width $\alpha$ and its maximum position $x_0$, while fixing the bump amplitude to $\epsilon=10^{-5}$. Narrow corrections to the potential are usually quite stable, unless the deviation is centered on the maximum of the Regge-Wheeler barrier. It actually appears that, no matter how sharp the deviation is, if the P\"oschl-Teller bump has its maximum coinciding with the Schwarzschild potential one then, the KdV integrals (beyond the first one) are unstable under small perturbations. On the other hand, when we move the maximum of the bump, $x_0$, farther away from the Regge-Wheeler maximum, it becomes easier to make the chosen KdV integral stable if the width of the potential bump is sufficiently small. A similar behavior can be observed for higher integrals as well. What it is important to remark here is that, despite this is a toy-model, the width of the P\"oschl-Teller deviation is actually relevant to properly assess the stability of the KdV integrals. This, at least, raises the doubt about whether such a parameter may be relevant as well in the study of the  instability of the QNMs via the computation of the pseudospectrum (see, e.g.~\cite{Jaramillo:2020tuu,Jaramillo:2021tmt,Jaramillo:2022kuv}).

\begin{figure}[h!]
\centering
\includegraphics[width=0.4\textwidth]{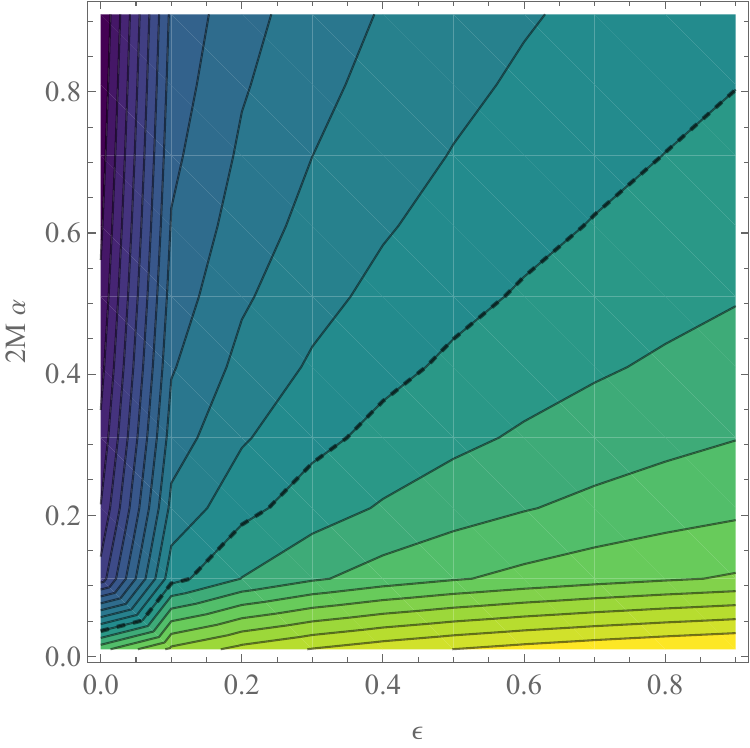}
\includegraphics[width=0.08\textwidth]{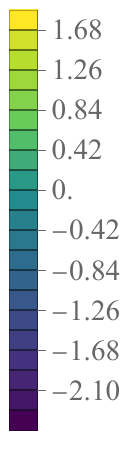}
\includegraphics[width=0.4\textwidth]{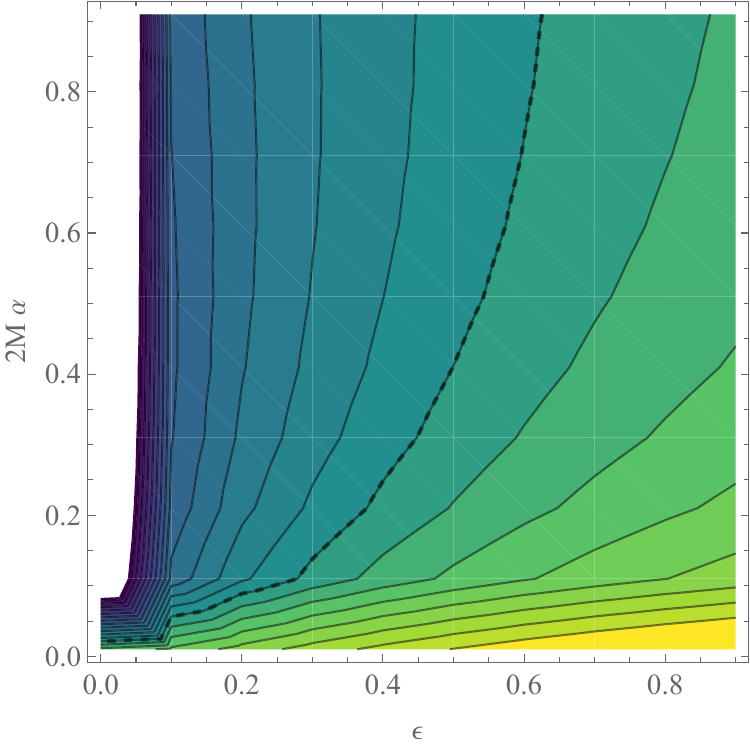}
\includegraphics[width=0.08\textwidth]{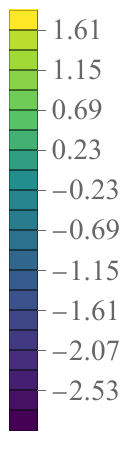}
\caption{Relative error $\log_{10}\delta \mathcal{K}_{3}$~\eqref{KdV-rel-error} (left panel) and $\log_{10}\delta \mathcal{K}_{5}$ (right panel) for $x_0 = 80 r^{}_s = 160 M$ and for varying $\epsilon$ and $\alpha$. The black dashed line corresponds to the threshold value above which we have instability, i.e. $\delta \mathcal{K} = \epsilon$.}
\label{Fig:KdVbump_cp1_eps-alpha}
\end{figure}

\begin{figure}[h!]
\centering
\includegraphics[width=0.4\textwidth]{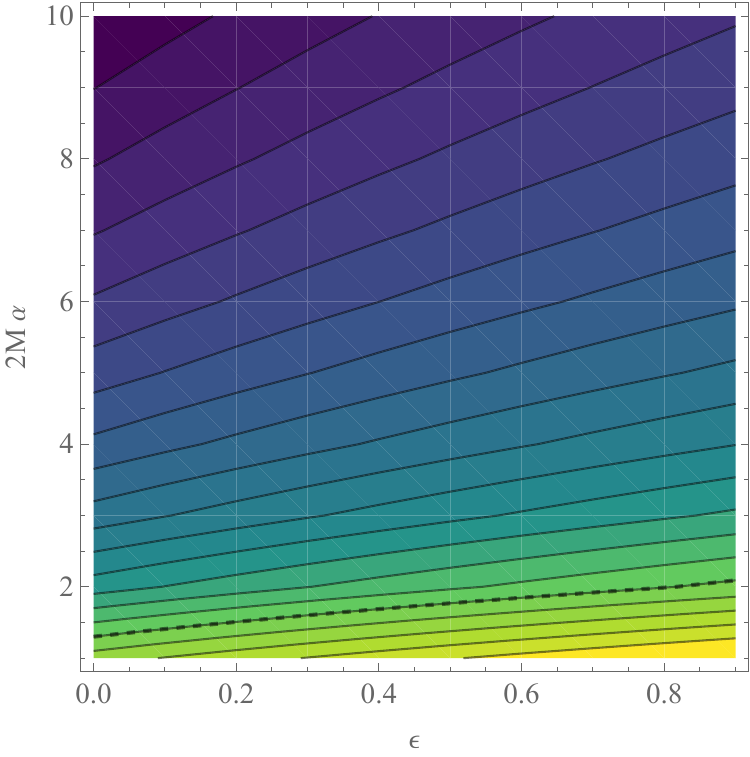}
\includegraphics[width=0.08\textwidth]{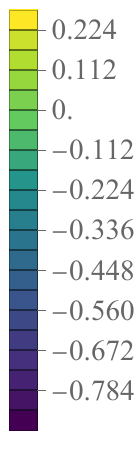}
\includegraphics[width=0.4\textwidth]{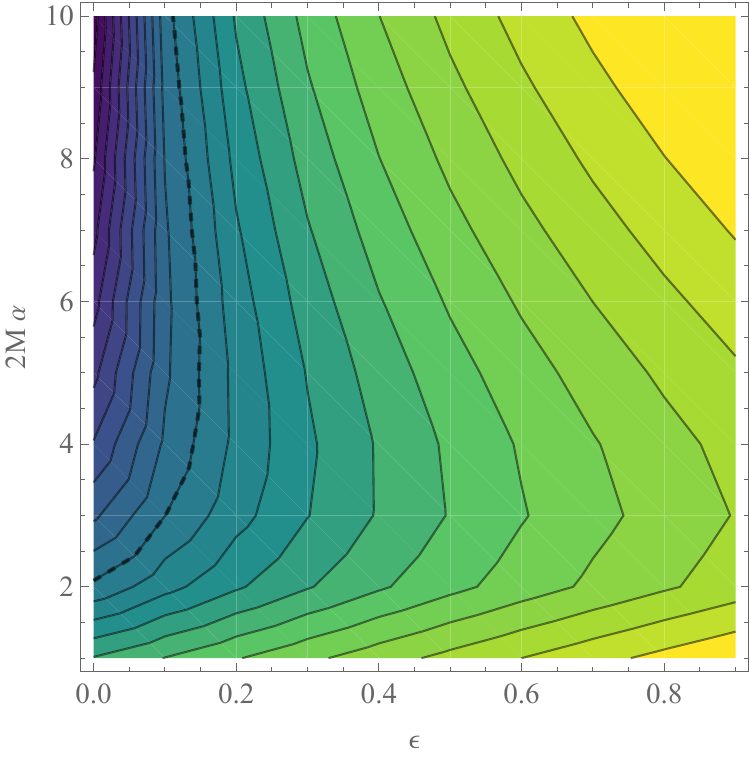}
\includegraphics[width=0.08\textwidth]{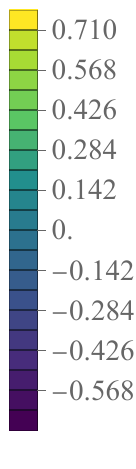}
\caption{Relative error $\log_{10}\delta \mathcal{K}_{3}$~\eqref{KdV-rel-error} (left panel) and $\log_{10}\delta \mathcal{K}_{5}$ (right panel) for $x_0 = 0 $ and for varying $\epsilon$ and $\alpha$. The black dashed line corresponds to the threshold value above which we have instability, i.e. $\delta \mathcal{K} = \epsilon$.}
\label{Fig:KdVbump_cp3_eps-alpha}
\end{figure}

In Figs.~\ref{Fig:KdVbump_cp1_eps-alpha} and~\ref{Fig:KdVbump_cp3_eps-alpha} we present color-map plots of the values of two selected KdV integrals against both the width $\alpha$ and the intensity $\epsilon$ of the bump, while fixing the location of its maximum to $x_0 = 80 r^{}_s = 160 M$ or to $x_0 = 0$. In contrast with the previous plots in Fig.~\ref{Fig:KdVbump_x0-alpha} (width $\alpha$ versus position of maximum $x_0$), now the behavior looks quite different for different KdV integrals. Indeed, for both $x_0 \simeq 0$ and $x_0 \gg 1$ we notice that there is already a quite substantial difference between the third and fifth KdV integrals. While the third KdV integrals are almost linearly unstable with varying $\epsilon$ (top panels in Figs.~\ref{Fig:KdVbump_cp1_eps-alpha} and~\ref{Fig:KdVbump_cp3_eps-alpha}), meaning that one can in principle tune the width $\alpha$ so that the integral is stable, already the fifth ones show a drastic change as stability can only be guaranteed for some values of the intensity $\epsilon$ (bottom panels in Figs.~\ref{Fig:KdVbump_cp1_eps-alpha} and~\ref{Fig:KdVbump_cp3_eps-alpha}), irrespectively of how thin the bump is. One can check that this behavior has the tendency to get worse for higher KdV integrals, which is also consistently shown in Fig.~\ref{Fig:KdVbump_cp_d}. Despite sharing this qualitative behavior, we also recover what we already observed in Fig.~\ref{Fig:KdVbump_cp_d}, namely that close to the maximum of the bump, i.e. $x_0 \simeq 0$, we find more severe instabilities than for values of the maximum of the bump far away from the BH potential maximum, that is, for  $x_0 \gg r^{}_s$.

A particular case that can be treated analytically is the one consisting of a \emph{small} narrow bump far away from the maximum of the BH potential barrier (we consider here the Regge-Wheeler potential only).   This is going to be defined by the following conditions
\begin{equation}
\epsilon \ll 1\,,\quad    
\hat{\alpha} = r^{}_s\, \alpha \gg 1\,, \quad
x^{}_0 \gg r^{}_s \,. 
\label{assumptions-RW-plus-bump}
\end{equation}
The case of the first KdV integral is already given in analytical form in Eq.~\eqref{K1-bump-exact}. For the other KdV integrals, which are all non-linear in the BH potential and its derivatives, factors that contain cross terms in the Regge-Wheeler and bump potentials (with different powers and different derivatives) can be generically neglected as guaranteed by the conditions in Eq.~\eqref{assumptions-RW-plus-bump}. Then, in general, we have
\begin{eqnarray}
\mathcal{K}^{\epsilon}_{n} = \mathcal{K}^{}_{n}[V^\mathrm{RW}+\epsilon\delta V] =  
\mathcal{K}^{\mathrm{RW}}_{n} + \mathcal{K}^{}_{n}[\epsilon\delta V] + \mathrm{cross~terms} \,.
    \label{Kn-bump-exact}    
\end{eqnarray}
This can be illustrated with the case of the next KdV integral, i.e. $\mathcal{K}^{\epsilon}_{3}$, for which we can write:
\begin{eqnarray}
\mathcal{K}^{\epsilon}_{3} = \mathcal{K}^{\mathrm{RW}}_{3} + \mathcal{K}^{}_{3}[\epsilon\delta V] - 2\epsilon \int^{+\infty}_{-\infty} dx \, V^{\mathrm{RW}} \delta V  \,.
    \label{K3-bump-expr}    
\end{eqnarray}
We can approximate the third term in this expression by taking into account the assumptions in Eq.~\eqref{assumptions-RW-plus-bump}. Indeed, the fact that the bump is far away from the maximum of the barrier and that the bump is relatively quite narrow means that this integral is dominated by the contribution around the bump. Around this region, the Regge-Wheeler potential does not change much so we can approximate the integral in~\eqref{K3-bump-expr} as:
\begin{eqnarray}
\int^{+\infty}_{-\infty} dx \, V^{\mathrm{RW}} \delta V \approx V^\mathrm{RW}(x^{}_0) \int^{+\infty}_{-\infty} dx \,  \delta V  \approx \frac{\ell(\ell+1)}{r^2_0} \frac{2}{r^{}_s\hat{\alpha}} \,. 
\end{eqnarray}
Here, $r_0$ is the value of the radial coordinate corresponding to the value $x_0$ of the tortoise coordinate. Since the bump is far away, we have approximated the Regge-Wheeler potential by its dominant terms as $r\rightarrow\infty\,$. Using the value of $\mathcal{K}^{}_{3}[\epsilon\delta V]$  (see Appendix~\ref{App:KdV-PT}) we finally have:
\begin{eqnarray}
\mathcal{K}^{\epsilon}_{3} & \approx & \mathcal{K}^{\mathrm{RW}}_{3} - 4\frac{\ell(\ell+1)}{r^2_0 r^{}_s}\frac{\epsilon}{\hat{\alpha}} + \frac{4}{3r^3_s}\frac{\epsilon^2}{\hat{\alpha}} 
\approx  \mathcal{K}^{\mathrm{RW}}_{3}  + \frac{4}{3r^3_s}\frac{\epsilon^2}{\hat{\alpha}}\,.
    \label{K3-bump-expr-2}    
\end{eqnarray}
This result can be extended to higher KdV integrals and one can see that, under the assumptions we have adopted [See Eq.~\eqref{assumptions-RW-plus-bump}], we obtain that
\begin{eqnarray}
\mathcal{K}^{\epsilon}_{n} \approx  \mathcal{K}^{\mathrm{RW}}_{n}  + \mathcal{K}^{}_{n}[\epsilon\delta V]\,,  \label{Kn-bump-expr}    
\end{eqnarray}
where $\mathcal{K}^{}_{n}[\epsilon\delta V]$, using the parametrization in Eq.~\eqref{bump-potential-expression}, and the results of Appendix~\ref{App:KdV-PT} for the KdV integrals of the P\"oschl-Teller potential, can be written as:
\begin{eqnarray}
\mathcal{K}^{}_{n}[\epsilon\delta V] = \frac{1}{r^n_s}\frac{P^{}_n(\epsilon)}{\hat{\alpha}}
\,,  \label{K-epsilon-n-bump-expr}    
\end{eqnarray}
where $P^{}_n(\epsilon)$ is a polynomial in $\epsilon$. This analytical approximation is consistent with the previous results on the stability of the KdV integrals by varying the width of the P\"oschl-Teller bump shown in Figs.~\ref{Fig:KdVbump_cp_d} and~\ref{Fig:KdVbump_cp3_eps-alpha} (left panels): the KdV integrals are strongly destabilized mostly when $\hat{\alpha}\rightarrow 0$, in which case the bump becomes extremely wide. However, it is not clear whether such configuration can be physically achieved in real scenarios. In any case, the important point here is that the terms coming from the bump scale with $1/\hat{\alpha}$, which is also true for the cross terms we have neglected according to our assumptions.

\subsection{Oscillatory Potentials}
\label{Ss:oscillatory-potentials}

\begin{figure}[h!]
\centering
\includegraphics[width=0.49\textwidth]{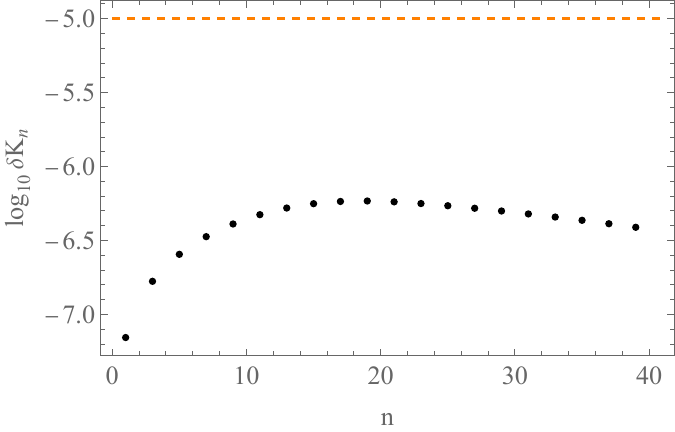}
\includegraphics[width=0.49\textwidth]{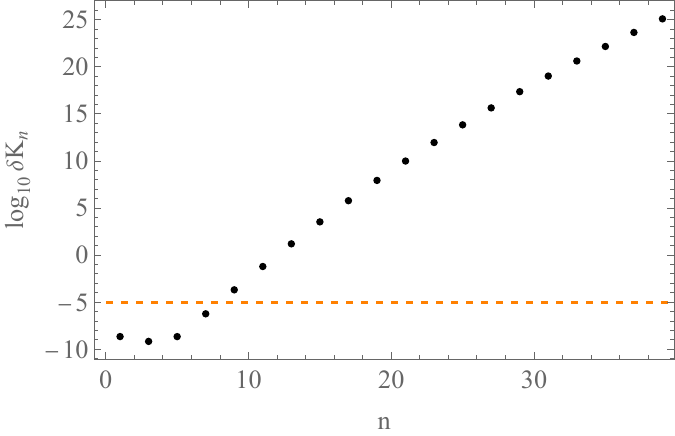}
\caption{Relative error $\log_{10}\delta \mathcal{K}_{n}$~\eqref{KdV-rel-error} for low frequency $\omega = 1/100$ (left panel) and high frequency $\omega = 301/100$ (right panel) for $\epsilon = 10^{-5}$. The dashed line correspond to the threshold value $\delta \mathcal{K}_{n}=\epsilon$.}
\label{Fig:KdVsin}
\end{figure}

In this section we consider the oscillatory potentials studied in Refs.~\cite{Jaramillo:2020tuu, Cardoso:2024mrw}, i.e.
\begin{eqnarray}
    \delta V = \frac{f(r) }{r^2}\sin \left(\frac{2 \pi  \omega r^{}_s}{r}\right)  \,.
    \label{sin-potential}
\end{eqnarray}
This gives a handle on the frequency dependence of the perturbation. 
In Fig.~\ref{Fig:KdVsin} we plot the relative error of the KdV integrals for low and high frequency of the sinusoidal potential. It is clear that low frequency modifications do not affect the stability of the KdV integrals while high frequency ones completely destabilize the higher integrals. This is again a qualitatively similar behavior to the one observed for the QNM spectrum~\cite{Jaramillo:2020tuu}. Moreover,  notice also that when the frequency is an integer then the correction to the first integral is zero by construction. For visual clarity, we do not include such values in the plots.

\begin{figure}[h!]
\centering
\includegraphics[width=0.45\textwidth]{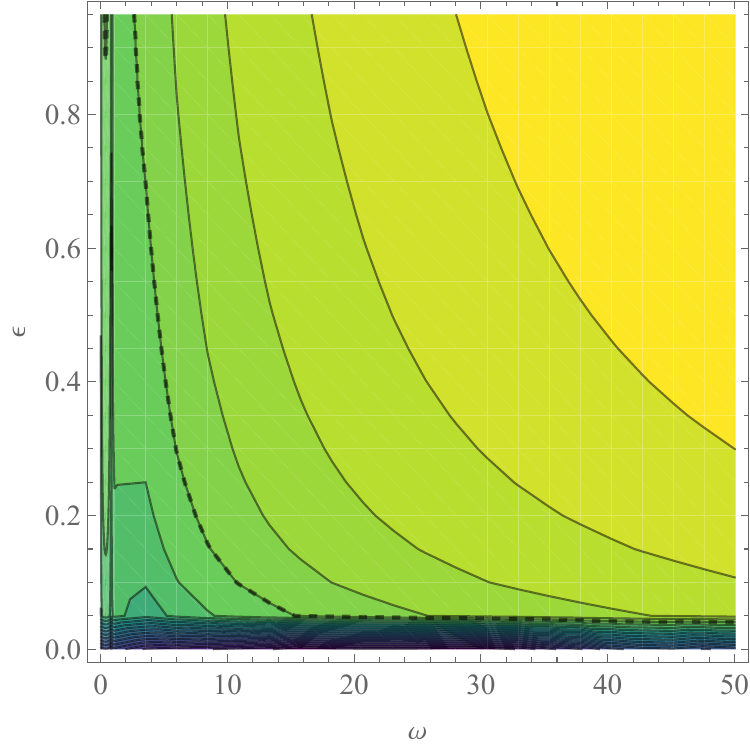}
\includegraphics[width=0.45\textwidth]{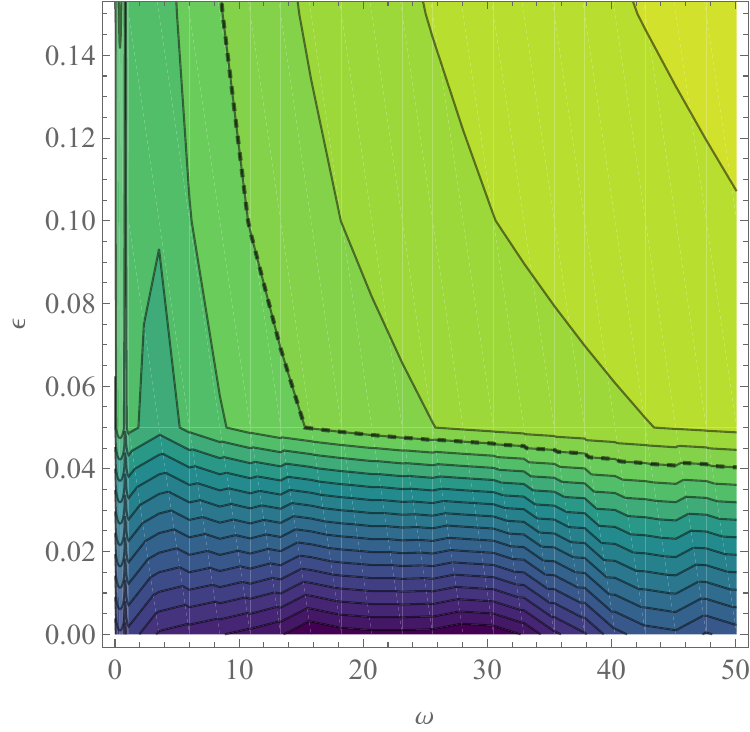}
\includegraphics[width=0.08\textwidth]{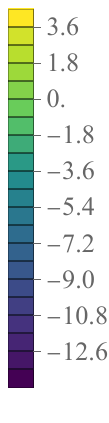}
\caption{Relative error $\log_{10}\delta \mathcal{K}_{7}$~\eqref{KdV-rel-error} for varying dimensionless frequency $\omega$ and intensity $\epsilon$ (left panel) and zoom over a smaller region (right panel). The dashed line correspond to the threshold value $\delta \mathcal{K}_{7}=\epsilon$.
}
\label{Fig:KdV7sin_eps-k}
\end{figure}

In Fig.~\ref{Fig:KdV7sin_eps-k} we present color-map plots of the relative error on the seventh KdV integral with respect to the dimensionless frequency $\omega$ and intensity $\epsilon$. 
The shape of the dashed black line, corresponding to the equality in Eq.~\eqref{stability-requirement}, stresses even more the fact that, while high frequency perturbations lead to severe instabilities, in the low frequency regime the instability region is only a small area corresponding to high intensity of the potential modification.

It is tempting to associate the destabilizing effects caused by the frequency of oscillation of the correction to the potential and the energy scale at which this correction is relevant in the underlying unknown theory. However, it is not completely clear what is the precise relation between the dimensionless frequency of the potential perturbation, as $\omega$ in Eq.~\eqref{sin-potential}, and the energy/frequency scale introduced for example by an EFT description.

\subsection{BH Potentials from Effective Field Theory Modifications of GR}
\label{Ss:EFT}

\begin{figure}[h!]
\centering
\includegraphics[width=0.8\textwidth]{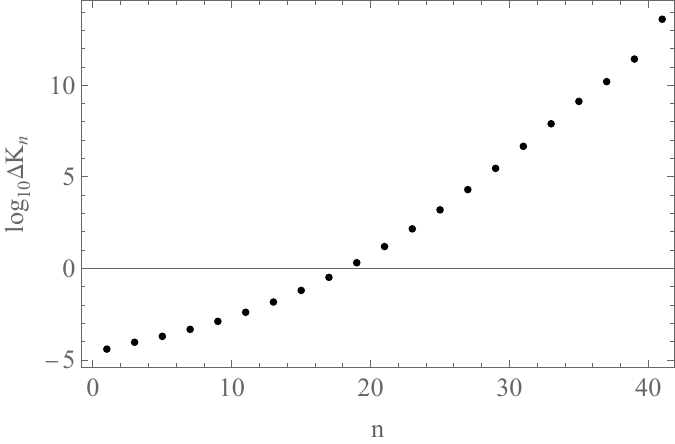}
\caption{Absolute difference $ \Delta \mathcal{K}_n = r^n_s\left|\mathcal{K}^{\epsilon,o}_n -\mathcal{K}^{\epsilon,e}_n\right|$ 
between the odd- and even-parity KdV integrals obtained from the two potentials in Eq.~\eqref{VEFT-odd-even} with $\epsilon = 10^{-5}$.}
\label{Fig:EFT_even_odd} 
\end{figure}

Let us now consider  EFT modifications of GR such that one can find the background BH solution and derive a master equation for the perturbations with deviations contained in the BH potentials~\cite{Endlich:2017tqa,Cardoso:2018ptl,Cano:2019ore, Silva:2024ffz}. This theory is constructed by adding cubic powers of the Riemann tensor to the Einstein-Hilbert Lagrangian, suppressed by the ratio of the EFT length-scale and the curvature scale. More details   on these EFT modifications of GR can be found in Refs.~\cite{deRham:2020ejn,Cano:2021myl,Cano:2023tmv,Cano:2023jbk}. The resulting potentials read as in Eq.~\eqref{modified-V} with (see Ref.~\cite{Silva:2024ffz} for details)
\begin{eqnarray}
\delta V^{\rm odd} = \frac{1}{r^2}\sum_{i=1}^{7} v^{\rm odd}_i \left(\frac{M}{r}\right)^i \,,
\qquad
\delta V^{\rm even} = \frac{4}{(\lambda r)^2}\sum_{i=1}^{10} v^{\rm even}_i \left(\frac{M}{r}\right)^i \,,
\label{VEFT-odd-even}
\end{eqnarray}
where $\lambda$ is defined in Eq.~\eqref{lambda-def-sch} and the coefficients $v^{\rm odd}_i$ and $v^{\rm even}_i$ are explicitly given in Appendix~\ref{App:EFT-coefficients}. The small parameter $\epsilon$ in this case depends on the ratio of the EFT length-scale and the curvature scale. As already mentioned, isospectrality of odd- and even-parity sectors is a special prediction of GR which is typically broken in most modifications of the theory or in effective descriptions of compact objects. This feature, in turn, provides an important and clear test of any theory when analyzing the ringdown part of GW signals. This is especially true in the case of future GW detectors~\cite{Sathyaprakash:2009xs,Barausse:2020rsu,LISA:2022kgy} which, thanks to the increased signal to noise ratio, improve considerably the chances of detecting QNMs and extracting with precision their frequencies. As already mentioned, isospectral potentials share the same KdV integrals. In Fig.~\ref{Fig:EFT_even_odd} we plot the absolute difference between the odd- and even-parity KdV integrals,  evaluated from Eq.~\eqref{VEFT-odd-even}, for the first 41 KdV integrals. It is clear from the picture that the KdV integrals for the two parities differ, even if only by slight deviations in some cases. The KdV integrals with even index are of course still vanishing thanks to the decaying properties of the corrections to the BH potential.

\begin{figure}[h!]
\centering
\includegraphics[width=0.8\textwidth]{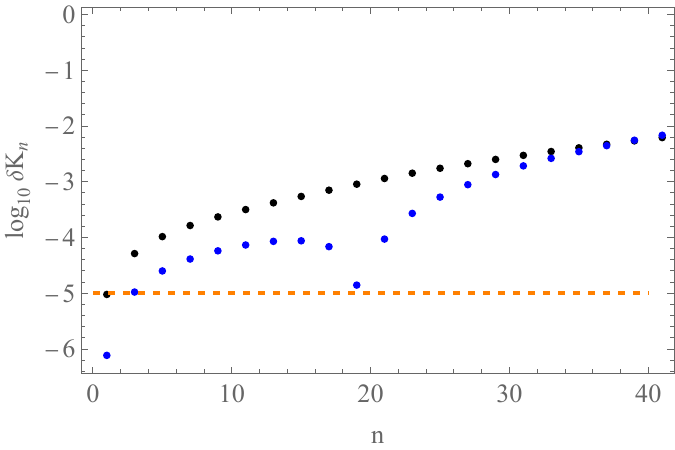}
\caption{Relative error $\delta \mathcal{K}_{2n-1}$~\eqref{KdV-rel-error} for $n=1,...,30$ for the odd (blue dots) and even (black dots) potential and the line corresponding to the value of $\epsilon$ (orange dashed line). In this case $\epsilon=10^{-5}$. The vertical axis is in logarithmic scale of base $10$.}
\label{Fig:KdV_EFT_n}
\end{figure}

In Fig.~\ref{Fig:KdV_EFT_n}, we show the behavior of the relative error~\eqref{KdV-rel-error} for the first twenty non-vanishing KdV integrals with EFT corrections with $\epsilon= 10^{-5}$. This plots shows in a very clear way that the only stable KdV integral is the first one, corresponding to the integral of the potential, while the higher ones quickly enter into the instability regime\footnote{These statements rely on the stability requirement in Eq.~\eqref{stability-requirement}, which only provides a “sharp” division between stable and unstable integrals. A more refined version would probably introduce intermediate behaviors such as meta-stability.}. It is also interesting to note that the EFT corrections to the even parity sector seem to be slightly more severe, even though the first KdV integral can be still considered stable. This trend seems to be broken when considering higher KdV integrals in the hierarchy since the relative errors shown in Fig.~\ref{Fig:KdV_EFT_n} for the odd (black dots) and even (blue dots) KdV integrals start crossing each other just before the fortieth KdV integral, suggesting that either the situation changes to this one and stays like this or we have an oscillatory behavior. Indeed, this is consistent with the modulation of isospectrality breaking we observe in Fig.~\ref{Fig:EFT_even_odd}. However, in order to make more precise statements on this behavior we would require analytical results or numerical calculation of a huge number of KdV integrals. 

\begin{figure}[h!]
\centering
\includegraphics[width=0.49\textwidth]{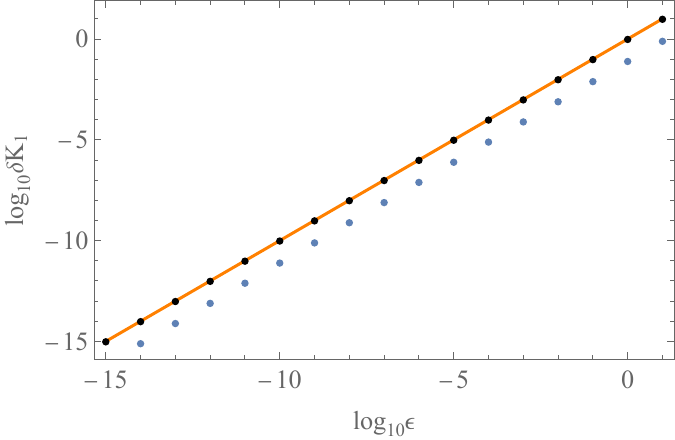} 
\includegraphics[width=0.49\textwidth]{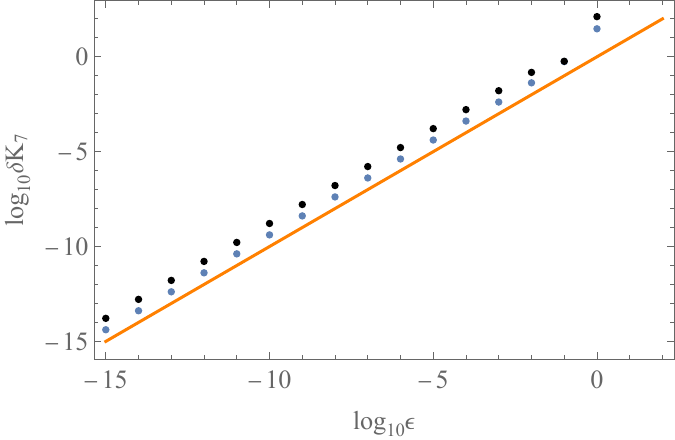}
\caption{Relative error $\delta \mathcal{K}_{1}$ (left panel) and $\delta \mathcal{K}_{7}$~\eqref{KdV-rel-error} (right panel) for $\epsilon$ ranging from $10^{-15}$ to $2$ for odd (blue dots) and even (black dots) parity. The diagonal line corresponds to plotting $\epsilon$ vs $\epsilon$ (orange line). The functions are in logarithmic scale of base $10$.}
\label{Fig:KdV_EFT_1_7}
\end{figure}

This trend is further supported by the results represented in Fig.~\ref{Fig:KdV_EFT_1_7}, where we show the (in)stability of $\mathcal{K}^{\epsilon}_1$ and $\mathcal{K}^{\epsilon}_7$ while varying $\epsilon$ over a broad range. Notice that we include in the plot values of the intensity $\epsilon>1$ only to check consistency, since it goes outside the range of an EFT modification and should not be trusted. The behavior of these first few KdV integrals seems to scale linearly with the intensity $\epsilon$ of the deformation. While by construction  this is true for the first KdV integral, it is not  in principle guaranteed for the higher ones. 

Overall, the behaviour of the KdV integrals for EFT corrected potentials seems to respect qualitatively the QNM instability expectations. Indeed, EFT corrections come from high-frequency modifications of the theory and this is what is expected to cause instability in the overtones, while maintaining the stability of the fundamental QNM~\cite{Jaramillo:2020tuu}. This enhanced sensitivity of overtones with respect to the fundamental QNM in EFT has been actually observed in Ref.~\cite{Silva:2024ffz}. Of course, any EFT comes with a lengthscale associated, which prevents us from considering arbitrarily high order KdV integrals as it prevents the evaluation of all the overtones~\cite{Silva:2024ffz}.

As a last comment to this section, let us compare the results obtained for EFT corrections and for P\"oschl-Teller “environmental” bumps in Sec.~\ref{Ss:PT-bump}. In fact, while the two scenarios come from very different physical motivations, one can find an interesting overlap in some cases. The orange points in Fig.~\ref{Fig:KdVbump_cp_d}, corresponding to a relatively sharp P\"oschl-Teller bump, close to the maximum of the Regge-Wheeler barrier, seems to have a very similar behavior to the odd EFT corrected KdV integrals, blue dots in Fig.~\ref{Fig:KdV_EFT_n}. This actually suggests that corrections close to the maximum of the potential are related to some high-frequency correction to the theory. Indeed, the change produced by both EFT corrects and the environmental bump (orange points in Fig.~\ref{Fig:KdVbump_cp_d}) are quite similar.

\section{Stability of the Greybody Factors: A Moment Problem Perspective}
\label{S:GF-stability}

In Sec.~\ref{S:GF-from-KdV} we reviewed the relation between the greybody factors and the KdV integrals which, as shown in Refs.~\cite{Lenzi:2022wjv,Lenzi:2023inn}, naturally appears in the form of a moment problem. In other words, apart from proportionality constants, the KdV integrals are the moments of a frequency-dependent distribution function that is logarithmic in the greybody factors. This establishes a solid mathematical ground, which we introduce to pave the way for future studies, to investigate the relation between the stability, under small perturbations, of the KdV integrals and the greybody factors. For instance, in Ref.~\cite{Lenzi:2022wjv} we used the fact that the existence and uniqueness conditions of the moment problem only depend on the moments to show that these are satisfied in the case of the Regge-Wheeler potential. In this sense, it is interesting to note the seemingly two-fold interpretation of the KdV integrals in this picture: the KdV integrals as first integrals of a Hamiltonian system, obtained by integration in configuration space of the associated conserved densities; or the KdV integrals as moments of a frequency-dependent distribution, and therefore defined by frequency-domain integrals.
Indeed, from Eqs.\eqref{kdv-integrals} and~\eqref{moments-hamburger} we can write the following relationships:
\begin{equation}
\mathcal{K}^{}_{2n +1}  = \int^{\infty}_{-\infty} dx\, \kappa^{}_{2n+1}(x) = (-1)^{n+1} 2^{2n +1}\int^{\infty}_{-\infty} d k\, k^{2n} \frac{\ln{T(k)}}{2\pi}\,,
\end{equation}
with the $ \kappa^{}_{n}(x)$ functions being the KdV densities defined by the recurrence relation in Eq.~\eqref{kdv-densities-recurrence} as differential polynomials, i.e. polynomials in the potential and its derivatives. This is the point we used in the discussion of the stability, under modifications of the BH potential, of the KdV integrals with respect to the one of the QNM spectrum in Secs.~\ref{S:PT-fit} and~\ref{S:KdV-integrals-modified-potentials}.

\begin{figure}
\centering
\includegraphics[width=0.48\linewidth]{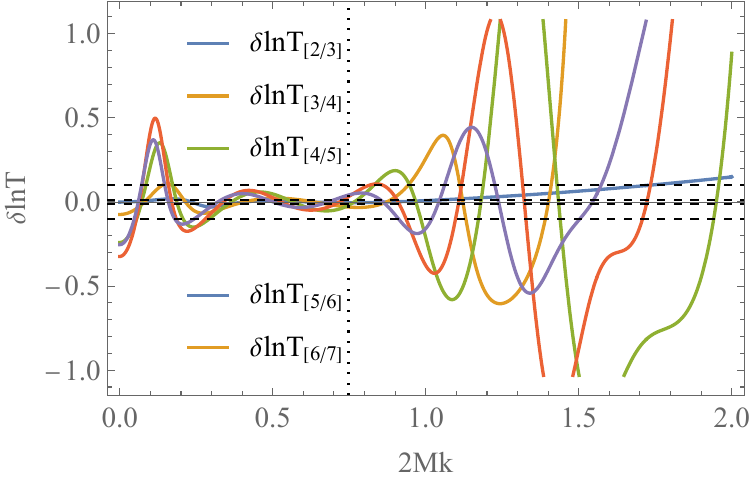}\,
\includegraphics[width=0.48\linewidth]{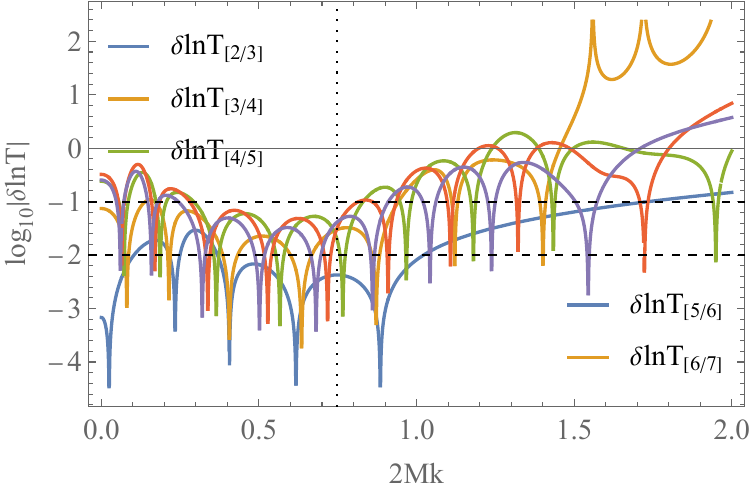}
\caption{Plot of the relative error on the distribution $\ln{T_{[N/M]}}$ (left panel) and the corresponding logarithmic plot (right panel) evaluated with the approximation in Eq.~\eqref{greybody-pade} for the P\"oschl-Teller bump correction~\eqref{bump-potential-expression} for $\alpha=1$ (narrow bump), $x_0=80 r_s$ and $\epsilon=10^{-2}$. The black dashed lines correspond to the limiting values $\epsilon = 10^{-2}$ and $\epsilon = 10^{-1}$. The black dotted line marks the value of the real part of the Schwarzschild QNM frequency. Note that we only plot the subdiagonal Pad\'e approximants ($M=N+1$) because the qualitative behavior is the same for the diagonal ones ($M=N$).
}
\label{Fig:lnT-stability}
\end{figure}

The moment problem perspective, see Eqs.~\eqref{moments-hamburger} and~\eqref{moments-hamburger-kdv}, already gives a first insight on the stability of the logarithmic distribution of the greybody factors. Indeed, as shown in Fig.~\ref{Fig:lnT-stability}, the $\ln{T}$ distribution is mostly stable (within one order of magnitude) for frequencies below (the real part of) the Schwarzschild fundamental QNM frequency~\footnote{The oscillations in the low frequency limit are accounted by the approximation procedure [see Eq.~\eqref{greybody-pade}] and improve when increasing the order of the Pad\'e approximants~\cite{Lenzi:2023inn}.}. The instability becomes relevant for higher frequencies. This behavior perfectly reflects the fact that the lower order KdV integrals/moments tends to be (almost) stable even for narrow bump perturbations, while the higher KdV integrals/moments becomes strongly unstable in this case (see Fig.~\ref{Fig:KdVbump_cp_d}), thus explaining the instability/stability of the distribution at high/low frequencies. 

\begin{figure}
\centering
\includegraphics[width=0.8\linewidth]{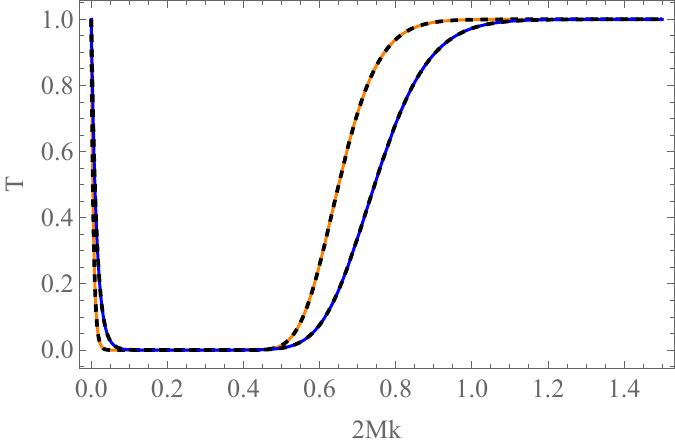}
    \caption{Plot of the greybody factors for the P\"oschl-Teller bump corrections for $\alpha=1/50$, $x_0=0$ and $\epsilon=10^{-5}$, evaluated with the approximation in Eq.~\eqref{T01} (orange line) and in Eq.~\eqref{T11} (blue line) versus the corresponding approximations for the Regge-Wheeler potential (black dashed lines). }
    \label{Fig:GF-bump}
\end{figure}

\begin{figure}
\centering
\includegraphics[width=0.49\textwidth]{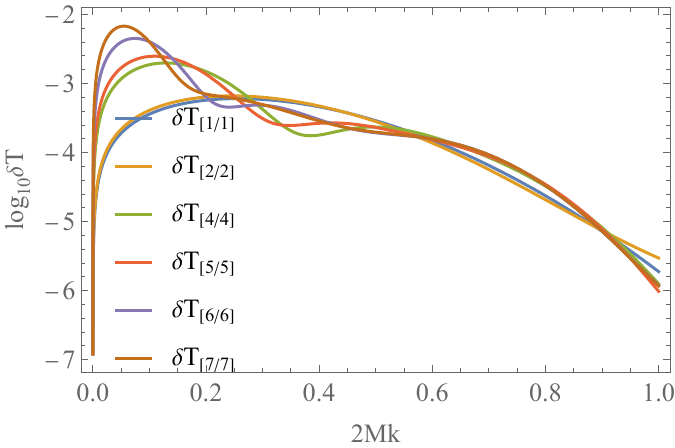}
\includegraphics[width=0.49\textwidth]{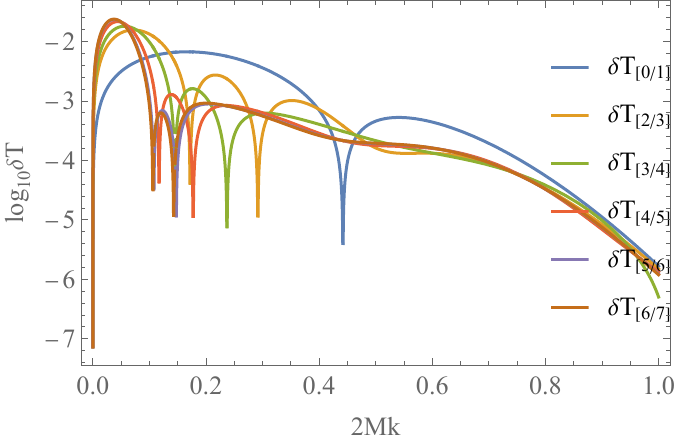}
\caption{Plot of relative error~\eqref{GF-rel-err} for the P\"oschl-Teller bump corrections for $r^{}_s\,\alpha=1/50$, $x_0=0$ and $\epsilon=10^{-5}$ for diagonal (left panel) and subdiagonal (right panel) Pad\'e approximants. }
\label{Fig:GF-diag-subdiag}
\end{figure}

\begin{figure}
\centering
\includegraphics[width=0.8\linewidth]{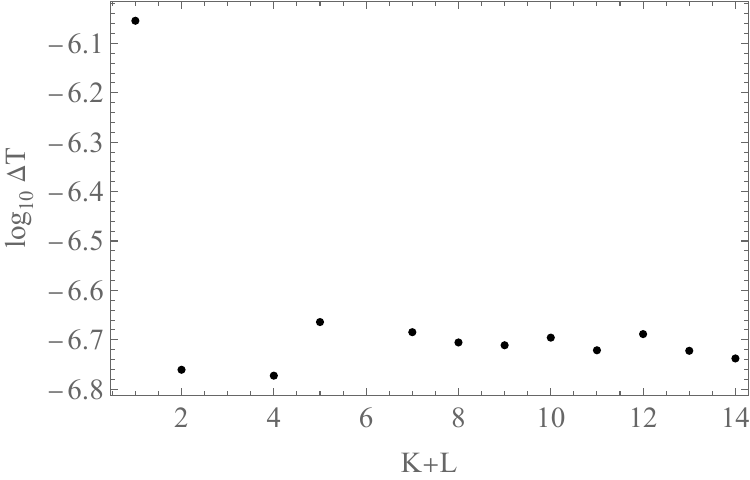}
\caption{Plot of the integrated error~\eqref{GF-integ-err} for the P\"oschl-Teller bump corrections for $r^{}_s\,\alpha=1/50$, $x_0=0$ and $\epsilon=10^{-5}$ for the Pad\'e approximants considered in the previous plots. On the x-axis we have $K+L$, tracking the order of the Pad\'e approximants in the notation of Eq.~\eqref{greybody-pade}.}
\label{Fig:GF-error-integ}
\end{figure}

The recent claims that ringdown signals may be encoded, at least partially, in the greybody factors~\cite{Oshita:2022pkc,Oshita:2023cjz,Okabayashi:2024qbz}, in addition to the frequently used sum of QNMs oscillations, enrich the discussion on the relation between the QNMs instability and the stability of the GW signal in the time domain. 
This has then led to consider stability properties of the greybody factors~\cite{Oshita:2024fzf,Rosato:2024arw} in contrast to those of the QNMs~\cite{Jaramillo:2020tuu,Jaramillo:2021tmt}. Our results on the instability of the KdV integrals under perturbations opens the door to a systematic study of the stability of the greybody factors  within the framework of the moment problem introduced in Refs.~\cite{Lenzi:2022wjv,Lenzi:2023inn} and briefly reviewed in Sec.~\ref{S:GF-from-KdV}. While such formal and analytic study inevitably requires a completely new investigation, which we postpone to future developments, we can already introduce some elements of analysis and prediction, serving as a guide for future works on the greybody factor stability. In fact, a simple intuition can be obtained from the first greybody factor approximants defined in Eq.~\eqref{greybody-pade}. Indeed, let us first recall the analytic expressions for the greybody factors obtained in Ref.~\cite{Lenzi:2023inn}, in particular those coming from the $[0/1]$ and $[1/1]$ Pad\'e approximants:
\begin{eqnarray}
T^{}_{[0/1]}(k) 
& = & 
\exp{\left(-4\pi  \frac{m^{2}_0}{m^{}_1} M k\,  e^{- \frac{4 m^{}_0}{m^{}_1} (Mk)^2}  \right)} \,,
\\
T^{}_{[1/1]}(k) 
& = & 
\exp\left(-16\pi \frac{m^{3}_1}{m^{2}_2} M k\,  e^{-\frac{8 m^{}_1}{m^{}_2}  (Mk)^2} \right) \,,
\end{eqnarray}
where $m_j = (2M)^{2j+1} \mu_{2j}$ are just the rescaled dimensionless moments. The equivalent expressions that depend explicitly only on the KdV integrals reads
\begin{eqnarray}
\label{T01}
T^{}_{[0/1]}(k) 
& = & 
\exp{\left(4\pi\, k \frac{\mathcal{K}^{2}_1}{\mathcal{K}^{}_3}\,  e^{\frac{4 \mathcal{K}^{}_1}{\mathcal{K}^{}_3} k^2}  \right)} \,,
\\
\label{T11}
T^{}_{[1/1]}(k) 
& = & 
\exp\left(16\pi\,k \frac{\mathcal{K}^{3}_3}{\mathcal{K}^{2}_5}  \,  e^{\frac{8 \mathcal{K}^{}_3}{\mathcal{K}^{}_5}  k^2} \right) \,.
\end{eqnarray}
In Fig.~\ref{Fig:GF-bump} we plot $T^{}_{[0/1]}$ and $T^{}_{[1/1]}$, given by Eqs.~\eqref{T01} and~\eqref{T11}, for the environmental-like P\"oschl-Teller corrections to the Regge-Wheeler potential. Let us comment that the first approximation, $T^{}_{[0/1]}$, is only qualitatively good while already from the second one, $T^{}_{[1/1]}$, the precision increases significantly (see Ref.~\cite{Lenzi:2023inn} for details). On the other hand, leaving the details of the approximation aside, we want to study the robustness of greybody factors in the presence of instabilities in the KdV integrals. It is clear from Fig.~\ref{Fig:GF-bump} that the effects of these instabilities do not qualitatively affect the greybody factors (as evaluated from the approximation in Eq.~\eqref{greybody-pade}). However, in order to address the stability of the greybody factors let us define the relative error
\begin{eqnarray}
\delta T^{}_{[N/M]}= \frac{\left| T^{\mathrm{bump}}_{[N/M]} -T^{\mathrm{RW}}_{[N/M]}\right|}{T^{\mathrm{RW}}_{[N/M]}} \,.
\label{GF-rel-err}
\end{eqnarray}
In Fig.~\ref{Fig:GF-diag-subdiag} we plot this function of the frequency for a wider number of approximations [see Eq.~\eqref{greybody-pade}]. While this seems to “locally” break the stability condition, we note that the integrated error actually satisfies it. Indeed, we can introduce the integrated error as
\begin{eqnarray}
    \Delta T^{}_{[N/M]}= \frac{\int_{0}^{\infty}dk \left| T^{\mathrm{bump}}_{[N/M]} -T^{\mathrm{RW}}_{[N/M]}\right|}{\int_{0}^{\infty}dk\,T^{\mathrm{RW}}_{[N/M]}}
    \,.
    \label{GF-integ-err}
\end{eqnarray}
The values of this integrated error are plotted in Fig.~\ref{Fig:GF-error-integ} for different approximants. As we can see, they  are smaller than $\epsilon = 10^{-5}$, as required. 
It is not clear whether such behavior is to be attributed to the breaking of our approximation for very small frequencies (see Ref.~\cite{Lenzi:2023inn}) or to the fact that higher KdV integrals tend to be more unstable as we have seen in this work. 

\begin{figure}
\centering
\includegraphics[width=0.49\textwidth]{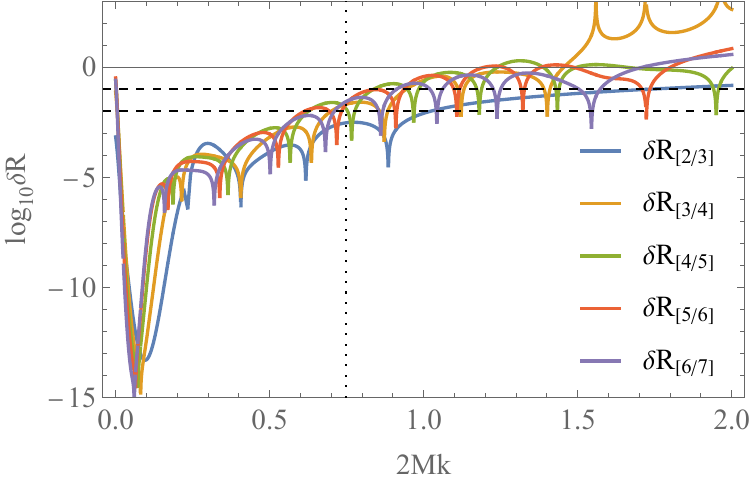}
\includegraphics[width=0.49\textwidth]{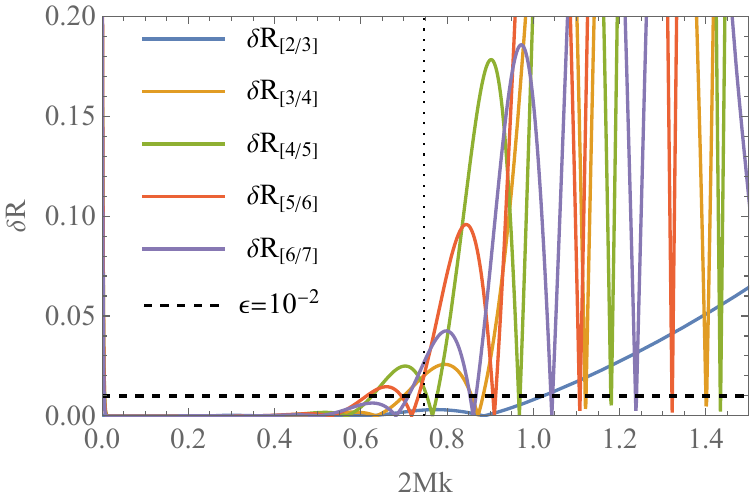}
\caption{Plot of the relative error on $R_{[N/M]}$ (left panel) and the corresponding logarithmic plot (right panel) evaluated with the approximation in Eq.~\eqref{greybody-pade} for the P\"oschl-Teller bump correction~\eqref{bump-potential-expression} for $\alpha=1$ (narrow bump), $x_0=80 r_s$ and $\epsilon=10^{-2}$. The black dashed lines in the left panel correspond to the limiting values $\epsilon = 10^{-2}$ and $\epsilon = 10^{-1}$. The black dotted line marks the value of the real part of the Schwarzschild QNM frequency. Again, we only plot the subdiagonal Pad\'e approximants ($M=N+1$) because the qualitative behavior is the same for the diagonal ones ($M=N$).}
\label{Fig:error-R}
\end{figure}

\begin{figure}
\centering
\includegraphics[width=0.49\textwidth]{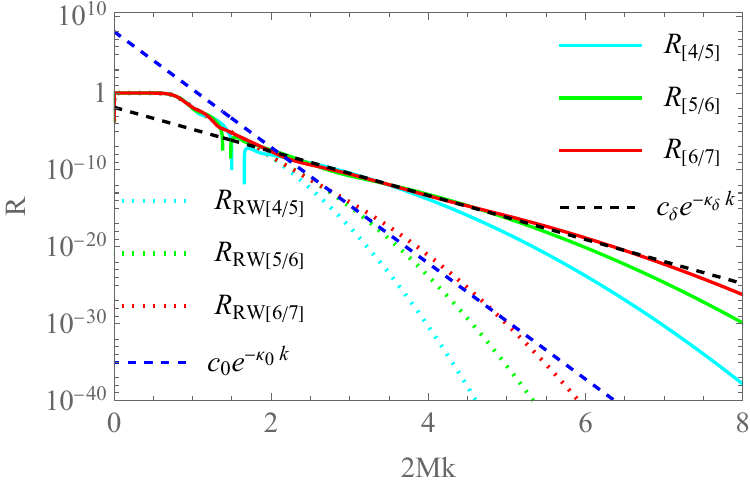}
\includegraphics[width=0.49\textwidth]{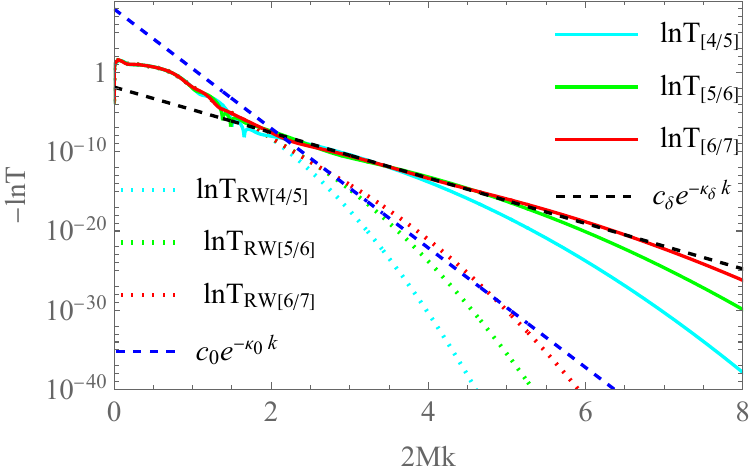}
\caption{Plot of $R_{[N/M]}$ (left panel) and $\ln{T_{[N/M]}}$ (right panel) evaluated with the approximation in Eq.~\eqref{greybody-pade} for the P\"oschl-Teller bump correction~\eqref{bump-potential-expression} for $\alpha=1$ (narrow bump), $x_0=80 r_s$ and $\epsilon=10^{-2}$. Here $c_{\delta} = 0.0014\,, r_s\kappa_{\delta} = 6.59$ and $c_{0} = 10^8\,,r_s\kappa_0 = 17.35$. We only plot the subdiagonal Pad\'e approximants ($M=N+1$) because the qualitative behavior is the same for the diagonal ones ($M=N$).}
\label{Fig:R-lnT}
\end{figure}

In relation to this, in Refs.~\cite{Oshita:2022pkc,Oshita:2023cjz,Okabayashi:2024qbz} the ringdown spectral amplitude is proportional to the reflection coefficient (reflectivity), i.e.
\begin{eqnarray}
|h_{\ell,m}(k)|\propto \sqrt{1-T_{\ell,m}(k)}= \sqrt{R_{\ell,m}(k)}
\,.
\end{eqnarray}
It is therefore instructive to consider the reflectivity $R^{}_{[N/M]} = 1- T^{}_{[N/M]}$, facilitating a better comparison with the results of Refs.~\cite{Oshita:2024fzf,Rosato:2024arw}. In Fig.~\ref{Fig:error-R} we show the relative error of the $R^{}_{[N/M]}$. This plot reveals an instability of the reflectivity for frequencies above the real part of the Schwarzschild fundamental QNM, in accordance with the results of Ref.~\cite{Oshita:2024fzf}. This is known to be related to the fact that the reflectivity is exponentially suppressed at high frequencies~\cite{Oshita:2024fzf,Rosato:2024arw} so the relative error ends up growing in this regime. In particular, the exponential decay is governed by the WKB approximation for the Schwarzschild case, i.e.
\begin{eqnarray}
  R \sim e^{-\kappa_{0}k}\,,\quad   \kappa_{0} = 2\pi \sqrt{\frac{2V_{\rm RW}(x_*)}{-V_{\rm RW, xx}(x_*)}} \,,
\end{eqnarray}
where $x_*$ is the maximum of the Regge-Wheeler potential in the tortoise coordinate.
In Ref.~\cite{Oshita:2024fzf} the authors further showed that high frequency limit of the reflectivity for perturbed potentials follows a similar exponential decay but with the coefficient in the exponential fixed by the maximum at the bump position, $x=x_0$ in our notation (see Sec.~\ref{Ss:PT-bump}). In other words:
\begin{eqnarray}
  R \sim e^{-\kappa_{\delta}k}\,,\quad   \kappa_{\delta} = 2\pi \sqrt{\frac{2V(x_0)}{-V_{, xx}(x_0)}} \,,
  \label{exp-decay-R-bump}
\end{eqnarray}
where $V$ is the potential given in Eq.~\eqref{modified-V} with $\delta V$ given by Eq.~\eqref{bump-potential-expression} and $x_0$ is the maximum of the bump $\delta V$.
Fig.~\ref{Fig:R-lnT} (left panel) actually confirms this behavior as we observe that the approximation correctly reproduce the exponential decay~\eqref{exp-decay-R-bump} at high frequency. Of course, the exact asymptotic behavior can only be obtained in the limit of infinite KdV integrals (the convergence of the approximation is studied in Ref.~\cite{Lenzi:2023inn}). The right panel of Fig.~\ref{Fig:R-lnT} shows instead the plot for $-\ln{T}$, which is proportional to the moment problem distribution (see Eq.~\eqref{moments-hamburger-kdv}). It is quite interesting to note that the high frequency behavior seems to match the one of the reflectivity with the same asymptotic decay. This is because, for high frequency we have
\begin{eqnarray}
    \ln{T} =\ln{(1-R)}\simeq \ln{\left(1-e^{-\kappa k}\right)}\simeq e^{-\kappa k} \simeq R
    \,.
    \label{approx-lnT-R}
\end{eqnarray}
Taking into account that the instability behavior of $ \ln{T}$ is understood in terms of the instability of the higher KdV integrals for narrow bumps (see Fig.~\ref{Fig:lnT-stability}), this seems to suggest a strong relation between the moment problem perspective and the instability of the reflectivity $R=1-T$~\footnote{We thank an anonymous referee for pointing out Eq.~\eqref{approx-lnT-R} in relation to this.}.
However, let us stress that a full stability assessment of these quantities requires a deeper numerical and analytical study, which definitely deserves more extensive investigations.
Nevertheless, the results we have shown here can be considered as preliminary results and the motivation for a more detailed analysis in the future that has to take into account the following points: i) The conditions imposed by the moment problem on the moments in order for a solution to exist and be unique; ii) higher orders in the approximation to test the convergence properties, since the two approximations considered here are only qualitatively good; iii) analytical tools in the context of the moment problem.

\section{Conclusions and Discussion}
\label{S:Conclusions-and-Discussion}

In this work, we have investigated the connection between the integrable structures that appear in the study of the dynamics of perturbed BHs~\cite{Lenzi:2021njy, Lenzi:2022wjv, Lenzi:2023inn, Jaramillo:2024nvr}, related to the KdV equation and the Virasoro algebra, and the spectral properties of the physical system, namely the greybody factors (which can be expressed only in terms of the KdV integrals as shown in Refs.~\cite{Lenzi:2022wjv, Lenzi:2023inn}) and the QNMs. 
The study of the KdV integrals for various potentials leads to very interesting indications of a link between the KdV integral stability properties and both the QNM instability and the greybody factors stability. From this study, we can identify the main ways to modify the potential of the master equations describing the BH gravitational perturbations. First, within GR, the only possibility is to change the background geometry, i.e. the Schwarzschild metric. But given that we have a uniqueness theorem for Schwarzschild, the Birkhoff theorem, this essentially means to add matter to the background spacetime, either in the form of matter outside the horizon or by changing the BH model by the model of a compact object, for instance a typical Tolman-Oppenheimer-Volkoff (TOV) relativistic stellar model. On the other hand, one can consider that GR is not the ultimate theory of gravity and study the many modifications to GR that have been proposed in the literature and that admit a well-defined perturbation theory. Finally, one may also consider phenomenological modifications and exotic compact objects.

Let us now summarize the main results of this work. 
In Sec.~\ref{S:PT-fit} we have first treated some (more or less) known cases, namely the approximation of the Regge-Wheeler potential with a P\"oschl-Teller one, which allows us to exploit the analytical solvability of the second. Or in other words, that the master equation can be exactly solved for the P\"oschl-Teller potential in terms of hypergeometric functions. Then, we have applied the same method to the  Zerilli potential for even-parity perturbations, and have checked that the approximations for the two different parity potentials consistently give almost indistinguishable results. We have further obtained the full sequence, the recurrence algorithm, of KdV integrals for the P\"oschl-Teller potential in exact closed form for the first time to our knowledge.  This has allowed us to easily calculate an arbitrary number of KdV integrals associated to the two fits, the P\"oschl-Teller potential fitted to the Regge-Wheeler and to the Zerilli potentials. Interestingly, comparing the KdV integrals for the fits to the real Regge-Wheeler values, we found that the first KdV integral differs significantly, showing the high sensitivity of the first integral to the long-range modifications of the BH potential, due to the exponential decay of P\"oschl-Teller at $x=\pm \infty$ (in contrast with the $r^{-2}$ decay of the standard BH potentials). On the other hand, the strongest “instability” is related to the higher KdV integrals, something which is expected to be related to local modifications of the potential since the fit is based on fixing only up to the second derivative at the maximum, while higher integrals depend on higher derivatives. If we take the point of view that the KdV integrals are moments of a frequency-domain distribution function (see Sec.~\ref{S:GF-from-KdV}), we can reinterpret the instability of the first moment/KdV integral with the infrared sensitivity of the lower moments while higher-order moments would be sensitive to high frequency modifications. Finally, the KdV integrals given by the P\"oschl-Teller fits seem to approach the Regge-Wheeler values in the eikonal limit, as already observed for the QNMs.
It is quite exciting to realize that similar properties have been found in the study of the BH QNM stability, even if a fully analytical study or a detailed numerical exploration is required to assess a strong connection.

In Sec.~\ref{S:KdV-integrals-modified-potentials}, we studied the sensitivity of the KdV integrals to small modifications of the potential. In particular, in Sec.~\ref{Ss:PT-bump} we considered small “bump” perturbations which should approximately imitate astrophysical environmental effects. We found that the stability of the KdV integrals depend on three parameters: the intensity, the location of the maximum of the bump and its width. We find there is a general trend in the case of high-order KdV integrals, which are more sensitive to thin perturbations while the lower-order ones are destabilized by wide perturbations. This seems to be qualitatively consistent with the findings in the P\"oschl-Teller case. Actually, a large width of the P\"oschl-Teller bump introduces similar infrared effects when we modify the long range behavior of the BH potential by fitting the Regge-Wheeler potential to a P\"oschl-Teller one. Moreover, the case of a thin bump affects high-order KdV integrals (which depend on higher-order derivatives of the potential) since it introduces more severe local modifications, in a similar way with the case of the P\"oschl-Teller fits. On top of these effects, the position of the maximum appears to contribute to increase the instability of the KdV integrals of any order: the closer the maximum of the bump is to the Regge-Wheeler maximum, the more unstable are the KdV integrals. 

In Sec.~\ref{Ss:oscillatory-potentials}, we have considered an oscillatory sinusoidal potential to further confirm these findings. The results clearly show that low-frequency modifications do not really affect the values of the KdV integrals while high-frequency ones strongly destabilize the high order KdV integrals. This is because the low-frequency case does not significantly affect either local or long range properties, while the high-frequency perturbations introduce rapid oscillations and as such, stronger local variations in the potential and its derivatives. 

Finally, in Sec.~\ref{Ss:EFT}, we have considered the variations in the BH potential introduced by EFT modifications to GR. In this case, we first showed how the KdV integrals are an important and simple indicator of the loss of isospectrality between the odd- and even-parity sectors, something which typically happens when introducing modifications to GR. Moreover, the behavior of the KdV integrals, within each parity, is clearly consistent with the previous cases in which the higher-order KdV integrals are more unstable. Indeed, the EFT approach introduces local modifications to the potential while not changing the decaying behavior. Let us note however, that in the evaluation of the QNMs~\cite{Silva:2024ffz}, only a limited number of those are meaningful due to the cut-off of the theory. Here, such cut-off should be imposed on the KdV integrals if we consider them as moments of the frequency distribution. However, it is not clear how it should emerge in the traditional definition as spatial integrals of KdV conserved densities. It would be tempting to introduce a spatial cut-off associated to the frequency/energy one coming from the EFT, but this deserves further investigation along with other cases in which the domain of integration may be modified, as in the case of exotic compact objects, where we have a surface instead of a BH horizon.  
Starting from these results, we can think of relating the frequency of oscillation of the BH potential, as clearly defined in the oscillatory potential case (see Sec.~\ref{Ss:oscillatory-potentials}), and the frequency/energy scale naturally introduced when considering some “first-principles” derivations of the master equations, such as in the case of EFT modifications of GR (see Sec.~\ref{Ss:EFT}). However, such connection may not exist at all and further analytical studies are necessary to address this issue.

In Sec.~\ref{S:GF-stability}, we have adopted the BH moment problem perspective introduced in Refs.~\cite{Lenzi:2022wjv, Lenzi:2023inn} (and reviewed in Sec.~\ref{S:GF-from-KdV}), which allows us to obtain the greybody factors entirely in terms of the KdV integrals, to study the interplay between the stability properties of the KdV integrals and the stability of the greybody factors. This seems to be of particular relevance in the light of some recent works showing how the amplitude of the ringdown signal may be modeled with greybody factors as an alternative to an expansion in QNMs~\cite{Oshita:2022pkc, Oshita:2023cjz, Okabayashi:2024qbz}. Our study shows: i) that the stability/instability of low/high KdV integrals correctly accounts for the stability/instability of the logarithm of the greybody factors (the moment problem distribution) at low/high frequencies; ii) preliminary evidence that the instability of the (higher) KdV integrals only affects the greybody factors locally, while the integrated error we have defined is actually well below the instability zone; iii) that the moment problem perspective to the greybody factors in terms of the KdV integrals seems to account for the high frequency instability of the reflectivity and to relate it to the instability of the higher order KdV integrals. In conclusion, there seems to be a very interesting qualitative relation between the KdV integrals and the QNM instability while, at the same time, the moment problem perspective may be the mathematical/analytical tool to explain, starting from the KdV integrals, the stability of the greybody factors. However, further in-depth studies, both analytical and numerical, are required to make stronger statements and establish a solid link between the integrable structures appearing in the perturbative dynamics of BHs and their spectral properties.

To conclude, the aim of this work has been to introduce and explore the deep connection between the KdV integrals and BH greybody factors and QNMs. This leaves a number of interesting open questions for future investigations. Among them, the question of more structural character consists in the full analytical understanding of the relation between KdV integrals and the BH QNM spectrum, which would provide a more solid mathematical ground to the results presented in this work.   
Moreover, as we have already mentioned, the moment problem provides a powerful analytical tool to handle the relation between KdV integrals and greybody factors. This should be explored in the future, especially in connection with the stability of the GW ringdown signal in mergers of binary BHs. On the other hand, the number of direct applications is quite large. One could extend this to different type of potentials, such as quantum corrected potentials as in~\cite{Cao:2024oud}, or to any other case satisfying the required minimal conditions. It could also be interesting to investigate in depth the relation with the eikonal limit defined by $\ell \to \infty$. Finally, another natural extension is to consider spacetimes with different asymptotics (like asymptotically anti-de Sitter or de Sitter spacetimes) and analyze the implications of a spatial “cut-off” as it would be the case for exotic compact objects.

\begin{acknowledgments}
ML and CFS have been supported by contract PID2019-106515GB-I00/AEI/10.13039/501100011033 (Spanish Ministry of Science and Innovation) and 2017-SGR-1469 (AGAUR, Generalitat de Catalunya). 
This work was also partially supported by the program Unidad de Excelencia Mar\'{\i}a de Maeztu CEX2020-001058-M (Spanish Ministry of Science and Innovation). ML has also been supported by a Juan de la Cierva contract FJC2021-047289-I funded by program MCIN/AEI/10.13039/501100011033 (Spanish Ministry of Science and Innovation) and by NextGenerationEU/PRTR (European Union).
AMA was supported by the JAE-ICU program (2023-ICE-07) to support the introduction to scientifc research, linked to the Mar\'{\i}a de Maeztu contract mentioned above;  by the \emph{Programme RI-24 - researcher in gravitational-wave astronomy} (Research and Innovation Program linked to Component 23, Investment 1, within the framework of the Recovery, Transformation and Resilience Plan, funded by the European Union – Next Generation EU, corresponding to the years 2024 and 2025); by the Spanish Agencia Estatal de Investigaci\'on grants PID2022-138626NB-I00, RED2024-153978-E, RED2024-153735-E, funded by MICIU/AEI/10.13039/501100011033 and the ERDF/EU; and by the Comunitat Aut\'onoma de les Illes Balears through the Conselleria d'Educaci\'o i Universitats with funds from the European Union - NextGenerationEU/PRTR-C17.I1 (SINCO2022/6719) and from the European Union - European Regional Development Fund (ERDF) (SINCO2022/18146).
We have used the Computer Algebra Systems Mathematica~\cite{Mathematica} to carry out most of the computations necessary for this paper.
\end{acknowledgments}

\appendix

\section{Scattering Problem for the P\"oschl-Teller Potential} 
\label{App:PT}

This is a problem in which the time-independent Schr\"odinger equation~\eqref{schrodinger} is exactly solvable in terms of hypergeometric functions. In the case of a P\"oschl-Teller potential barrier [see Eq.~\eqref{Vpt}]], the general solution reads (see Refs.~\cite{landau1981quantum,Cevik:2016mnr, Lenzi:2022wjv}):
\begin{eqnarray}
\nonumber
\psi
&=&
A\, 2^{\frac{ik}{\alpha}} \left[1 - \tanh^2(\alpha x)\right]^{-\frac{ik}{2\alpha}}
{}_{2}F_1 \left(\lambda -\frac{i k}{\alpha}, 1- \lambda -\frac{i k}{\alpha}  , 1-\frac{i k}{\alpha};\frac{1 - \tanh(\alpha x)}{2} \right)
\\
&+&
B\, \left[1 - \tanh(\alpha x)\right]^{\frac{ik}{2\alpha}} 
\left[1 + \tanh(\alpha x)\right]^{-\frac{ik}{2\alpha}}
{}_{2}F_1 \left(\lambda, 1- \lambda , 1+\frac{i k}{\alpha};\frac{1 - \tanh(\alpha x)}{2} \right)
\,,
\end{eqnarray}
where
\begin{equation}
\lambda =\frac{1}{2} +  i\beta \,.    
\end{equation}
By analyzing the asymptotic behaviour of the general solution, both at $x \rightarrow \infty$ (spatial infinity) and at $x \rightarrow -\infty$ (BH horizon), we can find analytical
expressions for the reflection and transmission coefficients. First, one can obtain the following expressions for the Bogoliubov coefficients [as defined in Eq.~\eqref{plane-wave-ab}]:
\begin{eqnarray}
a(k) & = & \frac{\Gamma\left(1-ik\right) \Gamma\left(-ik\right)}{\Gamma\left(\frac{1}{2}-i(k-\beta)\right)\Gamma\left(\frac{1}{2}-i(k+\beta)\right)} \,,
\label{ak-pt}
\\
b(k) & = & \frac{\Gamma\left(1-ik\right) \Gamma\left(ik\right)}{\Gamma\left(\frac{1}{2}+i\beta)\Gamma(\frac{1}{2}-i\beta\right)} \,,
\end{eqnarray}
where, for the sake of simplicity, we have chosen $\alpha =1$. Then, we can find the transmission and reflection coefficients from Eq.~\eqref{reflection-transmission-coefficient}. 
The transmission probability can be evaluated by taking the inverse modulus square of Eq.~\eqref{ak-pt} for real $k$ and the result can be found in Eq.~\eqref{pt-transmission},
where we have used well-known properties of the Euler Gamma function $\Gamma(z)$~\cite{Abramowitz:1970as}. Once the analytic expressions for the Bogoliubov coefficients are given, it is then easy to find the QNMs, with the usual practical definition, i.e. as the zeros of the $a(k)$ coefficient in Eq.~\eqref{ak-pt}. The result is shown in Eq.~\eqref{QNM-PT}.

\section{KdV Integrals for the P\"oschl-Teller Potential} 
\label{App:KdV-PT}

The series expansion of $\ln a(k)$ for the Pöschl-Teller potential involves certain subtleties, which we outline below to give, at the end of this section, the closed form of this expansion. First, we start from the analytical expression for $\ln a(k)$:
\begin{equation}
    \ln a(k)=\ln\left[\frac{\Gamma(1-ik)\Gamma(-ik)}{\Gamma(\frac{1}{2}-i(k-\beta))\Gamma(\frac{1}{2}-i(k+\beta))}\right].
\end{equation}
Here, $\Gamma(z)$ denotes the Euler Gamma function, and $\beta$ represents the parameter associated with the high-barrier Pöschl-Teller potential.
We then use the expansion of the logarithm of the Euler Gamma function~\cite{Abramowitz:1970as}:
\begin{eqnarray}
\ln \left[\Gamma(z)\right] &\simeq&
\left(z-\frac{1}{2}\right)\ln(z)-z+\frac{1}{2}\ln(2\pi) + 
\sum_{n=1}^{\infty}\frac{B_{2n}}{2n(2n-1)z^{2n-1}},\quad \text{as}~|z|\rightarrow +\infty\,,
\end{eqnarray}
where $B_{n}$ denotes the Bernoulli numbers. From here we can find the expansion of $\ln a(k)$ for $|k| \rightarrow \infty$. After restoring the constant factor $\alpha$, we obtain:
\begin{eqnarray}
\ln a(k) &=& \sum_{m=1}^{\infty}i\left[\frac{B_{2m}(-1)^{m+1}}{2m(2m-1)} 
+ \sum_{n=1}^{m}\frac{B_{2n}(2m-2)!(-1)^{m+1}}{(2n)!(2m-2n)!} \right. \nonumber\\[2mm]
& + & \sum_{n=1}^{m}\sum_{j=0}^{m-n}\frac{B_{2n}(2m-2)!(-1)^{j+n}(2\beta)^{2m-2n-2j}}{(2n)!(2j)!(2m-2n-2j)!2^{2m-2n-1}}
+ \sum_{j=0}^{m-1}\frac{(2m-2)!(-1)^{j}((2\beta)^{2m-2j}}{(2j)!(2m-2j-1)!2^{2m-1}} \nonumber \\[2mm]
& + & \left. \sum_{j=0}^{m}\frac{(2m-1)!(-1)^{j+1}(2\beta)^{2m-2j}}{(2j)!(2m-2j)!2^{2m-1}}+\frac{(-1)^{m+1}}{2(2m-1)}+\frac{(-1)^{m}}{2m}\right]\frac{\alpha^{2m-1}}{k^{2m-1}} 
\end{eqnarray}
Then, the KdV integrals for the Pöschl-Teller potential can be read off from this expression and Eq.~\eqref{ln-a} to give the closed form shown in Eq.~\eqref{pt-kdv-exact}.
In turns out that the KdV integrals written in the Appendix of Ref.~\cite{Lenzi:2023inn} are not correct, although the results of the paper are correct because we used the right expressions in Mathematica. The mistake was a typo in transcribing them into the Appendix. Here, we list the first non-vanishing KdV integrals [see Eq.~\eqref{kdv-integrals}] for the P\"oschl-Teller potential:
\begin{eqnarray}
\mathcal{K}^{}_1 & = & \frac{\alpha}{2} \left(4 \beta^2+1 \right) \,, 
\\
\mathcal{K}^{}_3 & = & -\frac{\alpha^3}{12} \left(4 \beta^2+1 \right)^2 \,, 
\\
\mathcal{K}^{}_5 & = & \frac{\alpha^5}{30}  \left(4 \beta^2+1\right)^2 \left(4 \beta^2+3\right) \,, 
\\
\mathcal{K}^{}_7 & = & -\frac{\alpha^7}{168}  \left(4 \beta^2+1\right)^2 \left(48 \beta^4+88 \beta^2+51\right) \,, 
\\
\mathcal{K}^{}_{9} & = &  \frac{\alpha^9}{90}  \left(4 \beta^2+1\right)^2 \left(64 \beta^6+208 \beta^4+284 \beta^2+155\right) \,, 
\\
\mathcal{K}^{}_{11} & = & -\frac{\alpha^{11}}{132}  \left(4 \beta^2+1\right)^2 \left(256 \beta^8+1280 \beta^6+3040 \beta^4+3856 \beta^2+2073\right) \,,
\\
\mathcal{K}^{}_{13} & = & \frac{\alpha^{13}}{2730}  \left(4 \beta^2+1\right)^2 \left(15360 \beta^{10}+108800 \beta^8+393088 \beta^6+860576 \beta^4  + 1070668 \beta^2\right. \nonumber \\[2mm] 
& + & \left.573405\right)\,, 
\\
\mathcal{K}^{}_{15} & = & -\frac{\alpha^{15} }{720} \left(4 \beta^2+1\right)^2\! \left(12288 \beta^{12}\!+\!116736 \beta^{10}\!+\!593152 \beta^8\!+\!1965824 \beta^6\!+\!4210384 \beta^4\!+\!5212104 \beta^2 \right. \nonumber \\[2mm]
& + & \left. 2788707\right) \,, \\[2mm] 
\mathcal{K}^{}_{17} & = &\frac{\alpha^{17}}{1530} \left(4 \beta^2+1\right)^2 \left(81920 \beta^{14}+1003520 \beta^{12}+6804480 \beta^{10}+31612160 \beta^8+102323648 \beta^6 \right.\nonumber
\\
&+&  \left. 217896816 \beta^4 +269394108 \beta^2+144103095\right) \,,
\\ \nonumber
\mathcal{K}^{}_{19} & = & -\frac{\alpha^{19}}{7980} \left(4 \beta^2+1\right)^2 \left(1376256 \beta^{16}+21102592 \beta^{14}+183844864 \beta^{12}+1137156096 \beta^{10} \right.
\\
&+& \left. 5153682944 \beta^8 + 16577459712 \beta^6+35249267776 \beta^4+43565974240 \beta^2 \right. \nonumber \\[2mm] &+& \left. 23302711005\right)\,.
\end{eqnarray}
%

\section{KdV Densities: Reduction of the Total Derivative Order} 
\label{App:KdV-densities}

The recurrence for the KdV densities is:
\begin{eqnarray}
\kappa^{}_1(x) & = & V(x) \,,\qquad \kappa^{}_2(x) =  - V'(x) \,,  \label{kdv-density-1} \\[2mm]
\kappa^{}_n(x) & = & - \kappa'^{}_{n-1}(x) - \sum_{m=1}^{n-2} \kappa^{}_{m}(x)\kappa^{}_{n-1-m}(x)\,, 
\label{kdv-density-n}
\end{eqnarray}
for any $n$ such that $n>2$. The KdV integrals are precisely the integrals of these KdV densities over the whole real line (i.e. for $x\in [-\infty,+\infty]$ as the relevant coordinate is the tortoise coordinate $x$):
\begin{equation}
\mathcal{K}^{}_n = \int^{+\infty}_{-\infty} dx\, \kappa^{}_n(x) \,,\quad n=1,\ldots\,\,.     
\end{equation}
From this expression, and for the purpose of computing the KdV integrals, it is clear that the KdV densities are defined up to total derivatives.  Using this observation, we can define \emph{reduced} KdV densities by adding, to the original densities in Eqs.~\eqref{kdv-density-1}-\eqref{kdv-density-n}, total derivative terms in order to cancel all the possible terms so as to reduce the order of the highest-order derivative as much as possible. 
This procedure reduces significantly both the maximum order of the derivatives of the potential present in the KdV density as well as the total number of terms. As a consequence, the reduced integrals are very useful for numerical (and symbolic) computations of the KdV integrals, as they reduce significantly the computational burden. 
It is possible to use an algebra system like MATHEMATICA~\cite{Mathematica}, which provides very useful tools for pattern recognition in mathematical expressions, to carry out this reduction automatically. We denote the densities obtained in this way by $\tilde{\kappa}_n(x)$, and we denote their equivalence with the previous one using the symbol $\simeq$. Then, we have the obtain the following expressions
\begin{eqnarray}
\tilde{\kappa}^{}_1(x) & \simeq & V(x) \,, \qquad \tilde{\kappa}^{}_2(x)  \simeq  0\,,
\label{kdv-tilde-density-1} \\[2mm]
\tilde{\kappa}^{}_n(x) & \simeq & - \sum_{m=1}^{n-2} \kappa^{}_{m}(x)\kappa^{}_{n-1-m}(x)\quad (n>2)\,. 
\label{kdv-tilde-density-n}
\end{eqnarray}
Then, using the MATHEMATICA script mentioned above, and the following notation to reduce the length of the formulae
\begin{equation}
V^p_{(n)} \equiv \left(\frac{d^n V(x)}{dx^n}\right)^p\,,    
\end{equation}
we obtain the following expressions for the reduced KdV densities (we show only the odd KdV integrals, as the even ones get reduced to zero; we also remove the $x$-dependence for the sake of simplicity):
\begin{eqnarray}
\tilde{\kappa}^{}_1 & \simeq & V \,, \\[2mm]
\tilde{\kappa}^{}_3 & \simeq & -V^2 \,, \\[2mm]
\tilde{\kappa}^{}_5 & \simeq & 2 V^3+V_{\text{(1)}}^2 \,, \\[2mm]
\tilde{\kappa}^{}_7 & \simeq & -5 V^4-10 V V_{\text{(1)}}^2-V_{\text{(2)}}^2\,, \\[2mm]
\tilde{\kappa}^{}_9 & \simeq & 14 V^5 + 70 V^2 V_{\text{(1)}}^2+14 V V_{\text{(2)}}^2+V_{\text{(3)}}^2 \,,\\[2mm]
\tilde{\kappa}^{}_{11} & \simeq & -42 V^6-420 V^3 V_{\text{(1)}}^2+35 V_{\text{(1)}}^4-126 V^2 V_{\text{(2)}}^2+20 V_{\text{(2)}}^3-18 V V_{\text{(3)}}^2-V_{\text{(4)}}^2 \,,\\[2mm]
\tilde{\kappa}^{}_{13} & \simeq & 132 V^7+2310 V^4 V_{\text{(1)}}^2-770 V V_{\text{(1)}}^4+924 V^3 V_{\text{(2)}}^2-462 V_{\text{(1)}}^2V_{\text{(2)}}^2-440 V V_{\text{(2)}}^3+198 V^2 V_{\text{(3)}}^2 \nonumber \\[2mm] 
& + & 22 V-110 V_{\text{(2)}}^{\text{}} V_{\text{(3)}}^2 V_{\text{(4)}}^2 + V_{\text{(5)}}^2 \,,\\[2mm]  
\tilde{\kappa}^{}_{15} & \simeq & -429 V^8-12012 V^5V_{\text{(1)}}^2 +10010 V^2 V_{\text{(1)}}^4 -6006 V^4
   V_{\text{(2)}}^2+12012 V V_{\text{(1)}}^2 V_{\text{(2)}}^2 +5720 V^2 V_{\text{(2)}}^3 \nonumber\\[2mm]
& - & 1001 V_{\text{(2)}}^4 - 1716 V^3V_{\text{(3)}}^2 +858 V_{\text{(1)}}^2 V_{\text{(3)}}^2 +2860 V V_{\text{(2)}}^{\text{}} -286 V^2 V_{\text{(4)}}^2V_{\text{(3)}}^2 +182 V_{\text{(2)}}^{\text{}} V_{\text{(4)}}^2 
\nonumber \\[2mm]
& - & 26 V V_{\text{(5)}}^2-V_{\text{(6)}}^2 \,, \\[2mm]
\tilde{\kappa}^{}_{17} & \simeq & 1430 V^9 +60060 V^6 V_{\text{(1)}}^2 -100100 V^3V_{\text{(1)}}^4 +10010 V_{\text{(1)}}^6 +36036 V^5 V_{\text{(2)}}^2 - 180180 V^2 V_{\text{(1)}}^2 V_{\text{(2)}}^2\nonumber \\[2mm] 
& - & 57200 V^3V_{\text{(2)}}^3 + 52624 V_{\text{(1)}}^2 V_{\text{(2)}}^3 +30030 V V_{\text{(2)}}^4 +12870 V^4 V_{\text{(3)}}^2 - 25740 V V_{\text{(1)}}^2 V_{\text{(3)}}^2\nonumber \\[2mm] 
& - & 42900 V^2 V_{\text{(2)}}^{\text{}} V_{\text{(3)}}^2 +16406 V_{\text{(2)}}^2V_{\text{(3)}}^2 +5096 V_{\text{(1)}}^{\text{}} V_{\text{(3)}}^3 +2860V^3 V_{\text{(4)}}^2 - 1430V_{\text{(1)}}^2 V_{\text{(4)}}^2 \nonumber\\[2mm]
& - & 5460 V V_{\text{(2)}}^{\text{}}V_{\text{(4)}}^2 +252 V_{\text{(4)}}^3 +390 V^2 V_{\text{(5)}}^2 -280 V_{\text{(2)}}^{\text{}} V_{\text{(5)}}^2 + 30VV_{\text{(6)}}^2 +V_{\text{(7)}}^2 \,, \\[2mm]
\tilde{\kappa}^{}_{19} & \simeq & -4862 V^{10} -291720 V^7 V_{\text{(1)}}^2 +850850 V^4V_{\text{(1)}}^4 -340340 V V_{\text{(1)}}^6 -204204 V^6V_{\text{(2)}}^2 \nonumber \\[2mm]
& + & 2042040 V^3V_{\text{(1)}}^2 V_{\text{(2)}}^2  - 510510 V_{\text{(1)}}^4 V_{\text{(2)}}^2 +486200 V^4 V_{\text{(2)}}^3 -1789216 V V_{\text{(1)}}^2V_{\text{(2)}}^3 \nonumber\\[2mm]
& - & 510510 V^2V_{\text{(2)}}^4 +101660 V_{\text{(2)}}^5 -87516 V^5 V_{\text{(3)}}^2 + 437580 V^2 V_{\text{(1)}}^2V_{\text{(3)}}^2 + 486200 V^3V_{\text{(2)}}^{\text{}} V_{\text{(3)}}^2\nonumber \\
& - & 418132 V_{\text{(2)}}^{\text{}} V_{\text{(1)}}^2 V_{\text{(3)}}^2 -557804 V V_{\text{(2)}}^2
   V_{\text{(3)}}^2 -173264 V V_{\text{(1)}}^{\text{}} V_{\text{(3)}}^3 +16099 V_{\text{(3)}}^4 \nonumber \\ 
& - & 24310V^4 V_{\text{(4)}}^2 +48620 V V_{\text{(1)}}^2 V_{\text{(4)}}^2 +92820 V^2
   V_{\text{(2)}}^{\text{}} V_{\text{(4)}}^2 -38726 V_{\text{(2)}}^2 V_{\text{(4)}}^2 -32368V_{\text{(1)}}^{\text{}} V_{\text{(3)}}^{\text{}} V_{\text{(4)}}^2  \nonumber \\
& - & 8568 V V_{\text{(4)}}^3 - 4420 V^3 V_{\text{(5)}}^2 +2210 V_{\text{(1)}}^2 V_{\text{(5)}}^2 +9520 V V_{\text{(2)}}^{\text{}}V_{\text{(5)}}^2 -1428
   V_{\text{(4)}}^{\text{}} V_{\text{(5)}}^2 -510 V^2V_{\text{(6)}}^2 \nonumber \\[2mm] 
& + & 408 V_{\text{(2)}}^{\text{}} V_{\text{(6)}}^2-34 V V_{\text{(7)}}^2-V_{\text{(8)}}^2 \,. 
\end{eqnarray}
%

\section{Coefficients of the EFT Potentials} 
\label{App:EFT-coefficients}

The coefficients in the sums that defined the EFT-modified potentials in Eq.~\eqref{VEFT-odd-even} read as follows~\cite{Silva:2024ffz} for the odd-parity sector:
\begin{eqnarray}
&& v^{\rm odd}_1 = -\frac{5}{8}\ell\left(\ell+1 \right)  \,, \qquad
v^{\rm odd}_2 = -\frac{5}{4}\left(\ell^2 +\ell -3 \right) \,, \qquad
v^{\rm odd}_3 = -\frac{5}{2}\left(\ell^2 +\ell -3 \right) \,, \\[2mm]
&& v^{\rm odd}_4 = -5\left(\ell^2 +\ell -3 \right) \,, \quad
v^{\rm odd}_5 = 1430 \ell\left(\ell+1 \right)  -8610\,, \\[2mm]
&& v^{\rm odd}_6 = 41460 -3332\ell\left(\ell+1 \right)  \,,   \quad
v^{\rm odd}_7 = -48192\,.
\end{eqnarray}
and for even-parity sector, with $L= \frac{1}{2} (\ell-1) (\ell+2) $:
\begin{eqnarray}
v^{\rm even}_1 &=& -5 L^2 (L+1) \,, \\[2mm]
v^{\rm even}_2 &=& -5 L^2 (2 L+5)\,,  \\[2mm]
v^{\rm even}_3 &=&  -\frac{5}{2} L \left(4 L^2+10 L+9\right)\,, \\[2mm]
v^{\rm even}_4 &=&  -5 \left(8 L^3+20 L^2+18 L+9\right)\,, \\[2mm]
v^{\rm even}_5 &=& -10 \left(\frac{576 \left(\ell^2+\ell-6\right) L^3}{\lambda (r)}+8 L^3+20 L^2 + 18 L+9\right)\,,
\\[2mm]
v^{\rm even}_6 &=& 4\left(\frac{108 \left(15 (\ell+1)^2 \ell^2-336 (\ell+1) \ell+836\right) L^2}{\lambda (r)} + 176 L^3+116 L^2-90 L-45\right), ~~~~ \\[2mm]
v^{\rm even}_7 & = &  24 \left(\frac{30L}{\lambda (r)} \left(147 (\ell+1)^2 \ell^2-1304 (\ell+1) \ell + 2164\right) + 88 L^2-30 L-15\right) \,, \\[2mm]
v^{\rm even}_8 & = & \!144\! \left(\frac{6 L}{\lambda (r)}\! (1073 \ell (\ell+1)-3988) +44 L-5\right)\,, \\[2mm]
v^{\rm even}_9 & = & \frac{2}{\lambda (r)}\left(778608 \ell (\ell+1)-1938240\right)+6336 \,, \\[2mm]
v^{\rm even}_{10} &=& \frac{1759104}{\lambda (r)} \,.
\end{eqnarray}
%

\bibliographystyle{JHEP}

\providecommand{\href}[2]{#2}\begingroup\raggedright\endgroup

\end{document}